\documentclass[fleqn,10pt]{wlscirep}
\usepackage[utf8]{inputenc}
\usepackage[T1]{fontenc}
\usepackage{cite}
\usepackage{acronym}
\usepackage[caption=false,font=footnotesize]{subfig}
\usepackage{wrapfig}
\usepackage{multirow}
\usepackage{tabularx}
\newcolumntype{Y}{>{\centering\arraybackslash}X}
\newcolumntype{Z}{>{\raggedleft\arraybackslash}X}
\usepackage{booktabs}
\usepackage{threeparttable}
\usepackage[binary-units=true]{siunitx}
\DeclareSIUnit\nmi{nmi}

\renewcommand{\textcolor}[2]{#2}

\newcommand{\rev}[1]{\textcolor{magenta}{#1}}

\title{COVID-19 Impact on Global Maritime Mobility}

\author[1,+]{Leonardo~M.~Millefiori}
\author[1,+,*]{Paolo~Braca}
\author[2,3]{Dimitris~Zissis}
\author[3]{Giannis~Spiliopoulos}
\author[4]{Stefano~Marano}
\author[5]{Peter~K.~Willett}
\author[1]{Sandro~Carniel}

\affil[1]{NATO STO Centre for Maritime Research and Experimentation, Research Department, La Spezia, 19126, Italy}
\affil[2]{University of the Aegean, Department of Product and Systems Design Engineering,  Syros, 84100, Greece}
\affil[3]{MarineTraffic, Athens, 115 25, Greece}
\affil[4]{University of Salerno, Dipartimento di Ingegneria dell'Informazione ed Elettrica e Matematica Applicata (DIEM), Fisciano (SA), 84084, Italy}
\affil[4]{University of Connecticut, Department of Electrical and Computer Engineering, Storrs, 06269, USA}

\affil[*]{paolo.braca@cmre.nato.int}
\affil[+]{these authors contributed equally to this work}

\newacro{VLCC}{Very Large Crude Carrier}
\newacro{ULCC}{Ultra Large Crude Carrier}
\newacro{Eurostat}{European Statistical Office}
\newacro{LSCI}{Liner Shipping Connectivity Index}
\newacro{BDI}{Baltic Dry Index} 
\newacro{PVs}{Port Visits}
\newacro{TDM}{Traffic Density Map}
\newacro{TDMs}{Traffic Density Maps}
\newacro{ACT}{NATO Allied Command Transformation}
\newacro{AIS}{Automatic Identification System}
\newacro{AMS}{Alliance Maritime Strategy}
\newacro{CMRE}{Centre for Maritime Research and Experimentation}
\newacro{COG}{Course Over Ground}
\newacro{CU}{Covariance Union}
\newacro{DKOE}{Data Knowledge Operational Effectiveness}
\newacro{GLRT}{Generalized Likelihood Ratio Test}
\newacro{GPS}{Global Positioning System}
\newacro{GT}{gross tonnage}
\newacro{IMO}{International Maritime Organization}
\newacro{IOU}{Integrated Ornstein-Uhlenbeck}
\newacro{KF}{Kalman Filter}
\newacro{LLR}{Log-Likelihood Ratio}
\newacro{LS}{Least Squares}
\newacro{ML}{Maximum Likelihood}
\newacro{MM}{Method of Moments}
\newacro{MSA}{Maritime Situational Awareness}
\newacro{MSSIS}{Maritime Safety and Security Information System}
\newacro{MMSI}{Maritime Mobility Service Identity}
\newacro{NCV}{Nearly Constant Velocity}
\newacro{OOSM}{Out-of-Sequence Measurement}
\newacro{OU}{Ornstein-Uhlenbeck}
\newacro{MOU}{Mixed Ornstein-Uhlenbeck}
\newacro{PDF}{Probability Density Function}
\newacro{CDF}{Cumulative Density Function}
\newacro{PoL}{Pattern Of Life}
\newacro{ROC}{Receiver operating characteristic}
\newacro{SDE}{stochastic differential equation}
\newacro{SOG}{Speed Over Ground}
\newacro{SOLAS}{International Convention for the Safety of Life at Sea}
\newacro{STO}{NATO Science and Technology Organization}
\newacro{TREAD}{Traffic Route Extraction for Anomaly Detection}
\newacro{UTM}{Universal Transverse Mercator}
\newacro{VS-IMM}{Variable Structure Interactive Multiple Model}
\newacro{SD}{Standard Deviation}
\newacro{SaR}{Search and Rescue}
\newacro{SAR}{Synthetic Aperture Radar}
\newacro{CRLB}{Cram\'{e}r-Rao Lower Bound}
\newacro{TPR}{Technical Progress Report}
\newacro{FIM}{Fisher Information Matrix}
\newacro{SME}{Sample Mean Estimator}
\newacro{MSE}{Mean Squared Error}
\newacro{WHO}{World Health Organization}
\newacro{DMA}{Danish Maritime Authority}
\newacro{TEU}{Twenty-foot Equivalent Unit}
\newacro{DWT}{deadweight tonnage}
\newacro{CNM}{Cumulative Navigated Miles}
\newacro{UNCTAD}{United Nations Conference on Trade and Development}
\newacro{ULCV}{Ultra Large Container Vessel}
\newacro{nmi}{nautical miles}
\newacro{GHG}{greenhouse gas}
\newacro{MEPC}{Marine Environment Protection Committee}

\newacroplural{PoL}[PoL]{Patterns Of Life}

\begin{abstract}
To prevent the outbreak of the Coronavirus disease (COVID-19), \rev{many} countries  around the world went into lockdown and imposed unprecedented containment measures. These restrictions progressively produced changes to social behavior and global mobility patterns, evidently disrupting social and economic activities.
Here, using maritime traffic data collected via a global network of \ac{AIS} receivers, we analyze the effects that the COVID-19 pandemic and containment measures had on the shipping industry, which accounts alone for more than \SI{80}{\percent} of the world trade. We \rev{rely on multiple data-driven maritime mobility indexes} to quantitatively assess ship mobility in a given unit of time. \rev{The mobility analysis here presented} has a worldwide extent and is based on the computation of\rev{:} \ac{CNM} of all ships reporting their position and navigational status via \ac{AIS}, \rev{number of \emph{active} and \emph{idle} ships, and fleet average speed. To highlight significant changes in shipping routes and operational patterns, we also compute and compare global and local vessel density maps}. 
We compare 2020 mobility levels to those of previous years assuming that an unchanged growth rate would have been achieved, if not for COVID-19. Following the outbreak, we find an unprecedented drop in maritime mobility, across all categories of commercial shipping. 
\rev{With few exceptions, a generally} reduced activity is observable from March to June, when the most severe restrictions were in force. \rev{We quantify} a variation of mobility between \SI{-5.62}{\percent} and \SI{-13.77}{\percent} for container ships, between \rev{\SI[retain-explicit-plus]{+2.28}{\percent}} and \SI{-3.32}{\percent} for dry bulk, between \SI{-0.22}{\percent} and \SI{-9.27}{\percent} for wet bulk, and between \SI{-19.57}{\percent} and \SI{-42.77}{\percent} for passenger \rev{traffic}.
The \rev{presented study} is unprecedented for the uniqueness and completeness of the employed \ac{AIS} dataset, which comprises a trillion \ac{AIS} messages broadcast worldwide by \num{50000} ships, a figure that closely parallels the documented size of the world merchant fleet.

\end{abstract}

\begin{document}
	
\flushbottom
\maketitle
%
%
\thispagestyle{empty}

\section*{Introduction}

The coronavirus (COVID-19) pandemic has recently produced one of the worst global crises since World War II. 
\rev{As of January 15, 2021, almost 100~million} people have been infected worldwide, and over 
\rev{2~million} have passed away due to the disease. Consequences of the outbreak are impacting broadly all aspects of our society. In February 2020, the \ac{WHO} recommended containment and suppression measures to slow down the spread of the virus.~\cite{Anderson2020,Hellewell2020} Towards this direction and aimed at ``flattening the curve'' of infections so as to avoid overwhelming healthcare systems, many countries implemented unprecedented confinement measures, ranging from bans to travel and social gatherings, to the closure of many commercial activities. 
Evidence that the lockdown measures achieved a reduction of the rate of new infections \rev{is} gradually appearing in the scientific literature.~\cite{DehningScience2020,AdaptiveBayesian}

\begin{figure}
    \centering%
    \subfloat[][]{%
        \includegraphics[trim=0 0 30 30,clip,width=0.32\textwidth]{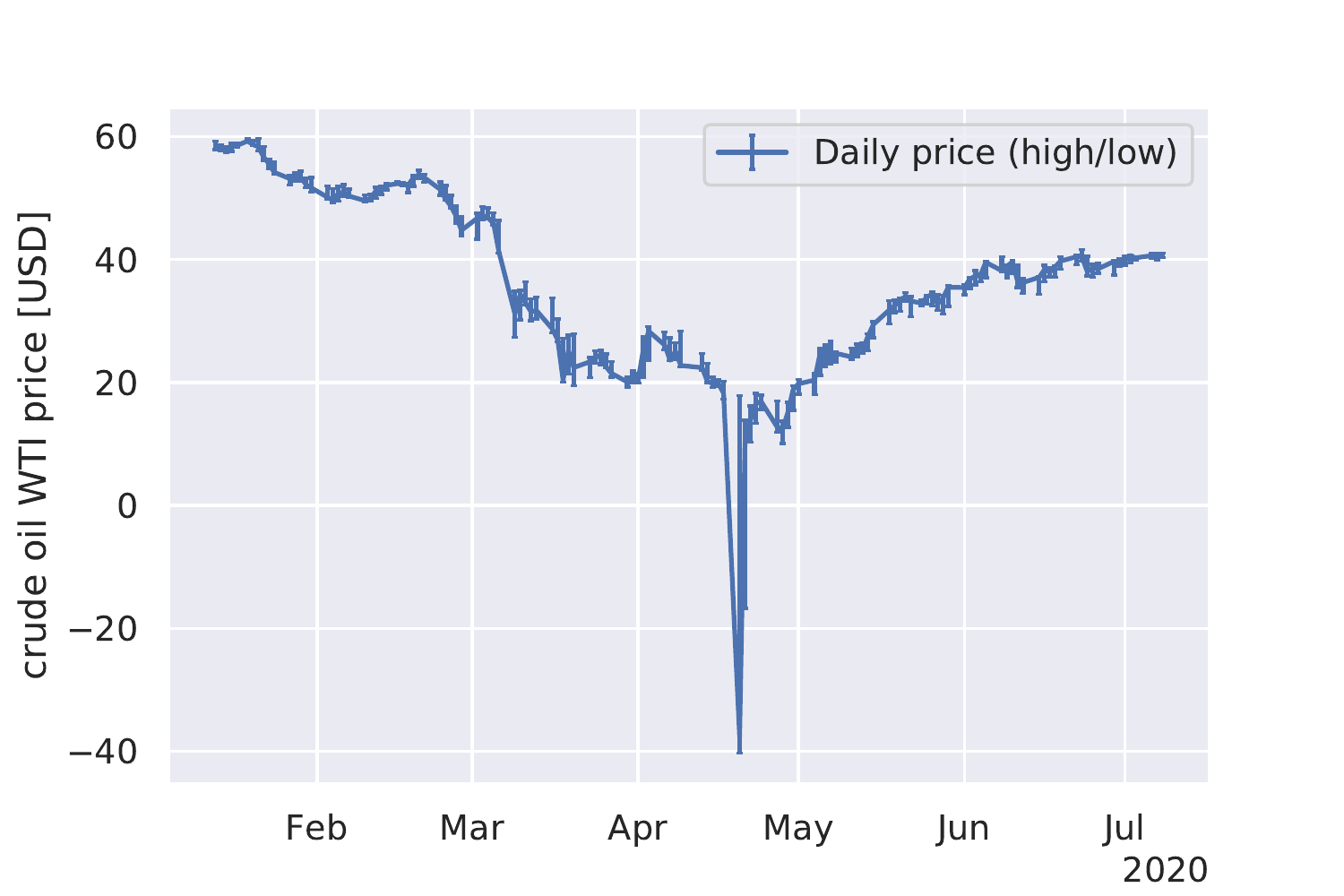}%
        \label{fig:Oil}}%
    \hfil %
    \subfloat[][]{%
        \includegraphics[trim=0 0 30 30,clip,width=0.32\textwidth]{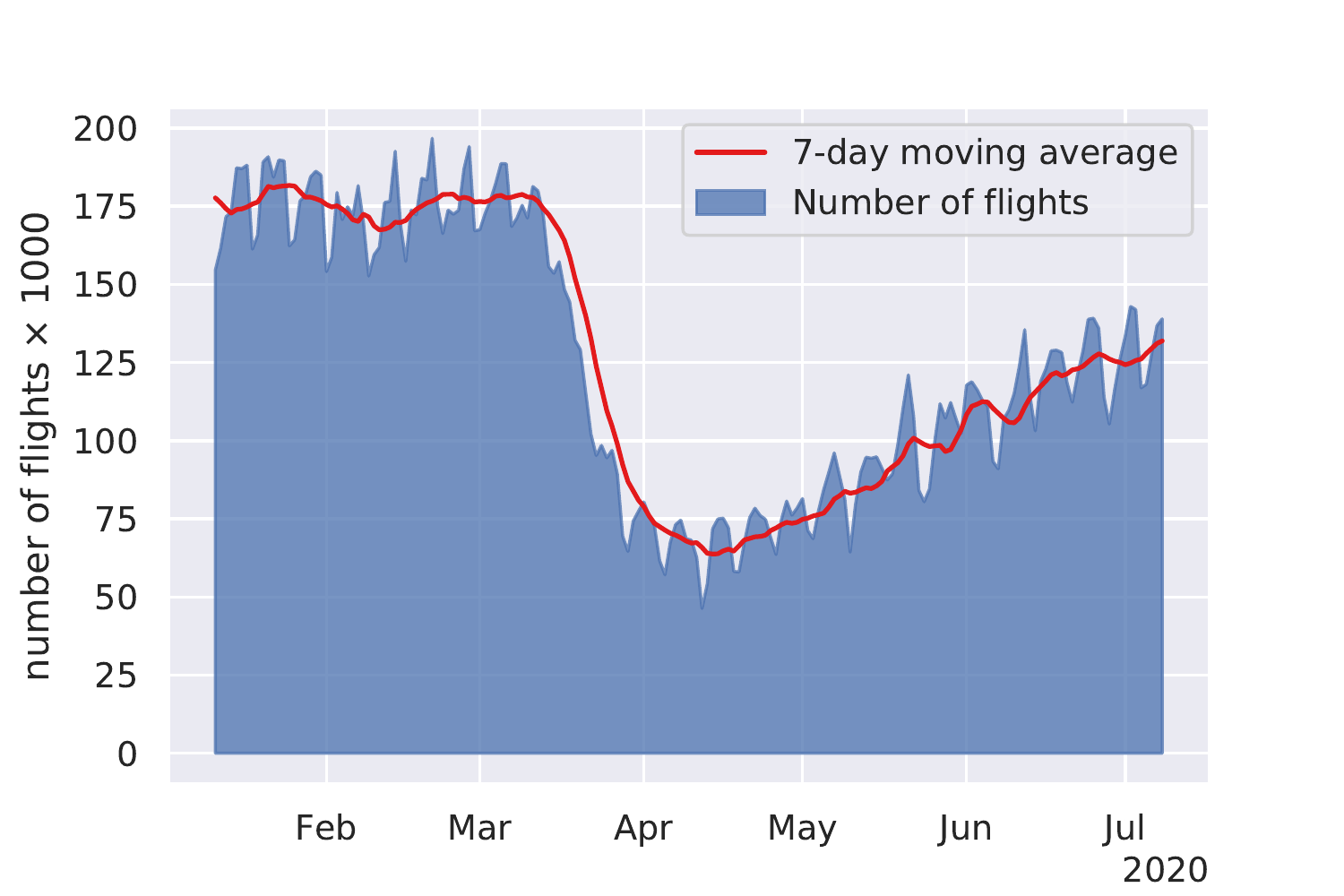}%
        \label{fig:flights}}%
    \hfil %
    \subfloat[][]{%
        \includegraphics[trim=0 0 30 32,clip,width=0.32\textwidth]{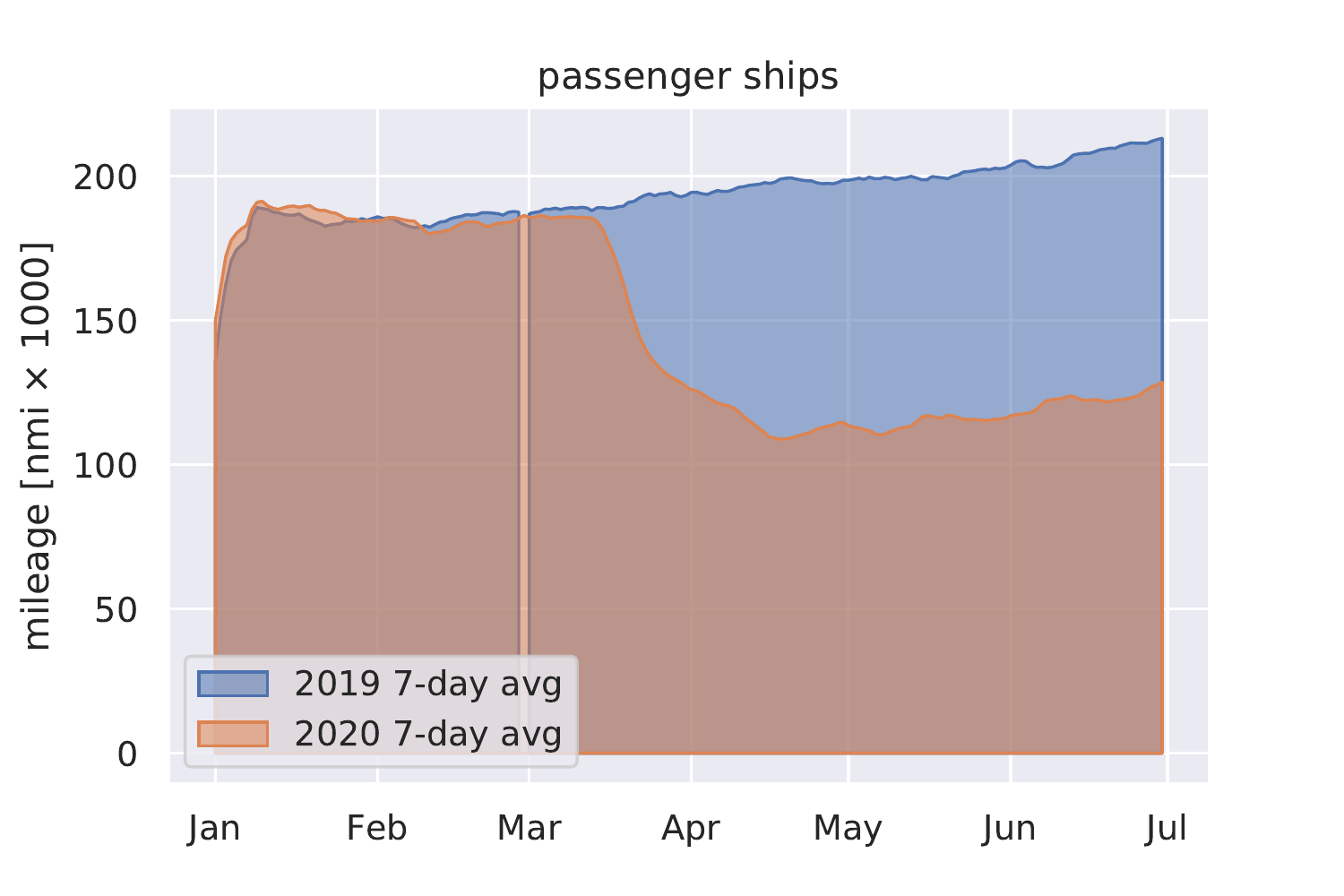}%
    \label{fig:dailymileagearea-mt-intro}}%
    \caption{(a)~Crude oil WTI price from January to July 2020.~
        \cite{Oilprice} Note the sudden and unprecedented negative price in April 2020. (b)~Total number of flights tracked by \emph{FlightRadar24} from January to July 2020.~\cite{FlightRadar} At the end of March there was an abrupt decrease of the number of flights ($\sim$ \num{100000} units) because of the lockdown restrictions. (c)~Daily navigated miles by passenger ships in 2020 compared with 2019.}%
\end{figure}

Many of the aforementioned restrictions are in contradistinction to ``normal'' routines. At a time when we are asked to come together and support one another in society, we must learn to do so from a distance. But the behavior changes have been deemed necessary, and some may provide useful insights regarding how we can facilitate \rev{transformation} toward more sustainable supply and production.~\cite{BraveNewWorld} The hope is that the macroeconomic system, global supply chains, and international trade relations will not revert back to ``normal'' and ``business-as-usual,'' and will allow the emergence and successful adoption of new types of economic development and governance models.~\cite{BraveNewWorld}

On the other hand, both the outbreak and the restrictions are revealing the fragility of the global economy, sparking fears of impending economic crisis and recession.~\cite{Maria2020} Social distancing, self-isolation and travel restrictions have led to workforce reductions across all economic sectors. Schools have closed down, and the demand for commodities and manufactured products has generally decreased. In contrast, the need for medical supplies has significantly risen. The food industry is also facing increased demand due to panic-buying and stockpiling.~\cite{Maria2020} 

\begin{wrapfigure}[33]{r}{0.5\textwidth}
\vspace{-2.75em}
\centering%
\includegraphics[width=0.48\textwidth]{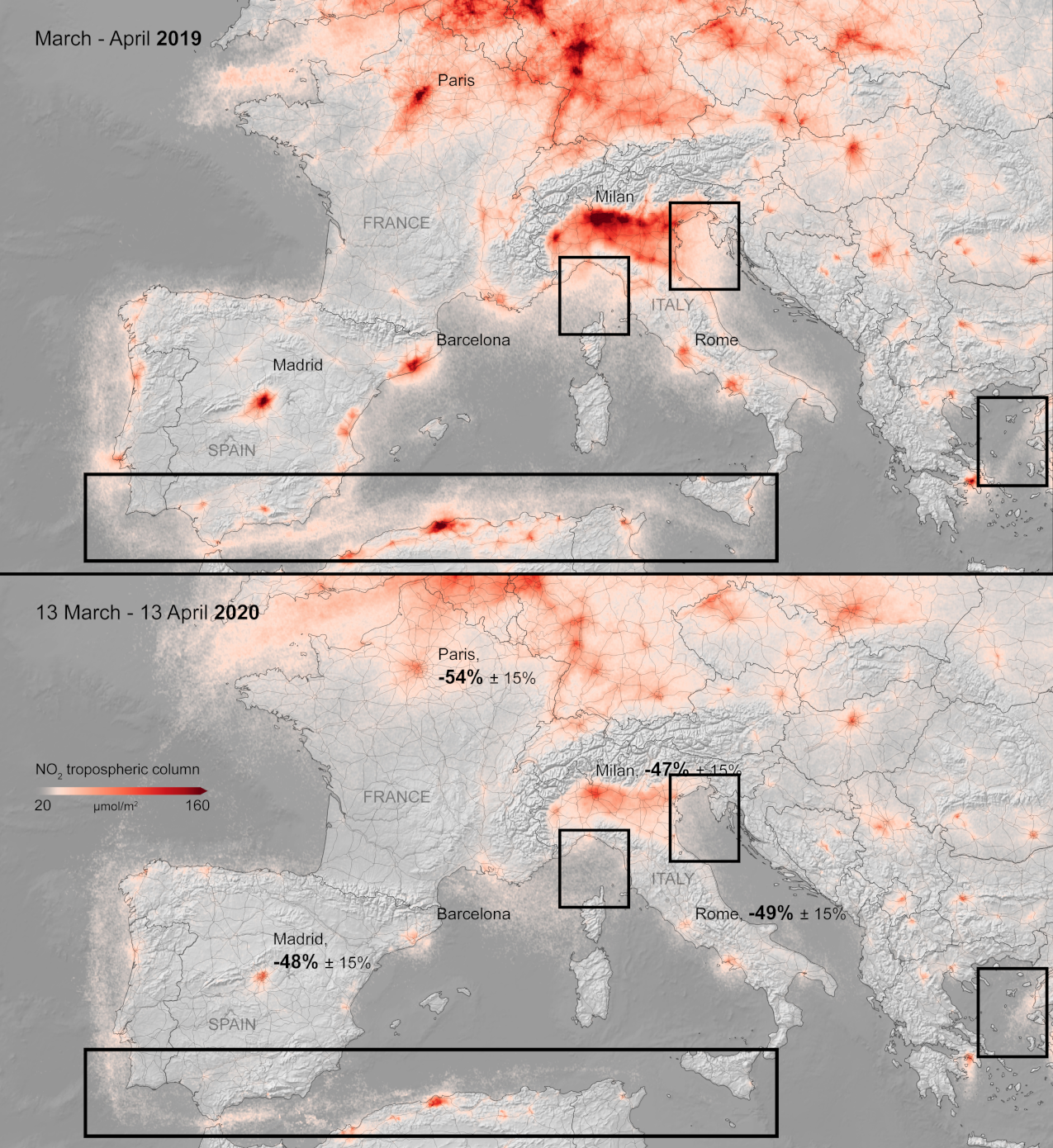}%
\caption{Average nitrogen dioxide (NO\textsubscript{2}) concentrations from 13 March to 13 April 2020, compared to the same period in 2019. The decrease of emission is evident, around \SI{-50}{\percent} in large European cities (Rome, Paris, Madrid and Milan). Highlighted with boxes there are regions of sea where it is evident a decrease of nitrogen dioxide concentrations, probably due to the reduced shipping activity. \ac{WHO} air quality guideline values quantify in \SI{40}{\micro\gram\per\cubic\metre} (annual mean) the NO\textsubscript{2} limit level for human health.~\cite{WHO_AirQuality} Reproduced with permission. \textcopyright~Contains modified Copernicus Sentinel data (2019-20), processed by KNMI/ESA.~\cite{ESA_NO2}}
\label{fig:NO2}
\end{wrapfigure}

Overall, world trade is expected to fall by between \SI{13}{\percent} and \SI{32}{\percent} in 2020 as the COVID-19 pandemic disrupts normal economic activity and life \rev{globally}.~\cite{WTO_2020} Business activity across the eurozone collapsed to a record low in March 2020, and US industrial production showed the biggest monthly decline since the end of World War II.~\cite{RAPACCINI2020} 
An example of the unprecedented financial changes is the price of oil dropping below zero due to expiry of delivery contracts and limited storage capacity to receive them, for the first time in history in April 2020,~\cite{REUTERS,Oilprice} as reported in Fig.~\ref{fig:Oil}. The connection between the recent spread of COVID-19, oil price volatility, the stock market, geopolitical risk and economic policy uncertainty in the US is studied by Sharif \textit{et al.}~\cite{SHARIF2020} 
Guan \emph{et al.}~\cite{GlobalSupplyChain} analyzed the supply-chain effects of a set of idealized lockdown scenarios, using the latest global trade modeling framework. Given that lockdowns be necessary, the authors demonstrate that they best occur early, \rev{and be} strict and short in order to minimize overall losses. 


Useful \rev{insight} related to the global supply chain can be gained by observing the impact of lockdowns on the mobility of goods. Indeed, in today's increasingly globalized economy, transport plays a central and critical role as the primary enabler of the flow of freight within and across borders. In late March 2020, several sources reported a dramatic decline in air traffic: as governments put travel restrictions in place, airlines halted flights and commercial air traffic quickly dropped significantly below 2019 levels, see Fig.~\ref{fig:flights}. However, as it accounts only for a small part of the global freight transport, a decline in air traffic, as significant as it may be, would not necessarily imply reduced goods mobility. 

With over \SI{80}{\percent} of global trade by volume and \rev{up to} \SI{70}{\percent} of its value being carried on board ships and handled through seaports worldwide, maritime transport for trade and development is of paramount importance.~\cite{unctad_2017,NATO_AMS_2011,MillefioriBracaTAES2016} Shipping can be viewed as a barometer for the global economic climate. \rev{According to the~\ac{UNCTAD}, in 2018 total volumes are estimated to have reached 11~billion tons, an all-time high.~\cite{unctad_2019}} \ac{UNCTAD} originally projected an annual average growth rate of \SI{3.4}{\percent} for the period 2019–-2024.~\cite{unctad_2019} However, this estimated growth will possibly need to be revised,~\cite{WTO_2020} as the coronavirus pandemic led to a \SI{3}{\percent} drop in global trade values in the first quarter of 2020. The downturn is expected to accelerate in the second quarter, according to \ac{UNCTAD} forecasts,~\cite{unctad_2020} which project a quarter-on-quarter decline of \SI{27}{\percent}. \rev{For the full year, \ac{UNCTAD} expects a
drop of \SI{20}{\percent}. The World Bank further noted that merchandise trade appeared to have bottomed out in April, falling nearly \SI{20}{\percent} year over year, after a \SI{10}{\percent} decline in March. The
trade contraction caused by COVID-19 is deeper than the one observed during the financial crisis of 2008-2009.~\cite{unctad_2020b}}

Similar to commercial aviation, the maritime tourism industry was the first and \rev{most-affected} traffic segment, with cases of COVID-19 among cruise ships passengers and crew members reported all around the world, from Yokohama (Japan), to Corfu (Greece) and Sydney (Australia).~\cite{moriarty2020public,DEPELLEGRIN2020140123} The effects on this market segment might also be more enduring than in other sectors, as psychological effects might come into play in addition to restrictive measures, with passengers being less inclined to travel on large crowded ships. In the second half of March 2020, most European cruise terminals partially or in some cases completely suspended operations~\cite{DEPELLEGRIN2020140123} (e.g., 19 March in Italy, 20 March in Croatia, 25 March in Spain). In order to limit and slow the spread of the infection, many seaports closed down, limiting---and sometimes banning---cruise traffic at their terminals. National and local restrictions concerning ship operations were enforced, often leading to delayed port clearance. Limitations included crew embarking and disembarking, cargo discharge and loading, imposition of quarantine, and eventually refusal of port entry and refueling. %
Other measures followed in other maritime sectors and port activities with the aim to ensure safety at terminals and associated logistic facilities of stevedores and other personnel.~\cite{EU_Fishery} The fishing and aquaculture sectors were also affected by containment measures, leading to, e.g., voluntary fishing cessation and suspension or reduction of fish farming, with evident effects on the supply chain of fish food products.~\cite{EU_Fishery}

Global maritime mobility reductions not only affect global trade and the economy, but also the environment: especially sea pollution~\cite{viatte2020air,meletearth} and incursions by invasive species~\cite{sardain2019global} are heavily influenced by ship activities; in a recent \ac{IMO} report~\cite{IMO_GHG4} that has been submitted to the \ac{MEPC}, \ac{GHG} emissions from shipping---expressed in carbon dioxide equivalent (CO\textsubscript{2}e)---increased increased \SI{9.6}{\percent} in 2018 with respect to 2012, and accounted for the \SI{2.89}{\percent} of global anthropogenic emissions, with container and bulk shipping being accountable for most of the total emissions. The nexus between COVID-19 and the environment already attracted enormous attention within the scientific community, and several works are already available that analyze the effects of the pandemic in four main areas:~\cite{SHAKIL2020} (1) environmental degradation, (2) air pollution, (3) climate/metrological factors and (4) temperature. An exemplary environmental consequence of lockdowns is that pollution levels dropped significantly; for instance, greenhouse gas emissions, nitrogen dioxide, black carbon and water pollution decreased drastically.~\cite{SHAKIL2020} 
In Fig.~\ref{fig:NO2} we report, using data from the Copernicus Sentinel-5P satellite,~\cite{ESA_NO2} the average nitrogen dioxide concentrations in Europe from 13 March to 13 April 2020, compared with the same period in 2019. The decrease of pollutants is evident, around \SI{-50}{\percent} in large European cities (Rome, Paris, Madrid and Milan). In the same figure, we have also highlighted with boxes sea regions where the decrease of nitrogen dioxide concentrations is noticeable and could be also due (even if only partially) by a decreased shipping activity. Indeed, it is interesting to observe that just along the first part of one of the main sea lanes in the Mediterranean Sea (Gibraltar-Suez), pollution in 2020 reduced with respect to 2019 levels. Recent work by Faber \textit{et al.}~\cite{Faber} suggests how emissions from ships could be reduced if they reduced their speed; our analysis of \ac{AIS} data shows that in all highlighted areas, on average, ships reduced their speed in March-April 2020 with respect to the same months in 2019. Specifically, in the highlighted regions of the Gibraltar-Suez route, Ligurian Sea, Northern Adriatic Sea, and Aegean Sea, we report average fleet speed variations of \SI{-5.1}{\percent}, \SI{-15.3}{\percent}, \SI{-6.0}{\percent} and \SI{-9.5}{\percent}, respectively. 

\begin{figure}
    \centering%
    \includegraphics[trim=0 140 0 0,clip,width=\textwidth]{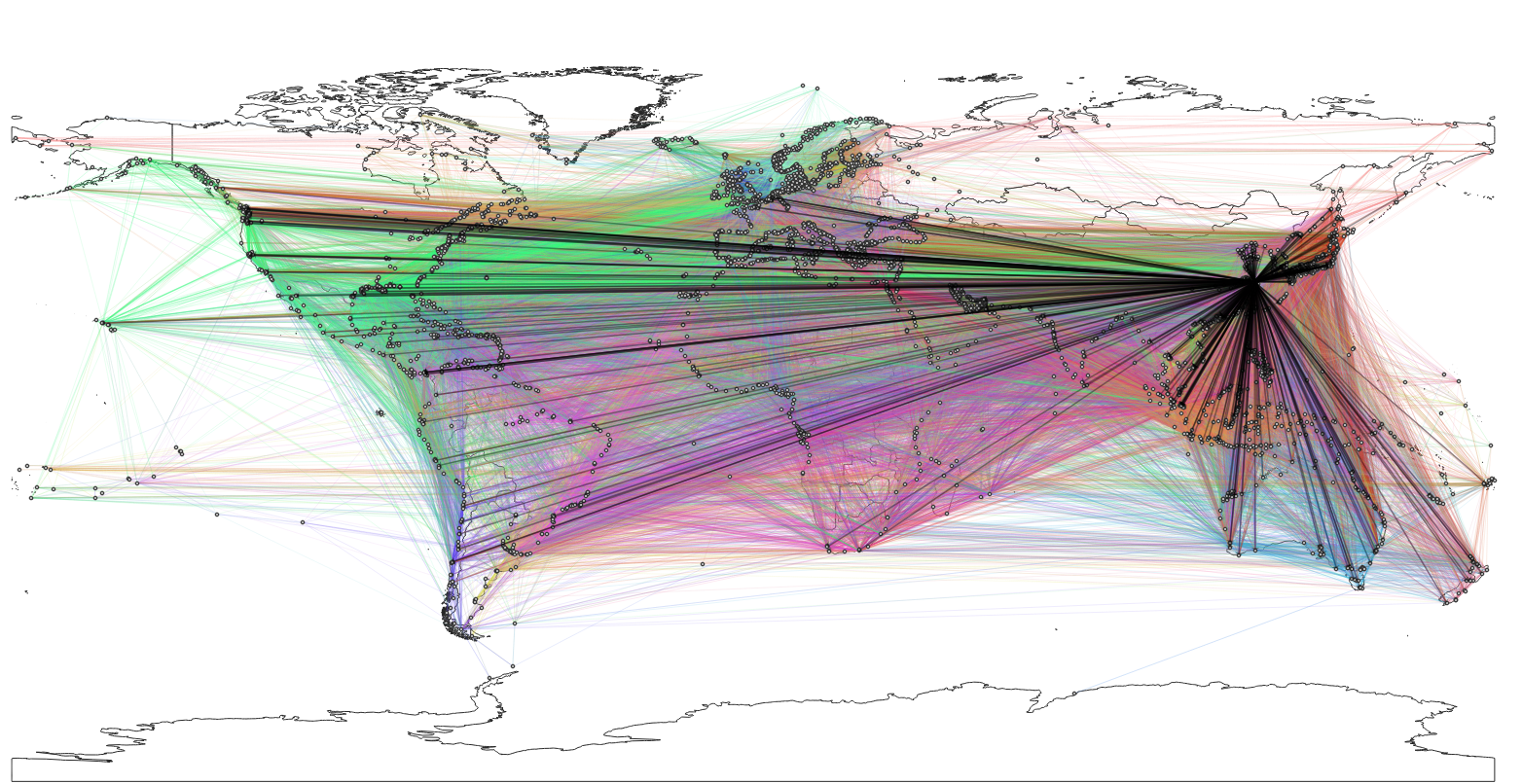}%
    \caption{Representation of Shanghai's ego network and its 3-step neighborhood: each port in the picture, \rev{represented by a circle on the coastline}, is reachable from Shanghai with at most three hops. An ego network~\cite{hanneman2005introduction} consists of a focal node (the \emph{ego}) and the nodes (the \emph{alters}) that are connected to it, either directly or within a fixed number of steps. \rev{The edge color is representative of the country of the port of origin.}}
    \label{fig:Shangai}
\end{figure}

The aim of this study is to analyze the short-term effects that the COVID-19 pandemic and containment measures had on global shipping \rev{mobility}. 
\rev{While a reduction of shipping activities is \emph{qualitatively} expected from several factors, e.g., from shipbuilders' reduced capacity and new ship building orders about \SI{75}{\percent} down,~\cite{Stopford_ThreeScenarios} we specifically aim at providing a \emph{quantitative} evidence-based assessment of such reduction. For this reason, we opted for purely data-driven indicators that could measure exhaustively the impact of the lockdown and containment measures on maritime traffic directly, also in ways that could not have been easily anticipated.} 
%
The analysis reported in this paper shows the combined effects of the coronavirus disease and the containment measures, in addition to the reported trade \rev{contraction,\cite{WTO_2020,unctad_2020,unctad_2020b}} had on the global maritime traffic. Maritime traffic reflects these effects and as such shows, for the first time and in specific sectors more than others, signs of slowing down, with possible negative future consequences on the entire global supply chain.




We propose the spatio-temporal analysis of positional \ac{AIS} messages to compute synthetic indicators capable of quantifying ship activities and highlighting changes in mobility patterns. 
The indicators are the \acf{CNM}, computed for each ship journey per category, the number of \emph{active} and \emph{idle} ships \rev{and their average speed}. \rev{To complement our analysis and highlight significant changes in shipping routes operational patterns, we also compute vessel density maps and show differences between the density of shipping traffic in 2019 and 2020}. In other words, we assess, with a \rev{thorough} data-driven approach, the global ship mobility for the traffic categories that account for the most traffic worldwide. \rev{As a matter of fact, the available statistics on maritime trade, or the related economic indicators, are not suitable to quantify exhaustively changes in shipping mobility patterns.}


 
\rev{Indeed, maritime trade statistics,~\cite{unctad_2019,unctad_2020,unctad_2020b,cerdeiro2020world} being focused on measuring the accessibility to global trade more than mobility patterns, are generally based on connectivity indexes, such as UNCTAD's \ac{LSCI}, and calculated at the country level. On the other hand, economic indicators, such as the \ac{BDI}, are meant to provide an assessment of freight cost on various (selected) routes. Even if they are both driven by the shipping mobility, none of them is actually suitable to assess the mobility of vessels globally. Given today's ubiquity of \ac{AIS}, both maritime trade and economic indexes are computed \textit{also} with the support of \ac{AIS}; however, the aim is to measure different entities than shipping mobility, and they cannot be computed from \ac{AIS} alone. For instance, one of the components of the \ac{LSCI} is the number of scheduled port calls, or \ac{PVs}, which can be computed automatically from \ac{AIS} data. However, the automatic computation of \ac{PVs} from \ac{AIS} data may cause inaccuracy. As an example, the \ac{PVs} may not get recorded because of low \ac{AIS} coverage (or because the \ac{AIS} device is switched off), or because the visited port is too small to appear in the ports' database. Care needs to be taken to ensure that all \ac{PVs} are captured in the data, and that all \ac{PVs} relate to actual operations and not, for instance, maintenance in a shipyard or refueling. It is noteworthy to mention Singapore, which will often appear as a destination for bulk carriers, even though there are no dry bulk terminals in the city state.} 
 
\rev{Given such limitations of \ac{AIS}-derived \ac{PVs}, which focus only on the end points of a voyage, we propose data-driven indicators that can look at a complete voyage and do not introduce additional ``uncertainty'' in the related calculations. Indeed, the \ac{CNM} indicator (which is highly correlated to \ac{PVs}, as we show in the Methods section) does not suffer from such limitations, as the ship trajectory can almost always be reconstructed in the data processing phase, even if pieces of it are missing.}

\rev{Moreover,} it is intuitive to acknowledge that there is a close relationship between mobility (e.g., the number of active or idle ships) and trade volumes; \rev{as such, shipping mobility, too, can be understood as a proxy for economic activity}. Yet, even today most statistics and economic forecasting indices focus only on the starting and finish lines of the supply chain, with little consideration of how goods arrive at their final destination. A more detailed look reveals a complex and dynamic network of ships and their cargo in constant motion across the world’s oceans. In this sense, maritime traffic can provide insights into the global supply and demand trends; thus considered as an indicator of future economic growth.



The results reported in this paper are based on a global dataset containing approximately a trillion \ac{AIS} messages collected between 2016 and 2020 indicating the movement of more than \num{50000} commercial ships across the globe, stored in a big data infrastructure of \SI{55}~TB. To give an idea of both the worldwide coverage of \ac{AIS} data and the capillarity of sea-routes network, we proffer in Fig.~\ref{fig:Shangai} the port of Shanghai's ego network~\cite{hanneman2005introduction} constructed based on \ac{AIS} data from September 2018--2019.

The global \ac{CNM} that we have computed from \ac{AIS} data indicates the scale of ship mobility. From January to June ships cumulatively travelled something around around \num{530000000}~\ac{nmi} in 2016, \num{580000000}~\ac{nmi} in 2019, and \num{575000000}~\ac{nmi} in 2020.
To aid understanding, the distance commercial shipping travelled in the first half of 2016 is comparable to travelling \num{6.5} times the mean distance between the Earth and sun, and in 2019  almost \num{7.2} times that.

Instead of increasing, \rev{as has happened} in all past years since 2016, the global ship mobility in terms of \ac{CNM}, in 2020 and for the first time, slightly decreased. The decrease in the first half of 2020, compared to the first half of 2019, amounted to a modest \SI{0.9}{\percent} and to over \SI{5}{\percent} in the period April-June 2020, compared to April-June 2019.  
The decline in the first half of 2020 is almost \SI{4}{\percent}, compared to the forecast values in 2020, and is over \SI{8}{\percent} in the period April-June. Moreover, there is great variation of this figure among traffic categories and different months; for instance, in June container ships decreased \SI{12}{\percent} compared to 2019, wet bulk ships of \SI{5}{\percent}, and passenger ships of \SI{42}{\percent}, while dry bulk ships slightly increased (\SI{1.7}{\percent}). 

\section*{Results}
    
The COVID-19 pandemic has \rev{led} to vast economic disruption across the world, \rev{and industry activity and confidence have collapsed.~\cite{RAPACCINI2020}} Overall, the shipping industry and seaborne trade followed this negative trend. In this paper, we compare global vessel mobility during the first half of 2020 to that of previous years, from 2016 to 2019, and the analysis confirms 
that shipping mobility has been negatively affected, but to different degrees in each market and depending on the size of vessels. Our results suggest that there is a substantial increase in idle ships across all types of ships/markets globally in the first six months of 2020, and a substantial decline in the vessel mobility measured in \ac{CNM} per unit of time. Given that, prior to the pandemic, the mobility---similar to global trade---was on an increasing trend, we compare the 2020 mobility with its forecast based on the analysis of previous years. Specifically, we assume an expected growth in 2020 given by the average growth observed in the same month of a few past consecutive years. 

\begin{figure}
       \centering%
       \subfloat[][Container]{%
         \includegraphics[trim=180 160 140 150,clip,width=0.98\textwidth]{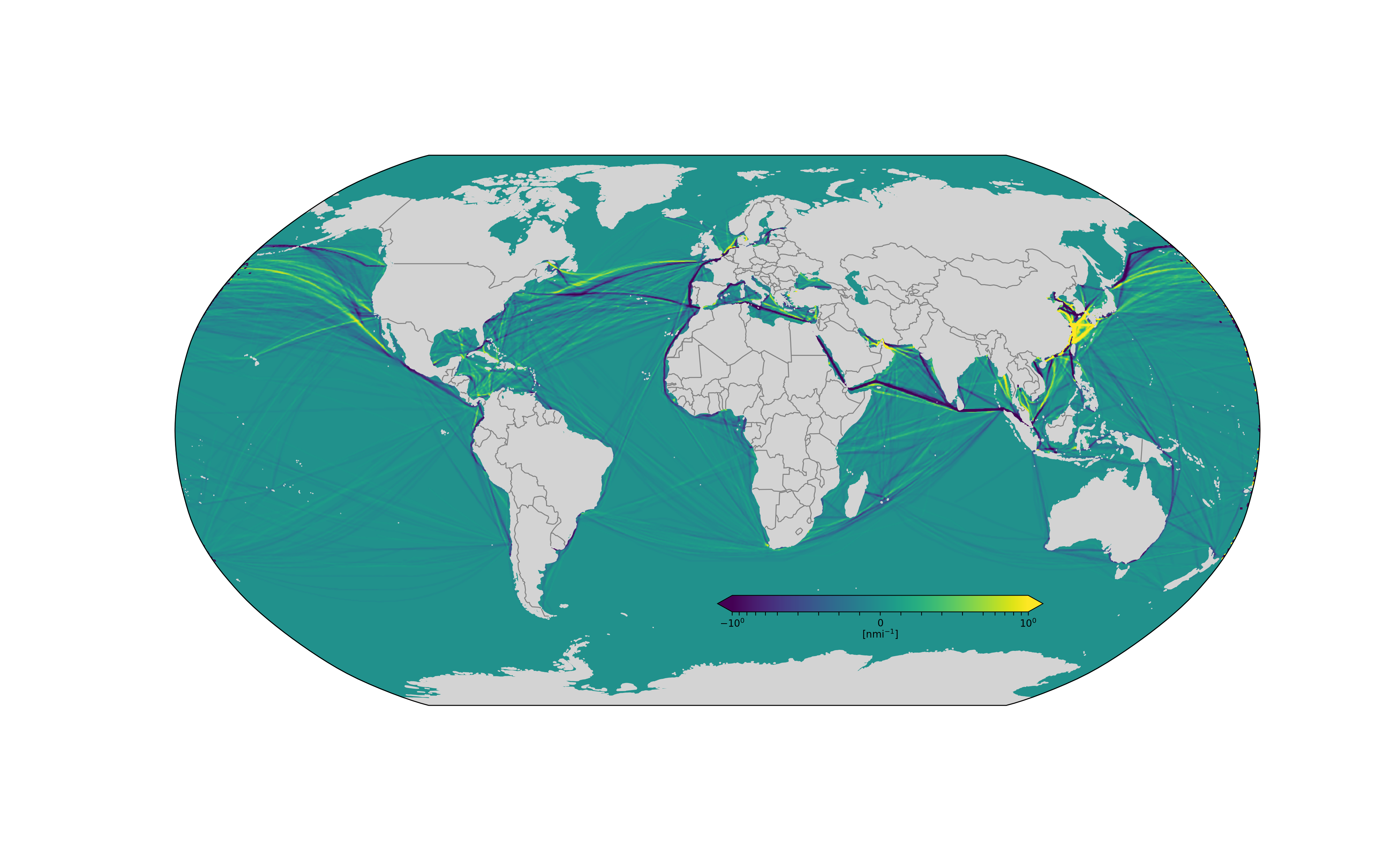}%
         \label{fig:density-map-container}%
       }%
       \\
       \subfloat[][Dry bulk]{%
         \includegraphics[trim=180 160 140 150,clip,width=0.98\textwidth]{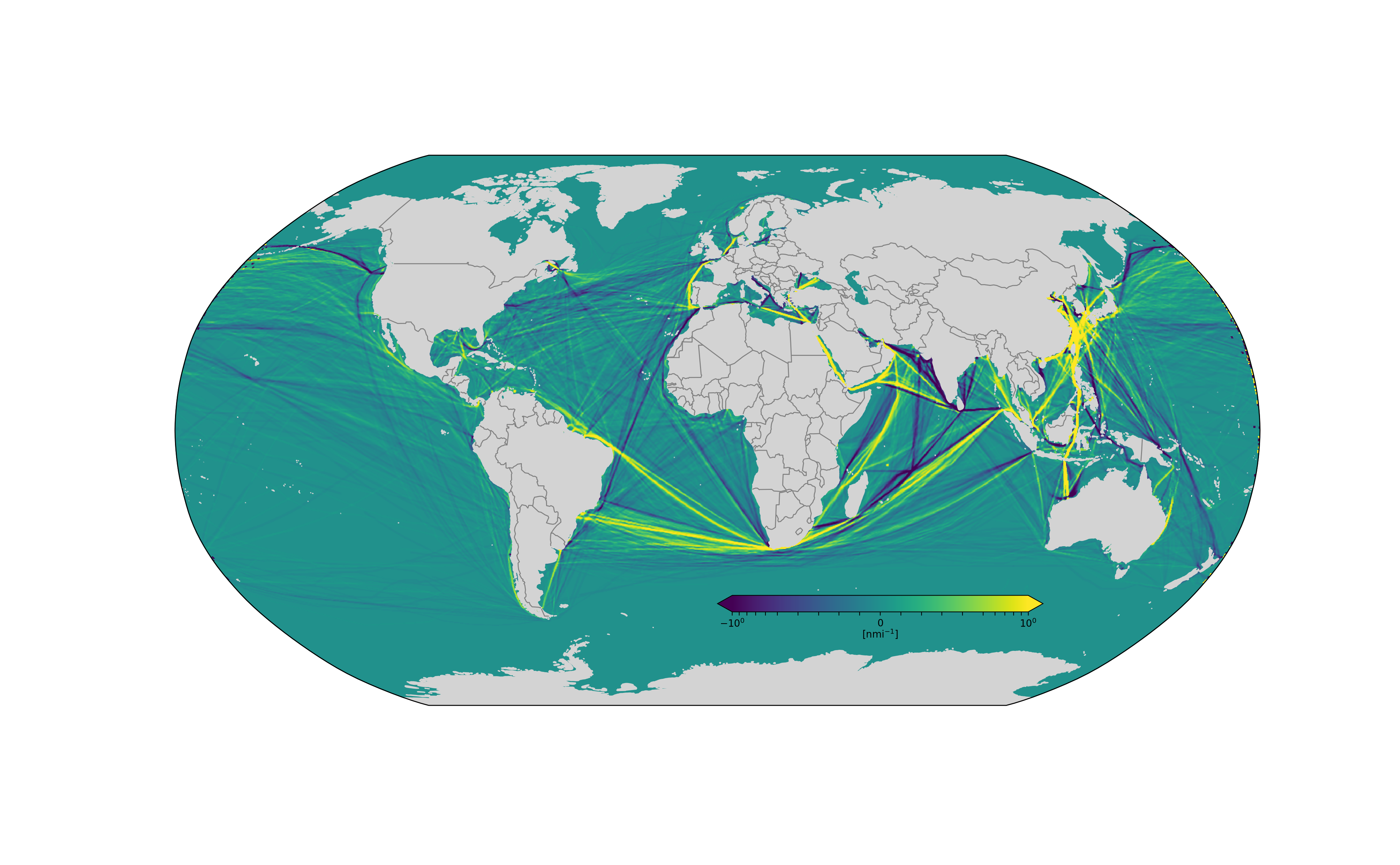}%
         \label{fig:density-map-dry-bulk}%
       }%
       \caption{\rev{Monthly \ac{CNM} density difference between 2020 and 2019 for container~\protect\subref{fig:density-map-container} and dry bulk~\protect\subref{fig:density-map-dry-bulk} shipping. The considered time period is from 13 March to 13 April. Each grid cell is colored based on the variation of the 2020 value with respect to 2019, ranging from dark purple, which represents a decrease of \ac{CNM} in 2020 with respect to 2019, to bright yellow, which represents instead an increase of navigated miles in 2020 with respect to the previous year.}}
       \label{fig:density-maps-1}
\end{figure}

\begin{figure}
       \centering%
       \subfloat[][Wet bulk]{%
         \includegraphics[trim=180 160 140 150,clip,width=0.98\textwidth]{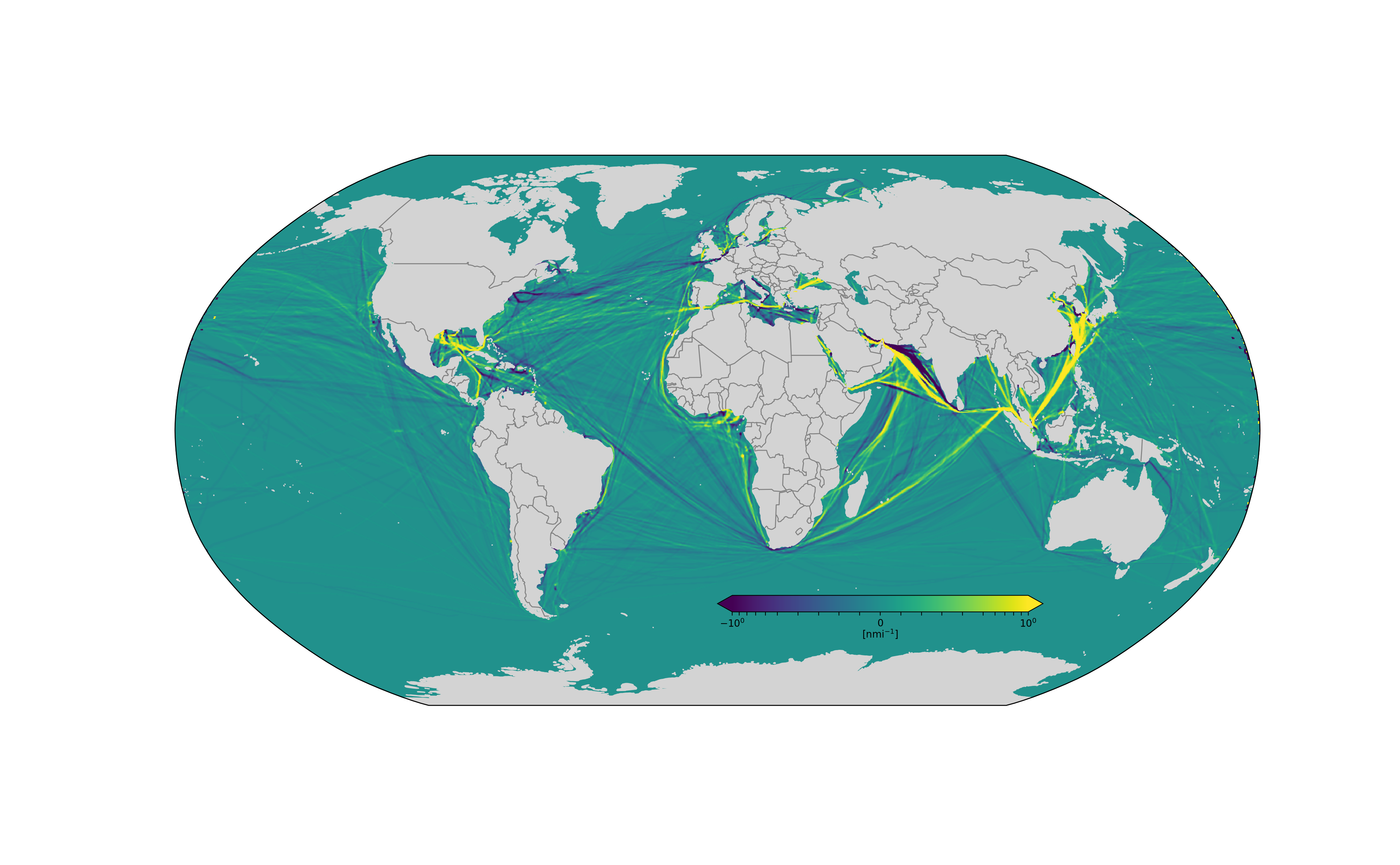}%
         \label{fig:density-map-wet-bulk}%
       }%
       \\
       \subfloat[][Passenger]{%
         \includegraphics[trim=180 160 140 150,clip,width=0.98\textwidth]{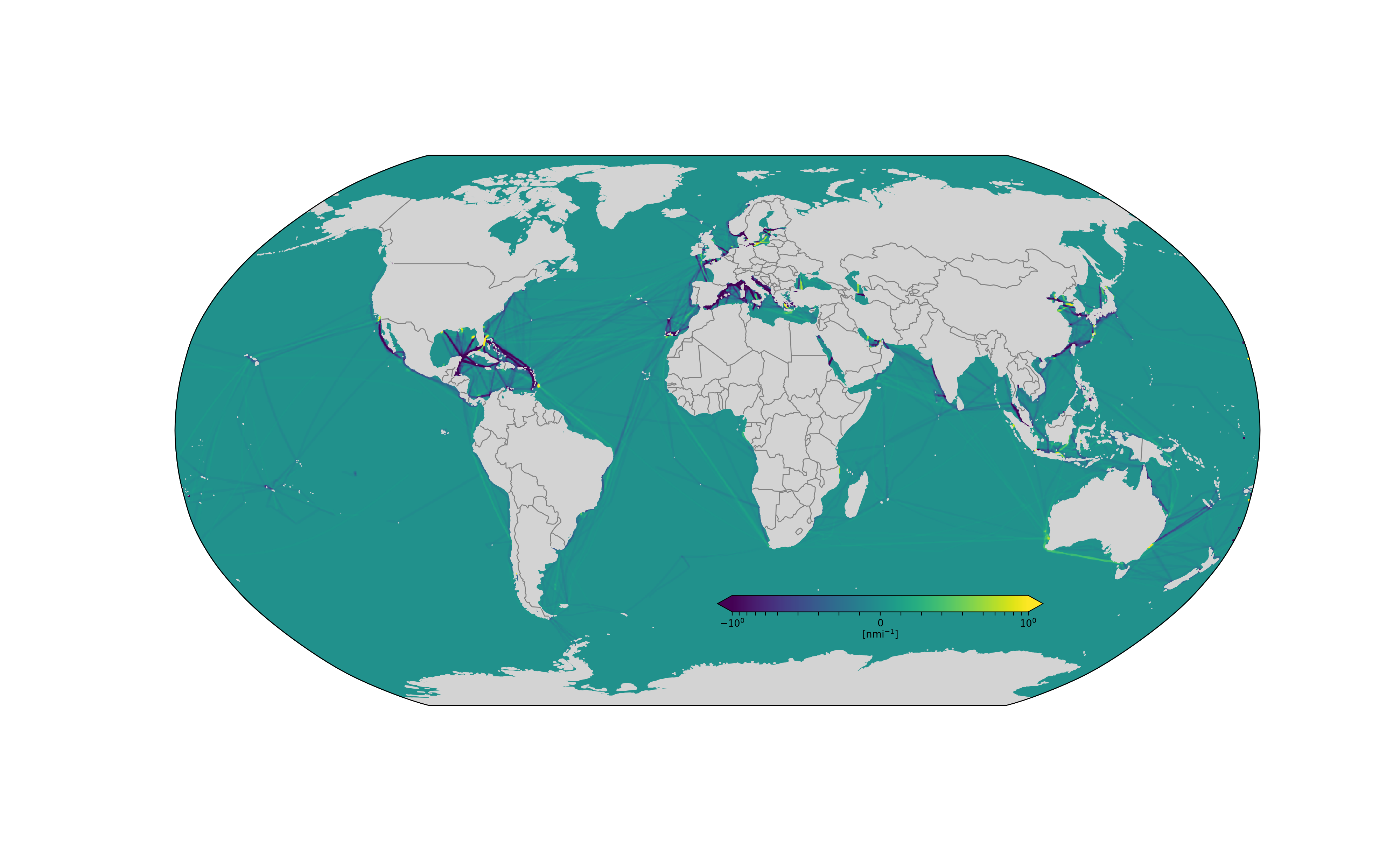}%
         \label{fig:density-map-passenger}%
       }%
       \caption{\rev{Monthly \ac{CNM} density difference between 2020 and 2019 for wet bulk~\protect\subref{fig:density-map-wet-bulk} and passenger~\protect\subref{fig:density-map-passenger} shipping. The considered time period is from 13 March to 13 April. Each grid cell is colored based on the variation of the 2020 value with respect to 2019, ranging from dark purple, which represents a decrease of \ac{CNM} in 2020 with respect to 2019, to bright yellow, which represents instead an increase of navigated miles in 2020 with respect to the previous year.}}%
       \label{fig:density-maps-2}
\end{figure}

\begin{figure}
       \centering%
       \subfloat[][Container]{%
         \includegraphics[trim=210 10 190  30,clip,width=0.3\textwidth]{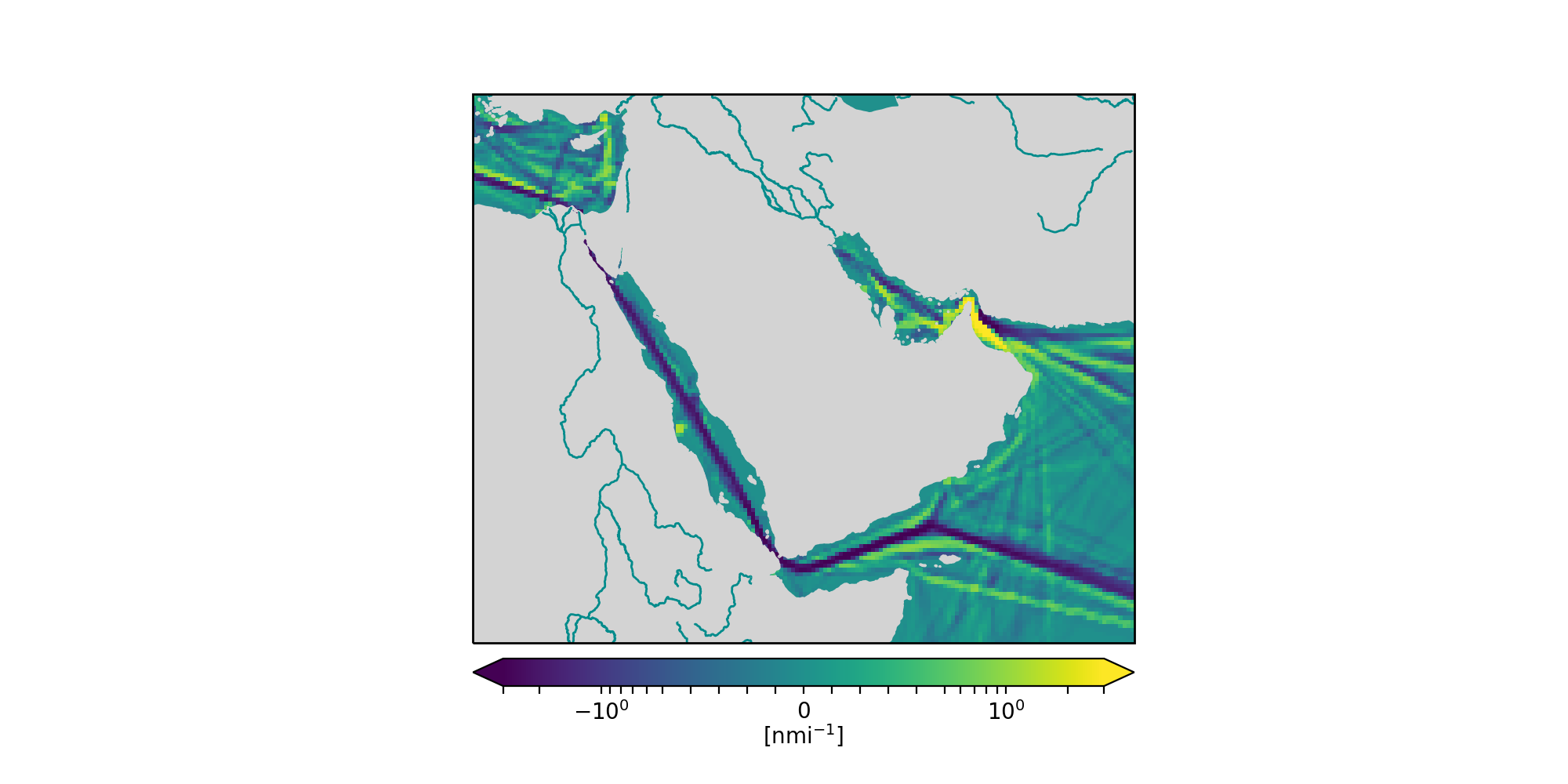}%
         \label{fig:density-map-suez-container}%
       }%
       \hfil%
       \subfloat[][Dry bulk]{%
         \includegraphics[trim=210 10 190  30,clip,width=0.3\textwidth]{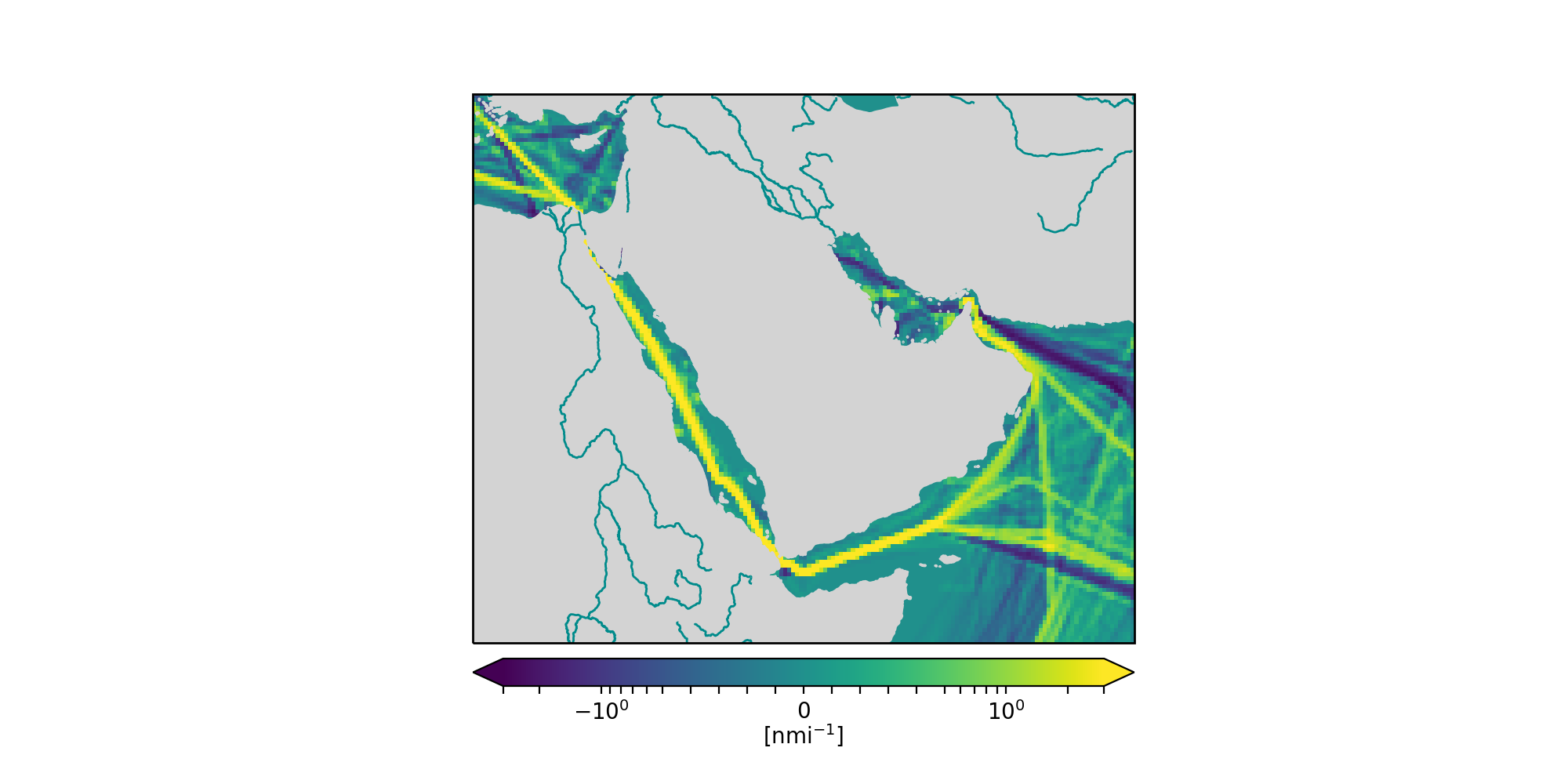}%
         \label{fig:density-map-suez-dry-bulk}%
       }%
       \hfil%
       \subfloat[][Wet bulk]{%
         \includegraphics[trim=210 10 190  30,clip,width=0.3\textwidth]{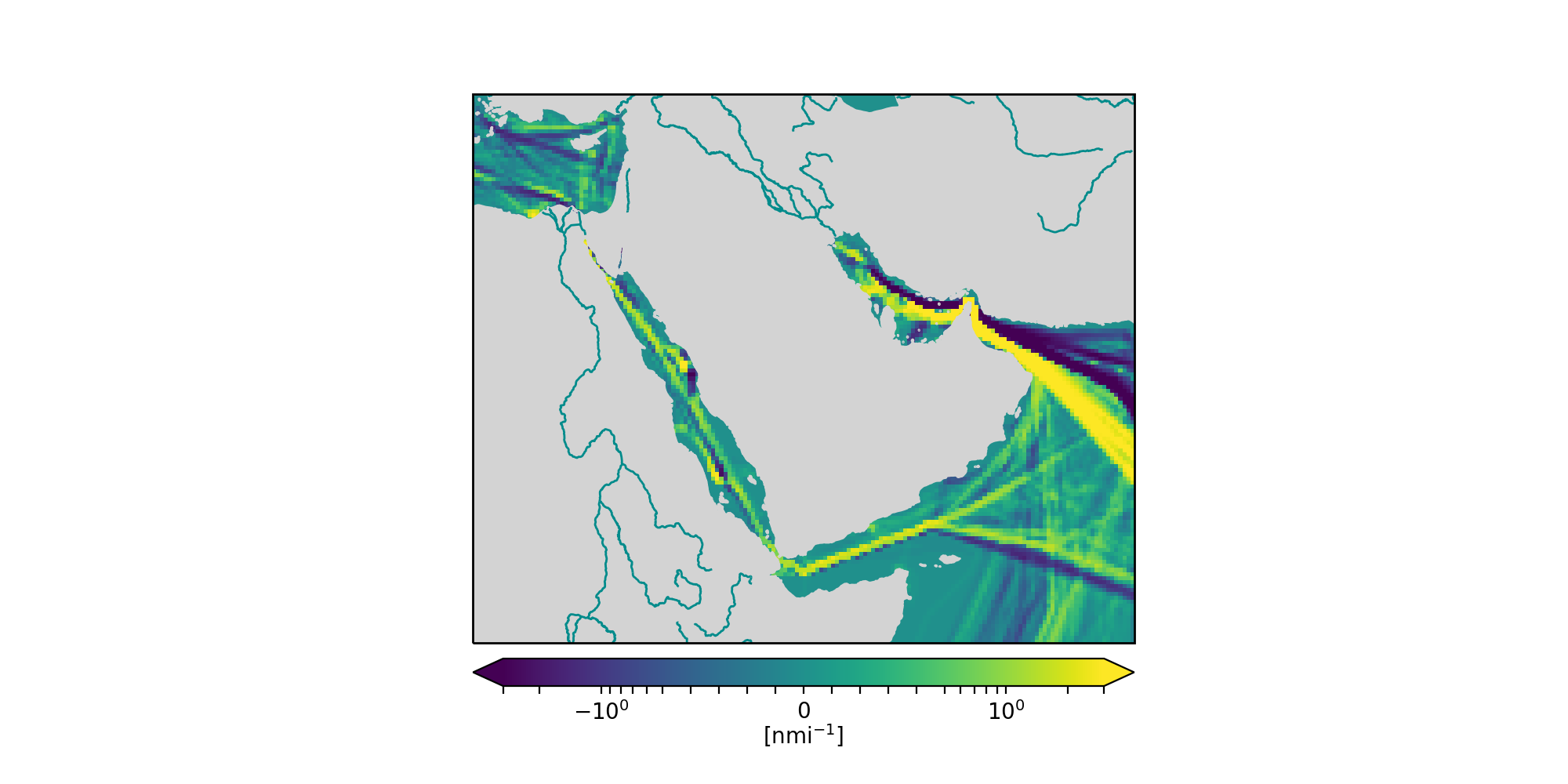}%
         \label{fig:density-map-suez-wet-bulk}%
       }%
       \caption{\rev{Monthly \ac{CNM} density difference in the area around the Suez Canal between 2020 and 2019 for container~\protect\subref{fig:density-map-suez-container}, dry~\protect\subref{fig:density-map-suez-dry-bulk} and wet bulk~\protect\subref{fig:density-map-suez-wet-bulk} shipping. The considered time period is from 13 March to 13 April. Each grid cell is colored based on the variation of the 2020 value with respect to 2019, ranging from dark purple, which represents a decrease of \ac{CNM} in 2020 with respect to 2019, to bright yellow, which represents an increase of navigated miles in 2020 with respect to the previous year.}}
       \label{fig:density-maps-suez}
\end{figure}

\begin{figure}[!t]
    \centering%
    \subfloat[][Container: daily mileage]{%
        \includegraphics[trim=40 20 60 37,clip,width=0.325\textwidth]{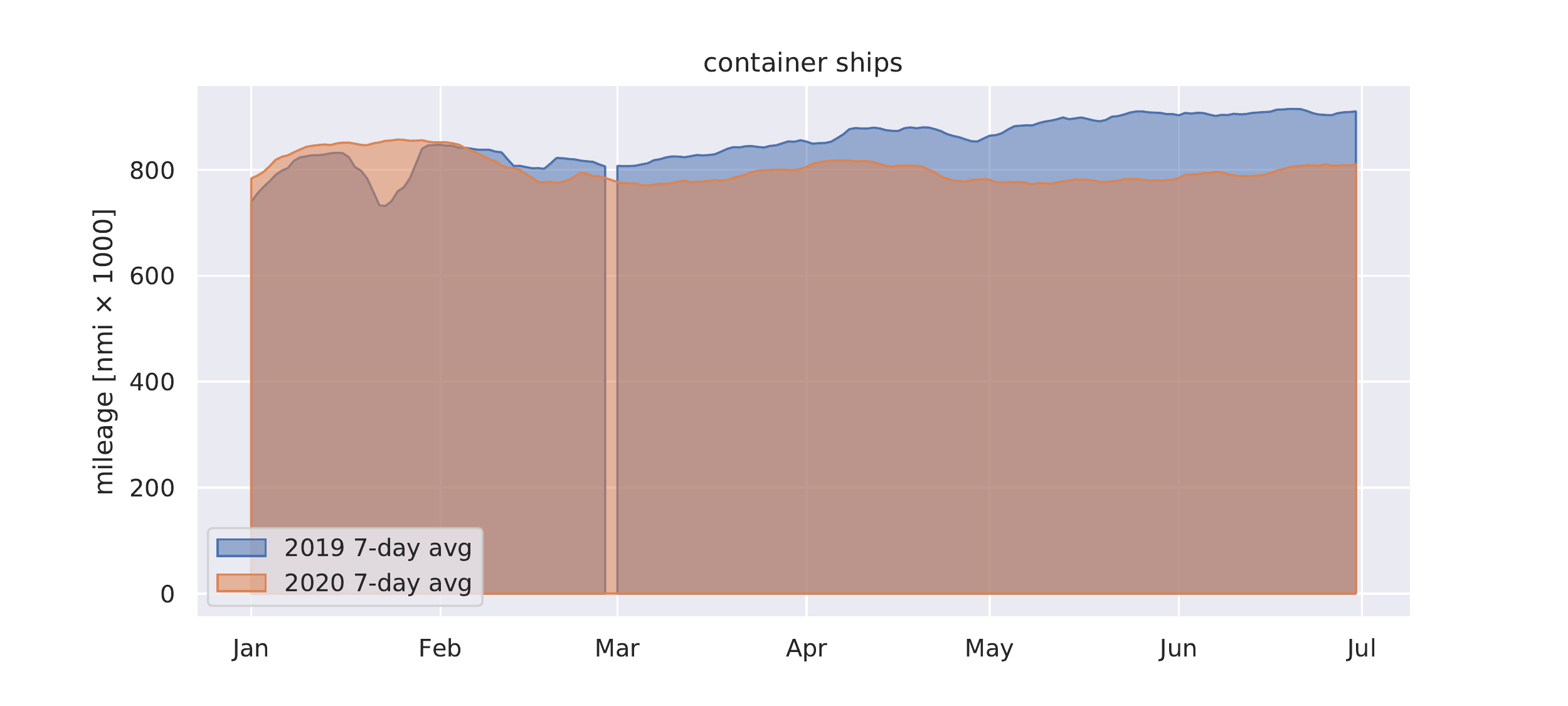}%
        \label{fig:container_dailymileagearea}%
        }%
    \hfil%
    \subfloat[][Container: monthly mileage]{%
        \includegraphics[trim=40 20 60 37,clip,width=0.325\textwidth]{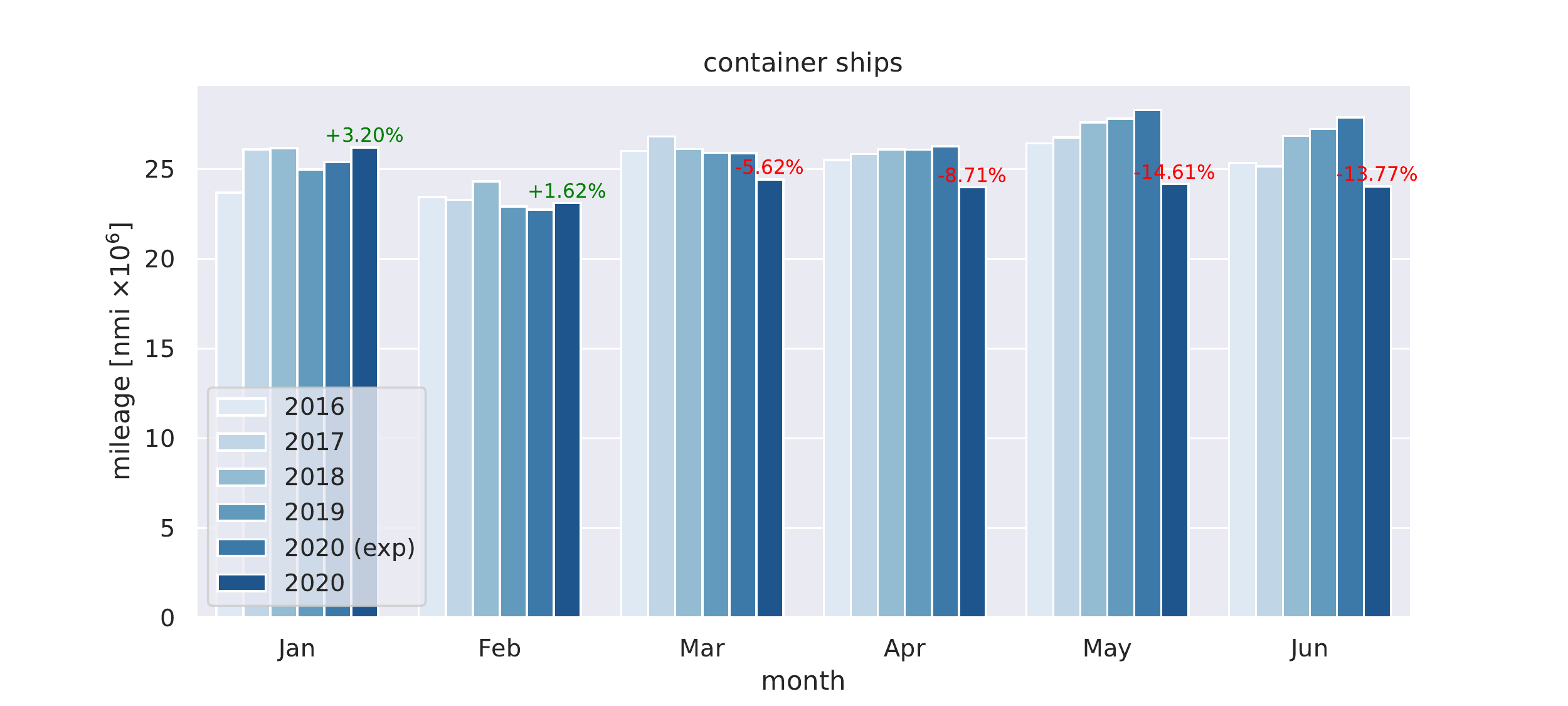}%
        \label{fig:container_mileage}%
        }%
    \hfil%
    \subfloat[][Container: active/idle ships]{%
        \includegraphics[trim=40 20 60 37,clip,width=0.325\textwidth]{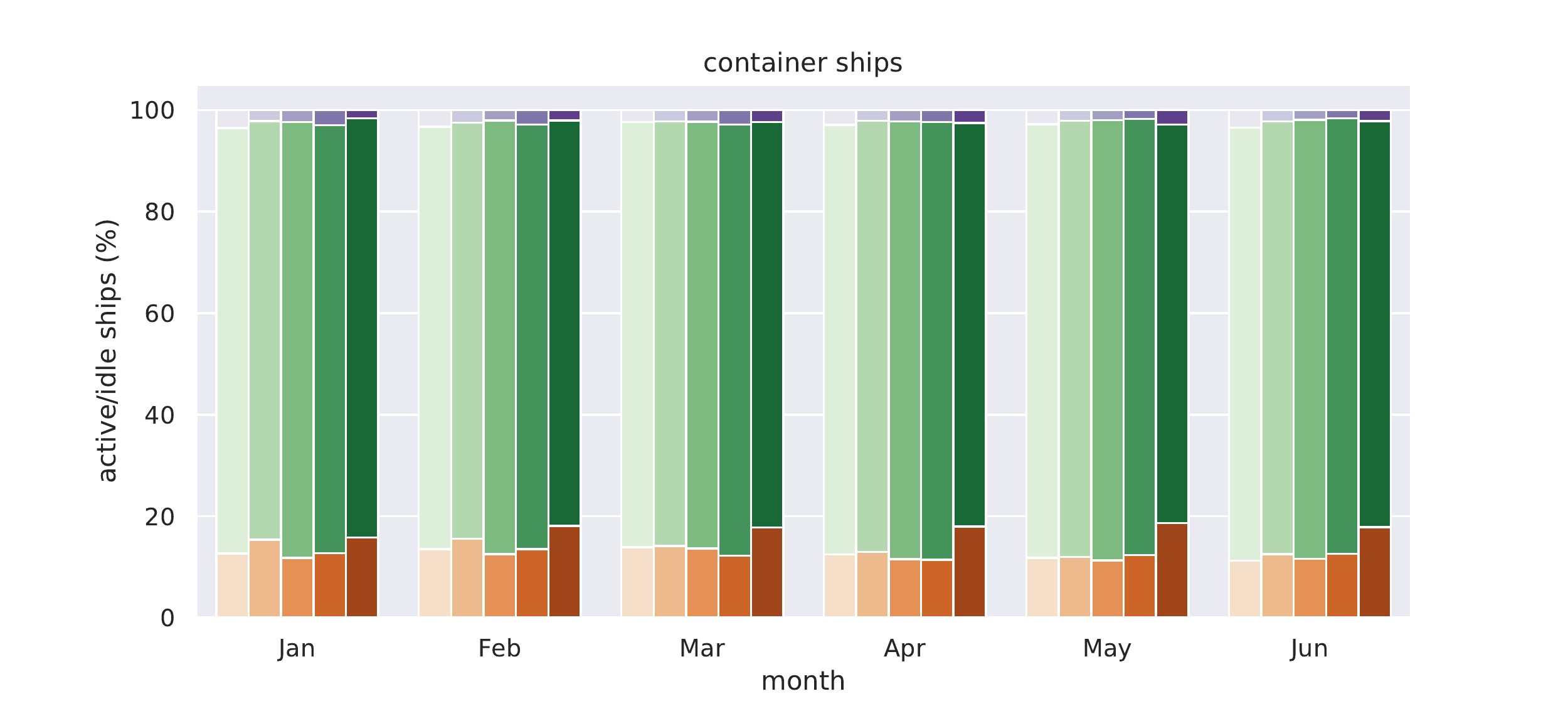}%
        \label{fig:container_activeidle}%
        }%
    \\
    \subfloat[][Dry bulk: daily mileage]{%
        \includegraphics[trim=40 20 60 37,clip,width=0.325\textwidth]{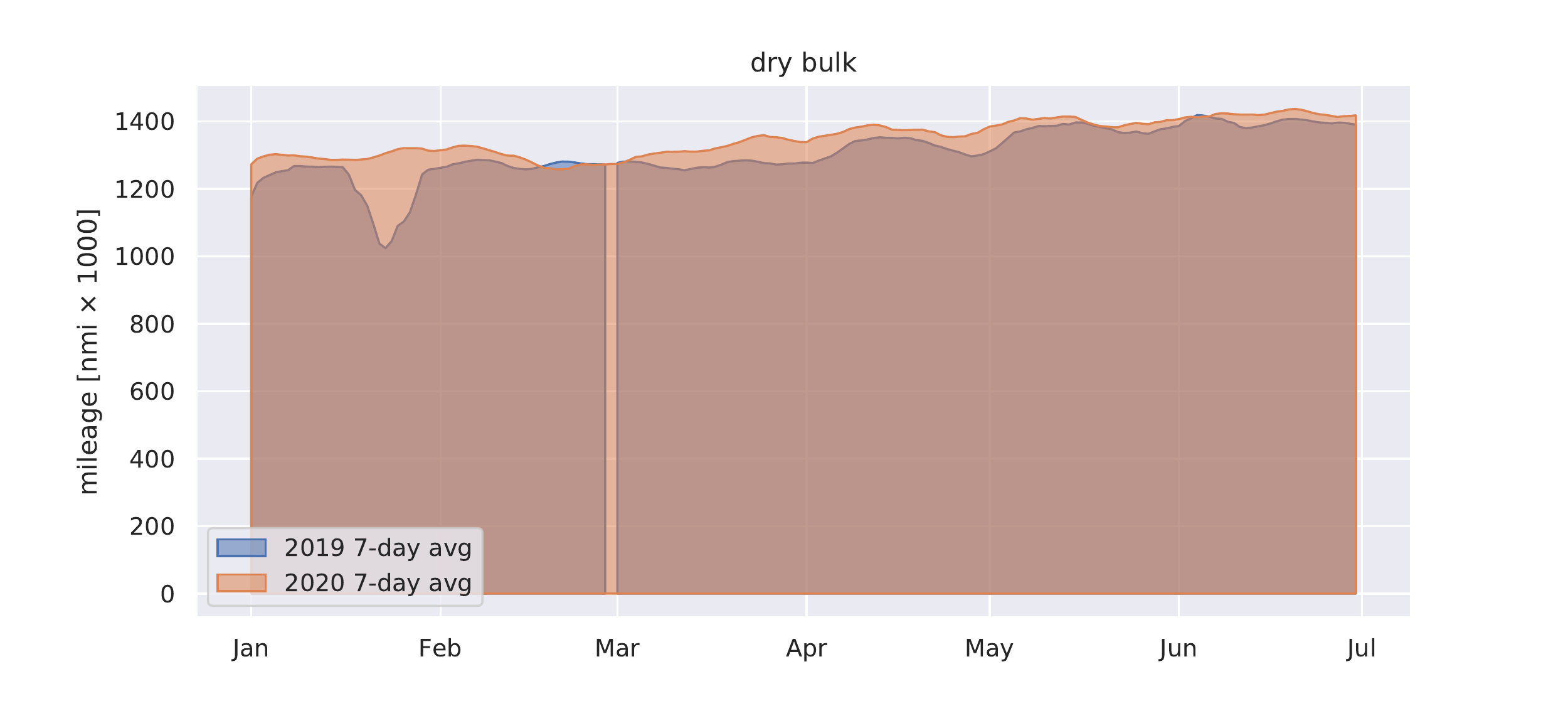}%
        \label{fig:dry-bulk_dailymileagearea}%
        }%
    \hfil%
    \subfloat[][Dry bulk: monthly mileage]{%
        \includegraphics[trim=40 20 60 37,clip,width=0.325\textwidth]{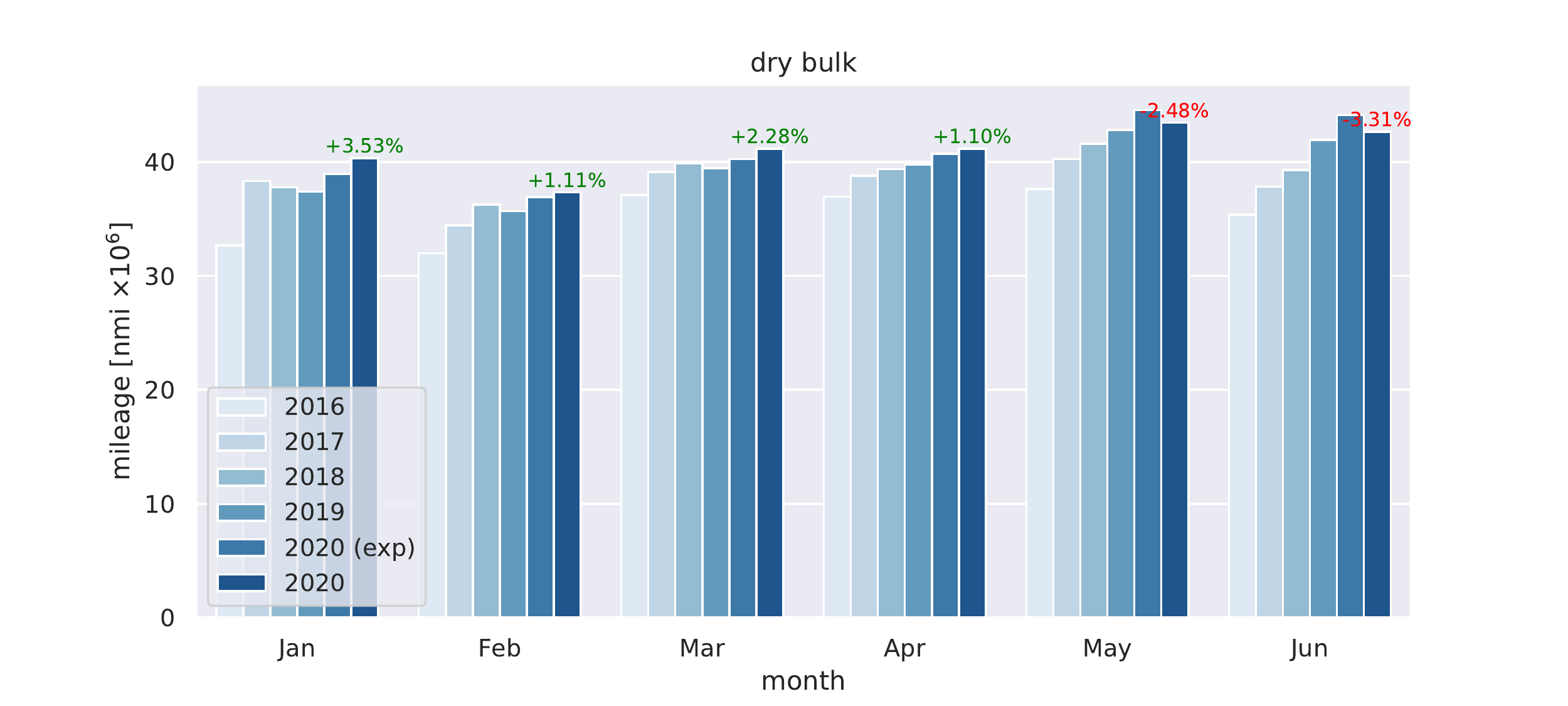}%
        \label{fig:dry-bulk_mileage}%
        }%
    \hfil%
    \subfloat[][Dry bulk: active/idle ships]{%
        \includegraphics[trim=40 20 60 37,clip,width=0.325\textwidth]{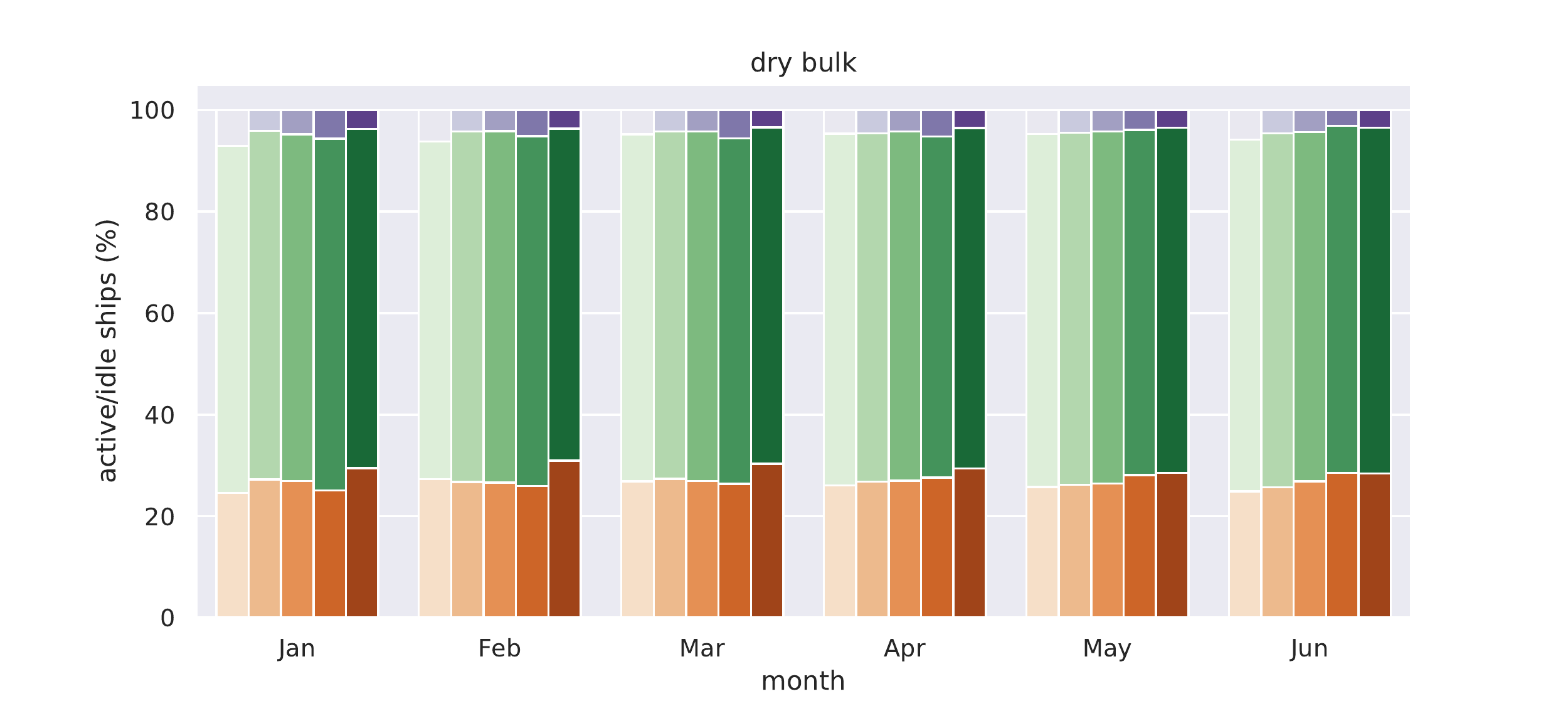}%
        \label{fig:dry-bulk_activeidle}%
        }%
    \\
    \subfloat[][Wet bulk: daily mileage]{%
        \includegraphics[trim=40 20 60 37,clip,width=0.325\textwidth]{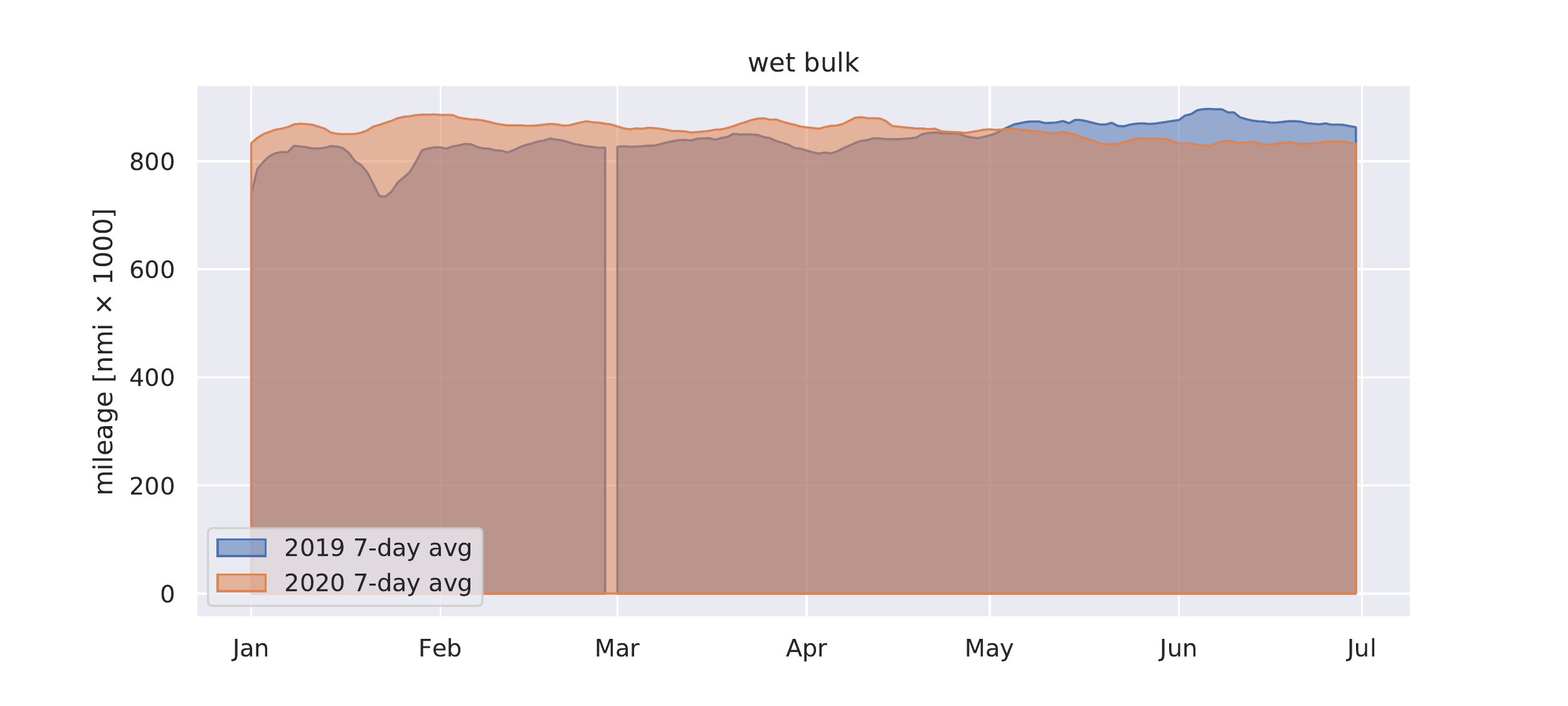}%
        \label{fig:wet-bulk_dailymileagearea}%
        }%
    \hfil%
    \subfloat[][Wet bulk: monthly mileage]{%
        \includegraphics[trim=40 20 60 37,clip,width=0.325\textwidth]{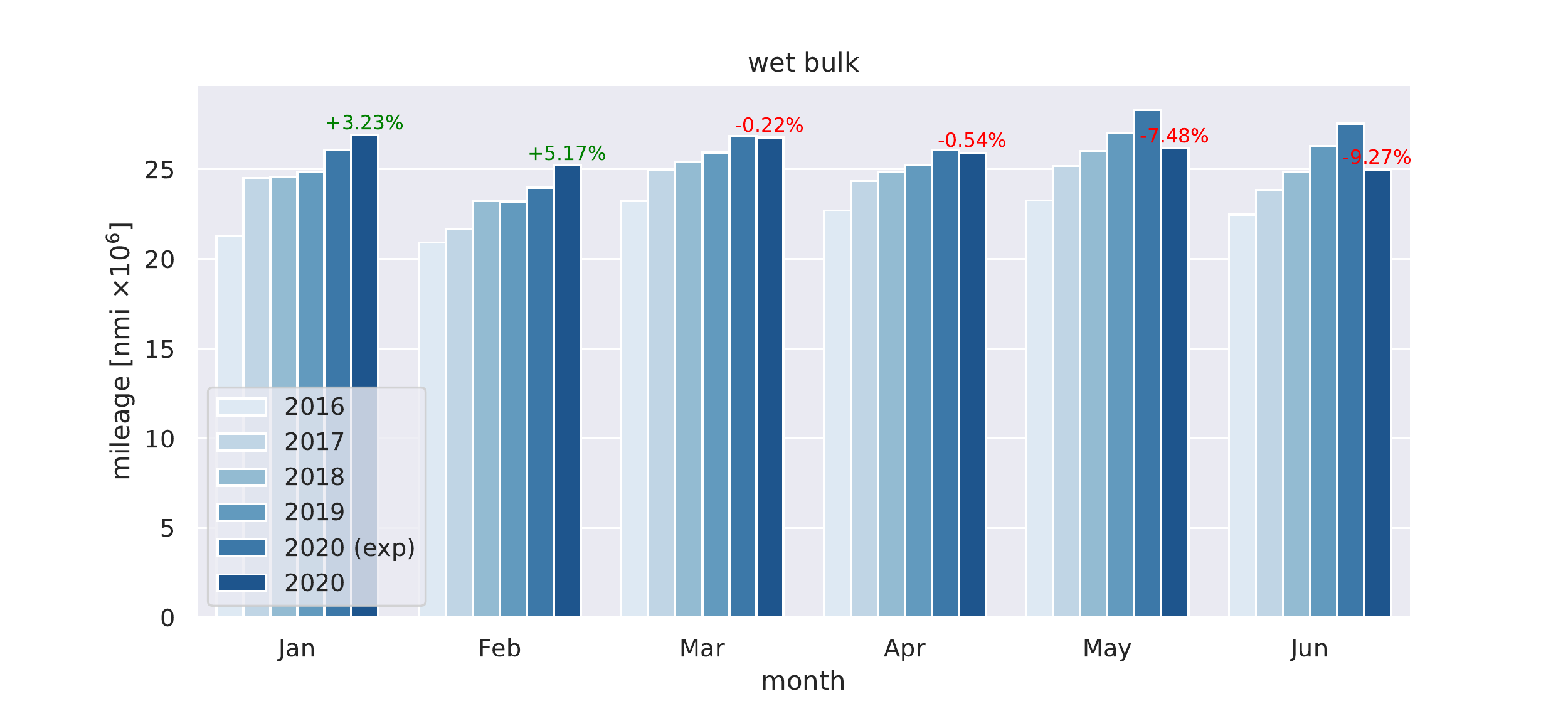}%
        \label{fig:wet-bulk_mileage}%
        }%
    \hfil%
    \subfloat[][Wet bulk: active/idle ships]{%
        \includegraphics[trim=40 20 60 37,clip,width=0.325\textwidth]{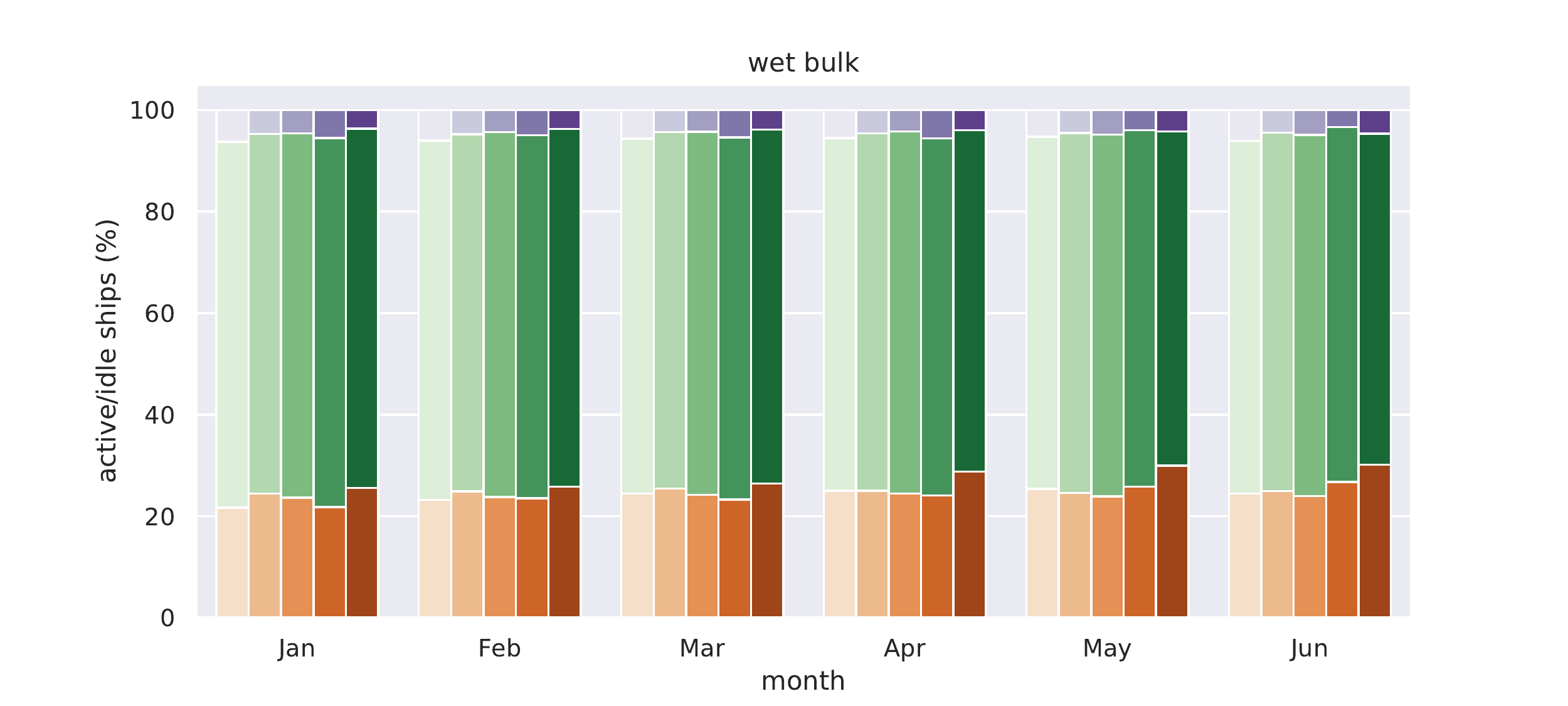}%
        \label{fig:wet-bulk_activeidle}%
        }%
    \\
    \subfloat[][Passenger: daily mileage]{%
        \includegraphics[trim=40 20 60 37,clip,width=0.325\textwidth]{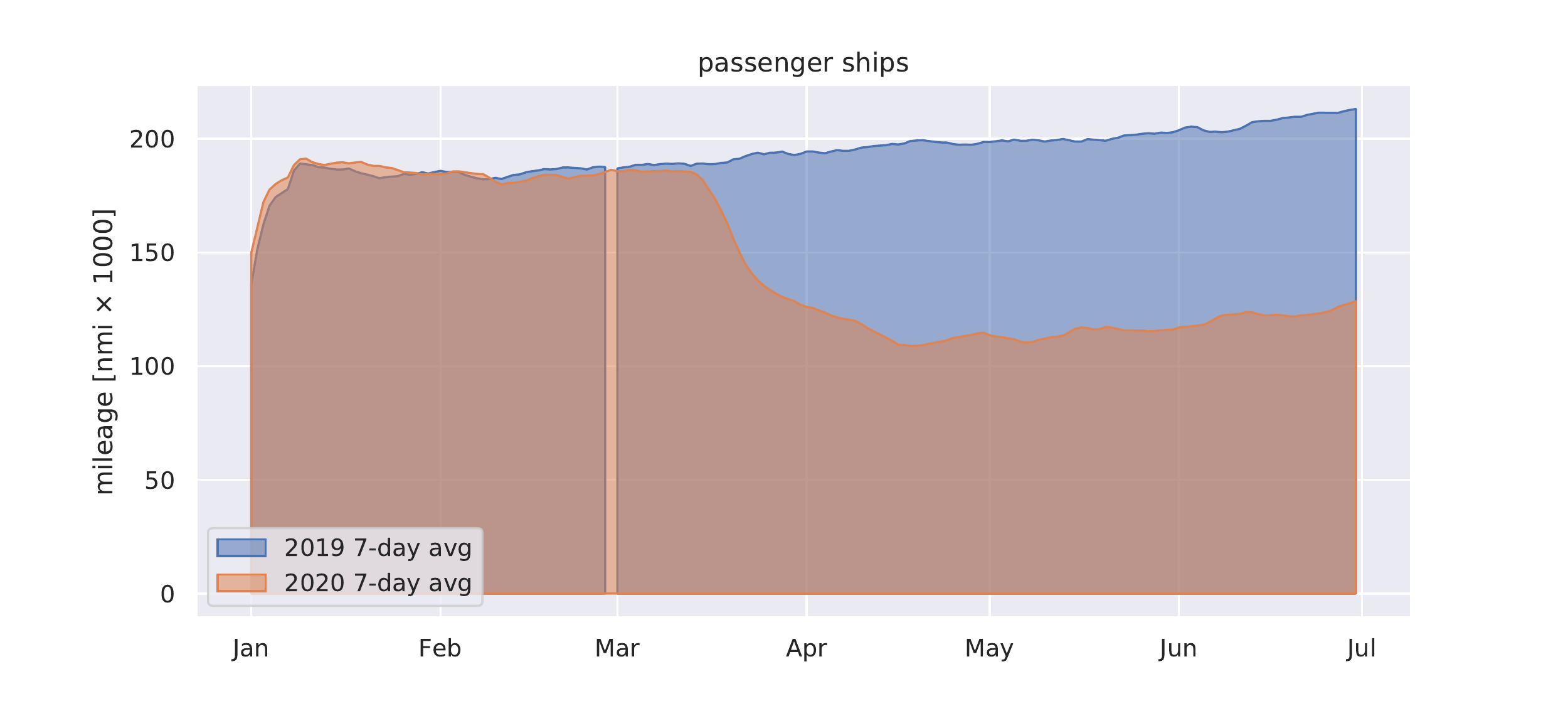}%
        \label{fig:passenger_dailymileagearea}%
        }%
    \hfil%
    \subfloat[][Passenger: monthly mileage]{%
        \includegraphics[trim=40 20 60 37,clip,width=0.325\textwidth]{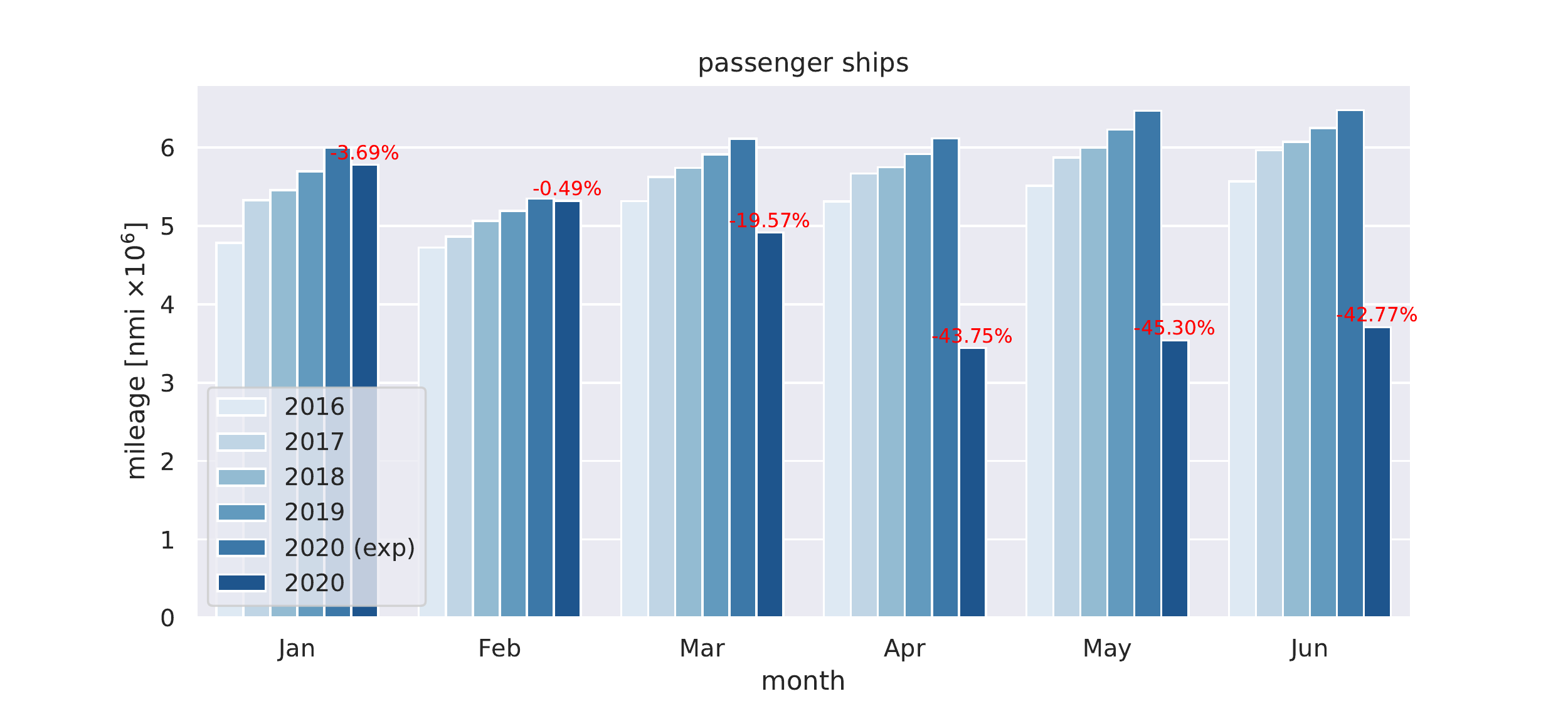}%
        \label{fig:passenger_mileage}%
        }%
    \hfil%
    \subfloat[][Passenger: active/idle ships]{%
        \includegraphics[trim=40 20 60 37,clip,width=0.325\textwidth]{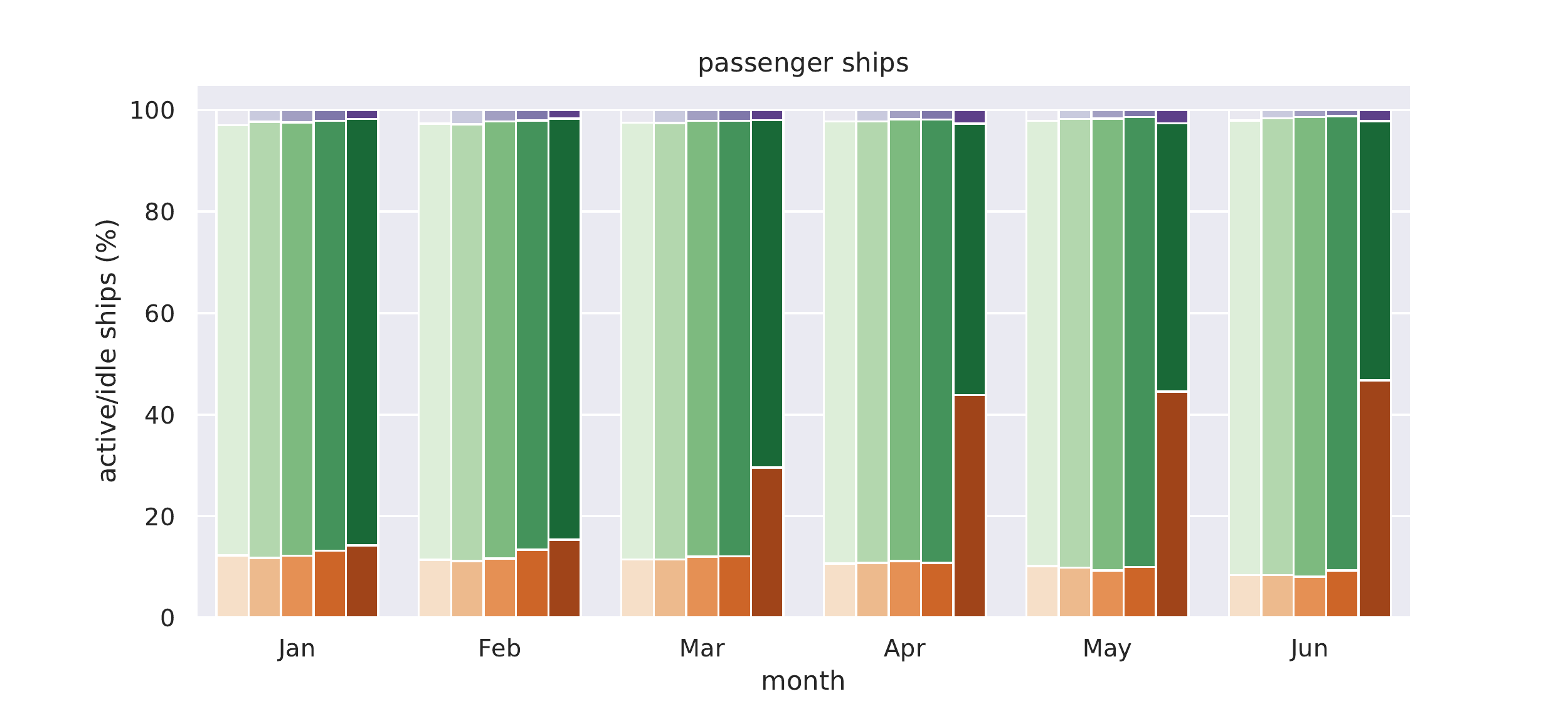}%
        \label{fig:passenger_activeidle}%
        }%
    \caption{\label{fig:traffic} Comparison of traffic indicators for several ship categories: container~\protect\subref{fig:container_dailymileagearea}--\protect\subref{fig:container_activeidle}, dry bulk~\protect\subref{fig:dry-bulk_dailymileagearea}--\protect\subref{fig:dry-bulk_activeidle}, wet bulk~\protect\subref{fig:wet-bulk_dailymileagearea}--\protect\subref{fig:wet-bulk_activeidle}, and passenger~\protect\subref{fig:passenger_dailymileagearea}--\protect\subref{fig:passenger_activeidle}. The left column shows daily navigated miles in 2019 (in blue) compared with 2020 (in orange); the two area charts are overlaid in transparency to highlight trend differences; the discontinuity in the blue data series corresponds to the leap day absence in 2019. The mid column shows monthly navigated since 2016 up to 2020; the second last bar in each group represents an estimation of navigated miles in 2020 given the growth rate observed in the previous years; the label on the 2020 bar quantifies the percentage increase or decrease w.r.t. the expected 2020 traffic volume. The last column shows the distribution of active (green) and idle (orange) ships over time, compared with past years, arranged by month with bars from left (2016) to right (2020); purple bars represent the (negligible) part of ships that could be labeled neither as active nor idle.}%
\end{figure}

\begin{figure}[t]
       \centering%
       \subfloat[][Container]{%
         \includegraphics[trim=120 120 100 120,clip,width=0.48\textwidth]{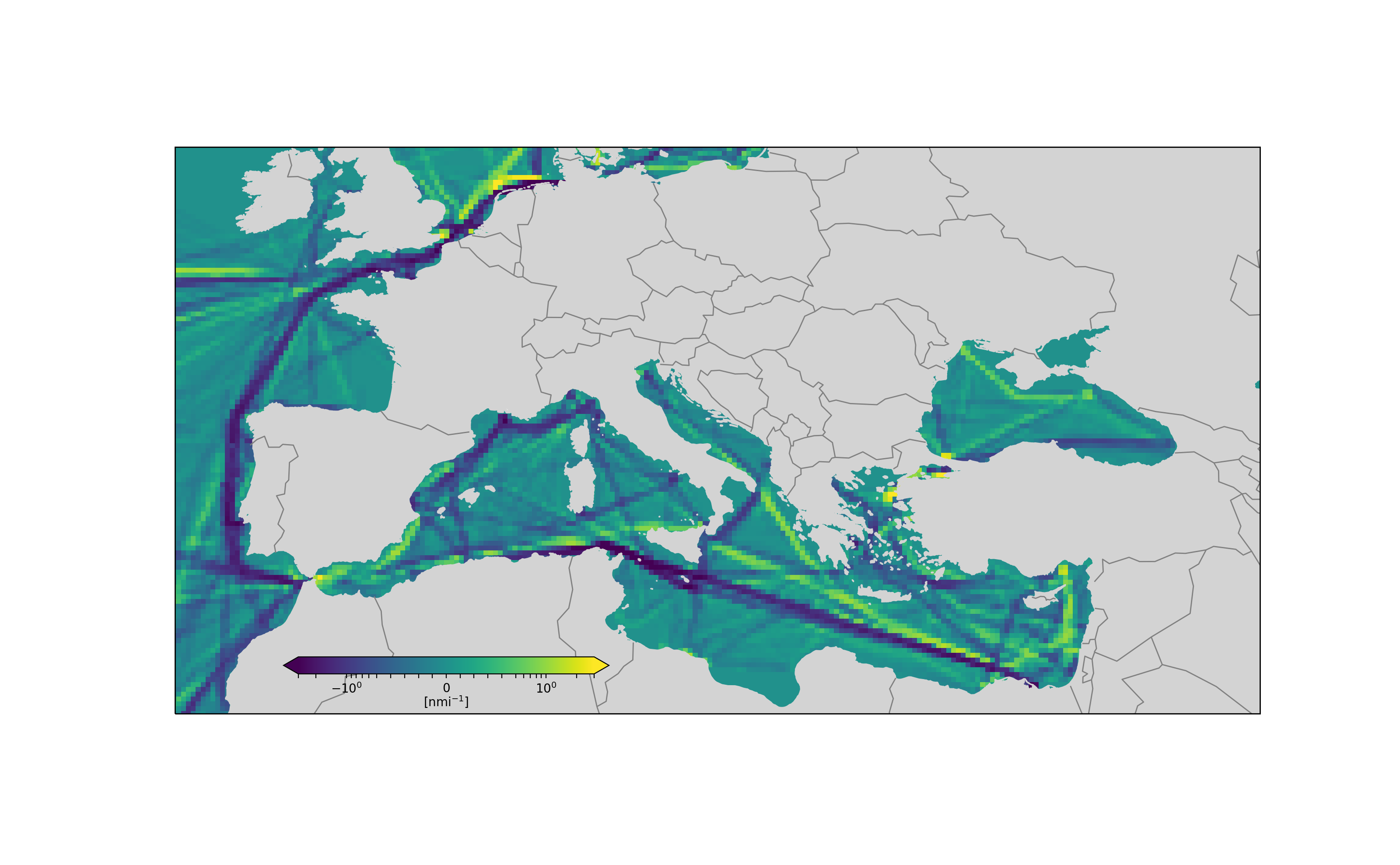}%
         \label{fig:density-map-med-container}%
       }%
       \hfil%
       \subfloat[][Dry bulk]{%
         \includegraphics[trim=120 120 100 120,clip,width=0.48\textwidth]{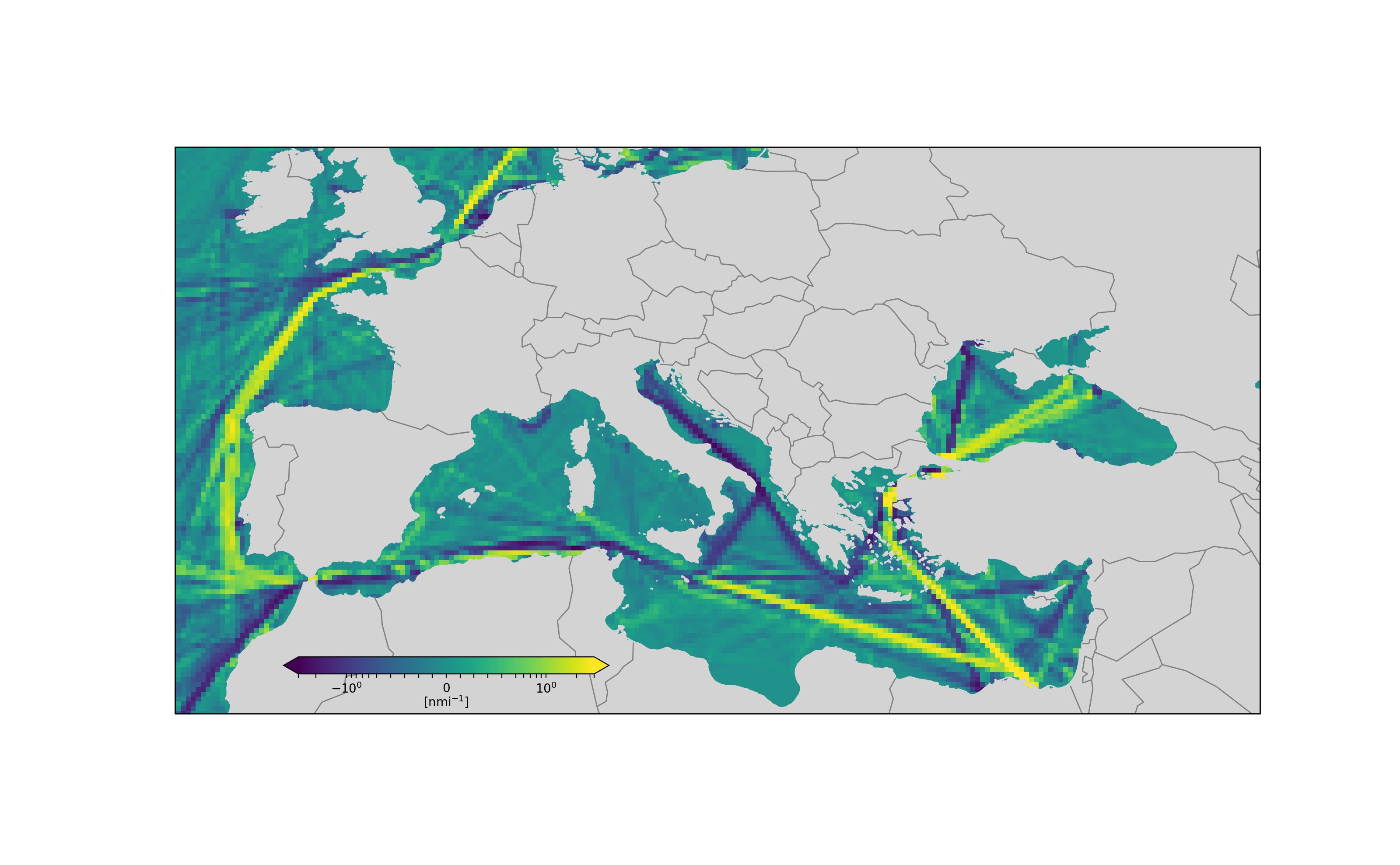}%
         \label{fig:density-map-med-dry-bulk}%
       }%
       \\%
       \subfloat[][Wet bulk]{%
         \includegraphics[trim=120 120 100 120,clip,width=0.48\textwidth]{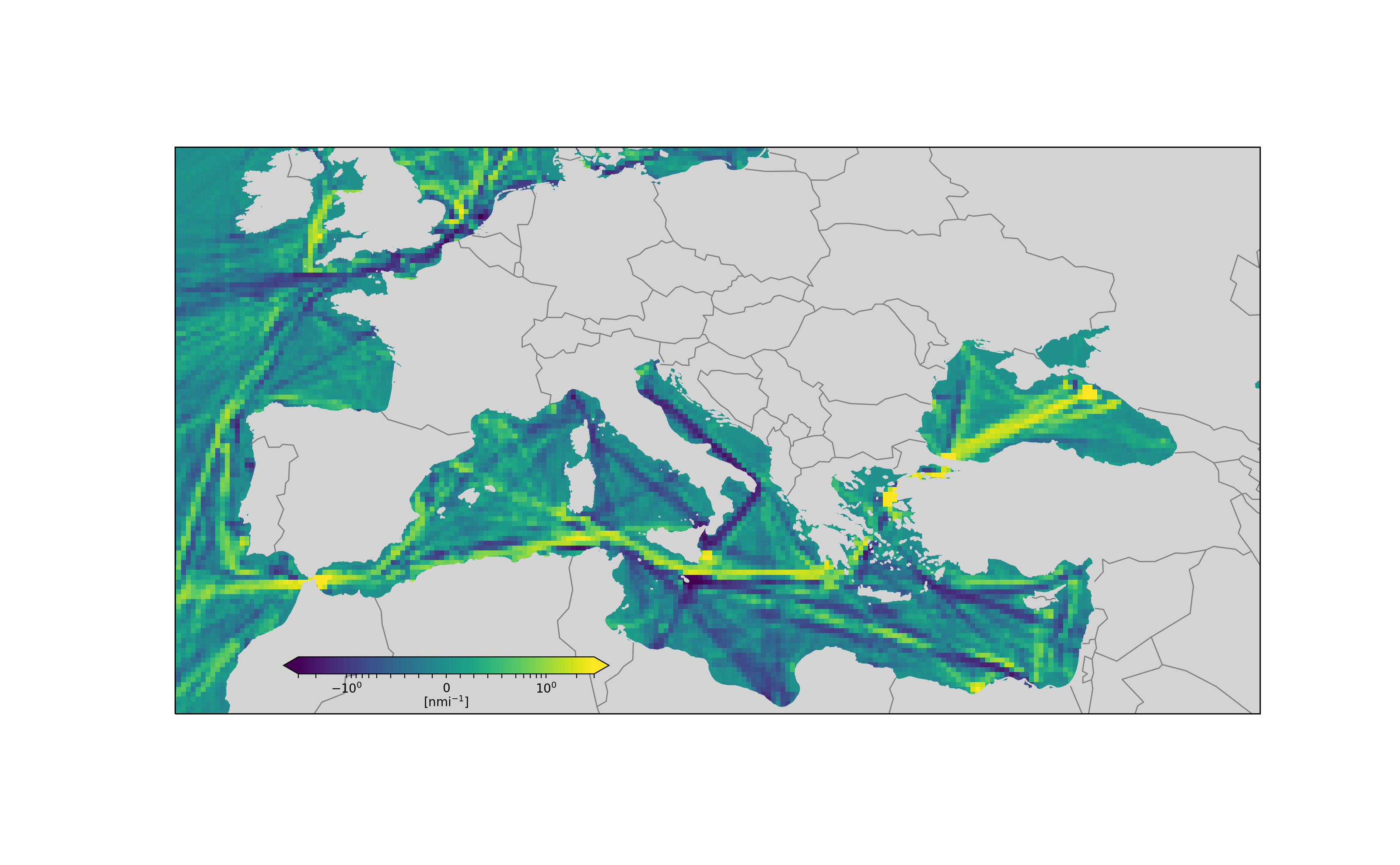}%
         \label{fig:density-map-med-wet-bulk}%
       }%
       \hfil%
       \subfloat[][Passenger]{%
         \includegraphics[trim=120 120 100 120,clip,width=0.48\textwidth]{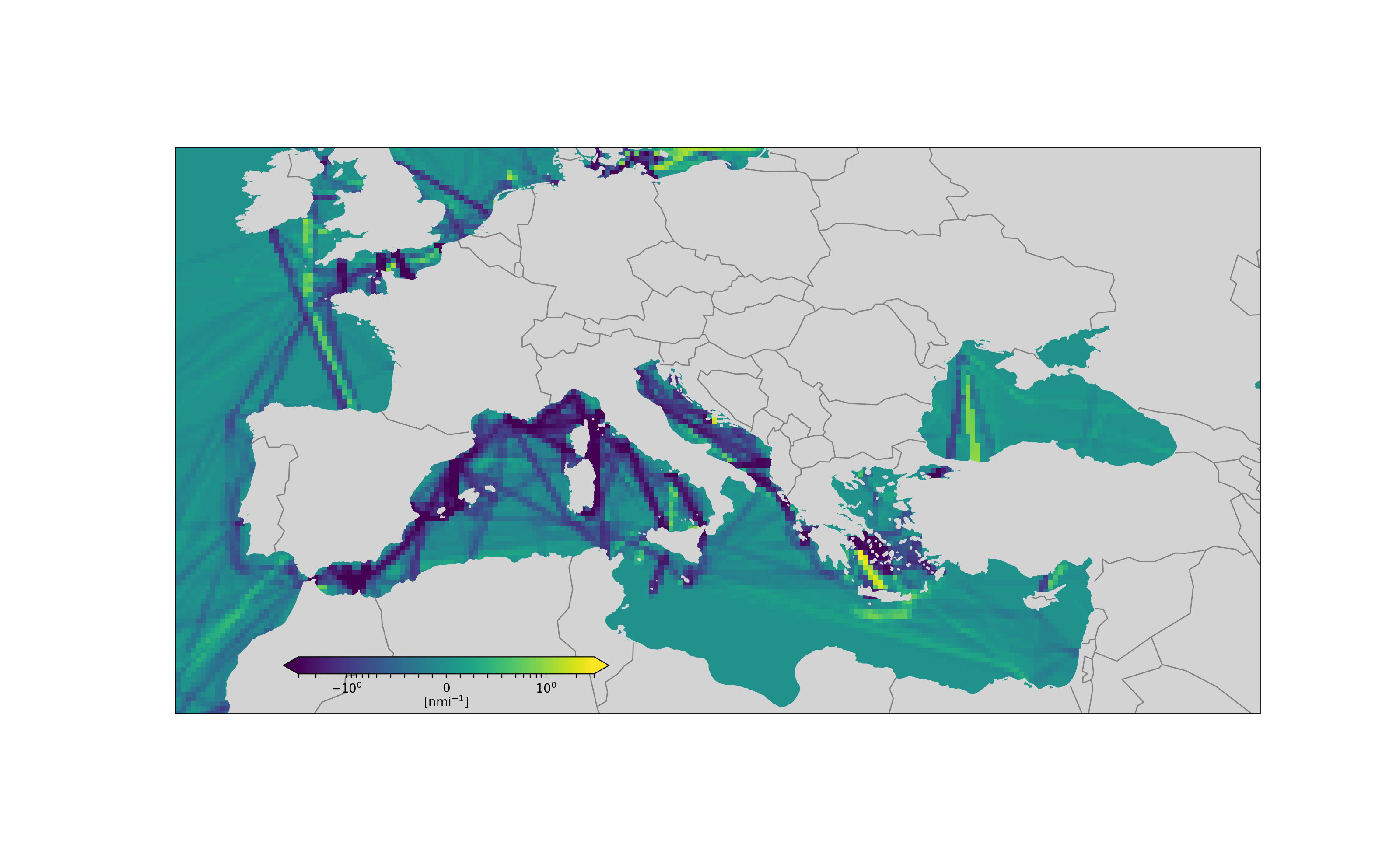}%
         \label{fig:density-map-med-pass}%
       }%
       \caption{\rev{Monthly \ac{CNM} density difference in the Mediterranean Sea between 2020 and 2019 for container~\protect\subref{fig:density-map-med-container}, dry~\protect\subref{fig:density-map-med-dry-bulk}, wet bulk~\protect\subref{fig:density-map-med-wet-bulk} and passenger~\protect\subref{fig:density-map-med-pass} shipping. The considered time period is from 13 March to 13 April. Each grid cell is colored based on the variation of the 2020 value with respect to 2019, ranging from dark purple, which represents a decrease of \ac{CNM} in 2020 with respect to 2019, to bright yellow, which represents an increase of navigated miles in 2020 with respect to the previous year.}}%
       \label{fig:density-maps-med}
\end{figure}

\rev{We have analyzed variations of shipping operational patterns computing the difference of spatial traffic density in 2019 and 2020, reported in  Figures~\ref{fig:density-maps-1}, \ref{fig:density-maps-2}, \ref{fig:density-maps-suez} and \ref{fig:density-maps-med}.}
\rev{A quantitative analysis} is represented in Fig.~\ref{fig:traffic}, which reports global vessel mobility indicators for four main types of ship traffic: container~\protect\subref{fig:container_dailymileagearea}--\protect\subref{fig:container_activeidle}, dry bulk~\protect\subref{fig:dry-bulk_dailymileagearea}--\protect\subref{fig:dry-bulk_activeidle}, wet bulk~\protect\subref{fig:wet-bulk_dailymileagearea}--\protect\subref{fig:wet-bulk_activeidle}, and passenger~\protect\subref{fig:passenger_dailymileagearea}--\protect\subref{fig:passenger_activeidle}. For each traffic category, we report the daily navigated miles from January to June in 2019 and 2020, the monthly navigated miles from 2016 to 2020 (with 2020 forecasts), and the monthly percentage of active/idle ships from 2016 to 2020. The global \ac{CNM} values in Fig.~\ref{fig:traffic} are also reported in Table~\ref{tab:mileage} for each category, where we provide all the total aggregate values.  

Interestingly, the mobility forecasts in January and February 2020 (before lockdowns) underestimate the growth of some markets, such as the container (Fig.~\ref{fig:container_mileage}), dry bulk (Fig.~\ref{fig:dry-bulk_mileage}), and wet bulk (Fig.~\ref{fig:wet-bulk_mileage}) markets, in the sense that the actual 2020 levels are between \SI[retain-explicit-plus]{+1.11}{\percent} (dry bulk in February) and \SI[retain-explicit-plus]{+5.17}{\percent} (wet bulk in February) with respect to the forecast level. Conversely, the forecast severely overestimates the mobility growth of passenger ships since January, \SI{-3.69}{\percent} w.r.t. the expected level, up to a dramatic $-45.3$\% in May, as depicted in Fig.~\ref{fig:passenger_mileage}. In general, the analysis reveals an overall decrease of mobility levels in 2020 compared both with their expected value (considering the growth observed in past years) and to 2019 levels, with only a few exceptions. 



\begin{wrapfigure}[25]{r}{0.5\textwidth}
\begin{center}
    \includegraphics[width=0.48\textwidth]{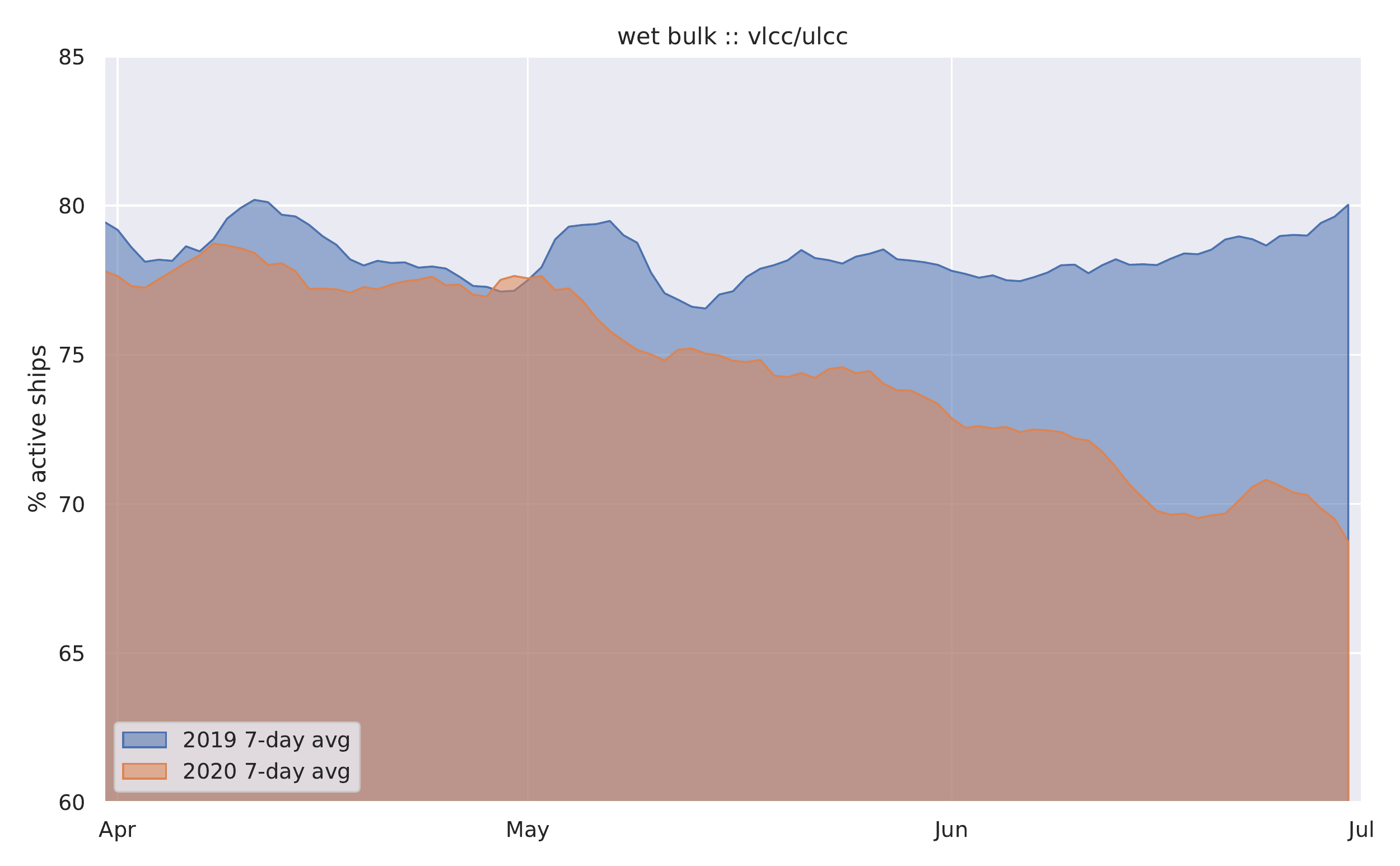}%
\end{center}
\caption{\rev{Daily active and idle ships in 2019 (in blue) and 2020 (in orange) of the supertankers, \ac{VLCC} and \ac{ULCC}. The two area charts are overlaid in transparency to highlight trend differences. It is evident a decrease (increase) of active (idle) ships for this category from April, 2020, not present in 2019. This is confirming the use of a significant subset of supertankers as oil storage.~\cite{REUTERS_storage}}}
\label{fig:vlcc_activeidle}
\end{wrapfigure}

We must note that there is a possible regional effect in the measured mobility levels; the mobility in specific regions might be impacted more than others. Specific sectors of the shipping industry were more resilient and continued delivering goods, while others have been much more vulnerable to such measures (e.g., cruise ships). Indeed, the essential supply chain, e.g., hospital and food, was guaranteed when the restriction measures were in place across the world.

\begin{table*}[!t]
\begin{threeparttable}
\centering%
\caption{Monthly \acf{CNM} [\ac{nmi} $\times$ \num{e6}].}
\label{tab:mileage}
\begin{tabularx}{\textwidth}{ZYYYYYYY}
\toprule
                            & Year & January & February & March & April & May & June \\\midrule
\multirow{6}{*}{Container}  & 2016 & \tablenum{23.68} & \tablenum{23.45} & \tablenum{26.01} & \tablenum{25.51} & \tablenum{26.42} & \tablenum{25.35} \\
                            & 2017 & \tablenum{26.10} & \tablenum{23.28} & \tablenum{26.82} & \tablenum{25.84} & \tablenum{26.76} & \tablenum{25.16} \\
                            & 2018 & \tablenum{26.17} & \tablenum{24.32} & \tablenum{26.12} & \tablenum{26.10} & \tablenum{27.61} & \tablenum{26.85} \\
                            & 2019 & \tablenum{24.96} & \tablenum{22.92} & \tablenum{25.91} & \tablenum{26.08} & \tablenum{27.82} & \tablenum{27.24} \\
                            & 2020\tnote{\textdagger} & \tablenum[round-mode=places,round-precision=2]{25.388896} & \tablenum[round-mode=places,round-precision=2]{22.741210} & \tablenum[round-mode=places,round-precision=2]{25.882280} & \tablenum[round-mode=places,round-precision=2]{26.271720} & \tablenum[round-mode=places,round-precision=2]{28.287694} & \tablenum[round-mode=places,round-precision=2]{27.876693} \\
                            & 2020\tnote{\textdaggerdbl} & \tablenum{26.20} & \tablenum{23.11} & \tablenum{24.43} & \tablenum{23.98} & \tablenum{24.15} & \tablenum{24.04} \\\midrule
\multirow{6}{*}{Dry bulk}   & 2016 & \tablenum{32.67} & \tablenum{31.99} & \tablenum{37.08} & \tablenum{36.94} & \tablenum{37.65} & \tablenum{35.38} \\
                            & 2017 & \tablenum{38.33} & \tablenum{34.43} & \tablenum{39.13} & \tablenum{38.79} & \tablenum{40.25} & \tablenum{37.83} \\
                            & 2018 & \tablenum{37.76} & \tablenum{36.27} & \tablenum{39.90} & \tablenum{39.36} & \tablenum{41.60} & \tablenum{39.29} \\
                            & 2019 & \tablenum{37.38} & \tablenum{35.69} & \tablenum{39.45} & \tablenum{39.77} & \tablenum{42.82} & \tablenum{41.93} \\
                            & 2020\tnote{\textdagger} & \tablenum[round-mode=places,round-precision=2]{38.946113} & \tablenum[round-mode=places,round-precision=2]{36.918515} & \tablenum[round-mode=places,round-precision=2]{40.247712} & \tablenum[round-mode=places,round-precision=2]{40.718030} & \tablenum[round-mode=places,round-precision=2]{44.549008} & \tablenum[round-mode=places,round-precision=2]{44.109726} \\
                            & 2020\tnote{\textdaggerdbl} & \tablenum{40.32} & \tablenum{37.33} & \tablenum{41.17} & \tablenum{41.17} & \tablenum{43.44} & \tablenum{42.65} \\\midrule
\multirow{6}{*}{Wet bulk}   & 2016 & \tablenum{21.28} & \tablenum{20.92} & \tablenum{23.24} & \tablenum{22.69} & \tablenum{23.26} & \tablenum{22.48} \\
                            & 2017 & \tablenum{24.50} & \tablenum{21.69} & \tablenum{24.99} & \tablenum{24.34} & \tablenum{25.18} & \tablenum{23.84} \\
                            & 2018 & \tablenum{24.55} & \tablenum{23.22} & \tablenum{25.40} & \tablenum{24.84} & \tablenum{26.02} & \tablenum{24.84} \\
                            & 2019 & \tablenum{24.88} & \tablenum{23.21} & \tablenum{25.94} & \tablenum{25.22} & \tablenum{27.04} & \tablenum{26.28} \\
                            & 2020\tnote{\textdagger} & \tablenum[round-mode=places,round-precision=2]{26.073592} & \tablenum[round-mode=places,round-precision=2]{23.977948} & \tablenum[round-mode=places,round-precision=2]{26.839182} & \tablenum[round-mode=places,round-precision=2]{26.058297} & \tablenum[round-mode=places,round-precision=2]{28.298099} & \tablenum[round-mode=places,round-precision=2]{27.546670} \\
                            & 2020\tnote{\textdaggerdbl} & \tablenum{26.92} & \tablenum{25.22} & \tablenum{26.78} & \tablenum{25.92} & \tablenum{26.18} & \tablenum{24.99} \\\midrule
\multirow{6}{*}{Passenger}  & 2016 & \tablenum{4.78} & \tablenum{4.73} & \tablenum{5.32} & \tablenum{5.31} & \tablenum{5.51} & \tablenum{5.57} \\
                            & 2017 & \tablenum{5.33} & \tablenum{4.87} & \tablenum{5.62} & \tablenum{5.67} & \tablenum{5.88} & \tablenum{5.96} \\
                            & 2018 & \tablenum{5.46} & \tablenum{5.06} & \tablenum{5.74} & \tablenum{5.75} & \tablenum{6.00} & \tablenum{6.08} \\
                            & 2019 & \tablenum{5.70} & \tablenum{5.19} & \tablenum{5.91} & \tablenum{5.92} & \tablenum{6.23} & \tablenum{6.25} \\
                            & 2020\tnote{\textdagger} & \tablenum[round-mode=places,round-precision=2]{6.002056} & \tablenum[round-mode=places,round-precision=2]{5.347609} & \tablenum[round-mode=places,round-precision=2]{6.112250} & \tablenum[round-mode=places,round-precision=2]{6.118868} & \tablenum[round-mode=places,round-precision=2]{6.471259} & \tablenum[round-mode=places,round-precision=2]{6.478738} \\
                            & 2020\tnote{\textdaggerdbl} & \tablenum{5.78} & \tablenum{5.32} & \tablenum{4.92} & \tablenum{3.44} & \tablenum{3.54} & \tablenum{3.71} \\\midrule
\multirow{7}{*}{Total}      & 2016 & \tablenum[round-mode=places,round-precision=2]{82.41761074558578} & \tablenum[round-mode=places,round-precision=2]{81.08812226235197} & \tablenum[round-mode=places,round-precision=2]{91.63705921960157} & \tablenum[round-mode=places,round-precision=2]{90.45518809236022} & \tablenum[round-mode=places,round-precision=2]{92.84581478024553} & \tablenum[round-mode=places,round-precision=2]{88.77606701588492} \\
                            & 2017 & \tablenum[round-mode=places,round-precision=2]{94.26175846038859} & \tablenum[round-mode=places,round-precision=2]{84.26295325364059} & \tablenum[round-mode=places,round-precision=2]{96.56115011804975} & \tablenum[round-mode=places,round-precision=2]{94.64396139378307} & \tablenum[round-mode=places,round-precision=2]{98.0662130201432} & \tablenum[round-mode=places,round-precision=2]{92.78926520970444} \\
                            & 2018 & \tablenum[round-mode=places,round-precision=2]{93.93648574400505} & \tablenum[round-mode=places,round-precision=2]{88.87832847512925} & \tablenum[round-mode=places,round-precision=2]{97.1561026571546} & \tablenum[round-mode=places,round-precision=2]{96.05176455461827} & \tablenum[round-mode=places,round-precision=2]{101.22439475810113} & \tablenum[round-mode=places,round-precision=2]{97.05888907466615} \\
                            & 2019 & \tablenum[round-mode=places,round-precision=2]{92.91239624590139} & \tablenum[round-mode=places,round-precision=2]{87.01099231894625} & \tablenum[round-mode=places,round-precision=2]{97.22033311197983} & \tablenum[round-mode=places,round-precision=2]{96.98898314466244} & \tablenum[round-mode=places,round-precision=2]{103.91599882393159} & \tablenum[round-mode=places,round-precision=2]{101.70288677573726} \\
                            & 2020\tnote{\textdagger} & \tablenum[round-mode=places,round-precision=2]{96.41} & \tablenum[round-mode=places,round-precision=2]{88.99} & \tablenum[round-mode=places,round-precision=2]{99.08} & \tablenum[round-mode=places,round-precision=2]{99.17} & \tablenum[round-mode=places,round-precision=2]{107.61} & \tablenum[round-mode=places,round-precision=2]{106.01} \\
                            & 2020\tnote{\textdaggerdbl} & \tablenum[round-mode=places,round-precision=2]{99.22018890754083} & \tablenum[round-mode=places,round-precision=2]{90.97668076613975} & \tablenum[round-mode=places,round-precision=2]{97.28886959192003} & \tablenum[round-mode=places,round-precision=2]{94.51000061640016} & \tablenum[round-mode=places,round-precision=2]{97.31813308616817} & \tablenum[round-mode=places,round-precision=2]{95.38911555657043} \\
                            \bottomrule
\end{tabularx}
\begin{tablenotes}
\item [\textdagger] Forecast mobility levels considering the average growth in past years 2016--2019.
\item [\textdaggerdbl] Actual mobility levels recorded in 2020.
\end{tablenotes}
\end{threeparttable}
\end{table*}

Figures~\ref{fig:container_dailymileagearea}--\ref{fig:container_activeidle} show an evident slowdown in the mobility of container ships with respect to previous years, with an increase of idle ships and a correspondent decrease of navigated miles. The slowdown becomes apparent in March, in comparison to both 2019 and 2020 \rev{forecasts}. Global navigated miles of container ships in June 2020 was on average \SI{10}{\percent} below the level in June 2019, and \SI{13.77}{\percent} below the 2020 forecast. Figures~\ref{fig:dry-bulk_dailymileagearea}--\ref{fig:dry-bulk_activeidle} inform on the effects of lockdowns on dry bulk shipping. 
Overall, there is a small increase in idle ships from January to April, and a decline of the navigated miles in May and June. This market appears to be less sensitive to the implemented containment measures, \rev{at least} in the short term. \rev{The reason is maybe that} dry bulk \rev{carriers} transport a wide range of cargo, including goods whose demand increased due to the pandemic, such as paper pulp, with several ports reporting record quantities. However, still there is a noticeable decrease, \SI{2.5}{\percent} and \SI{3.3}{\percent} in May and June, respectively, if compared to the forecast levels. %
Figures~\ref{fig:wet-bulk_dailymileagearea}--\ref{fig:wet-bulk_activeidle} show the mobility of wet bulk shipping. An evident rise of idle ships from the beginning of the year is apparent in this category, with a corresponding decrease of navigated miles from February to June. The reduced mobility of this shipping category is already observable in May and June, if compared with 2019 values, and \rev{this being} a growing market, the loss is even more pronounced if compared with the expected mobility levels in 2020, with a loss that can be assessed around \SI{7.5}{\percent} in May and \SI{9.3}{\percent} in June. %
Finally, Figures~\ref{fig:passenger_dailymileagearea}--\ref{fig:passenger_activeidle} report on the global mobility of passenger ships, the traffic category that was most affected by lockdown measures, with a dramatic collapse of monthly mileage recorded since March (more than \SI{40}{\percent} less than what expected by 2020 forecasts) and a corresponding increase of idle vessels. 

\begin{figure}
    \centering%
    \subfloat[][Passenger ships: \acs{GT} $\leq$ 10K]{%
        \includegraphics[trim=15 10 10 15,clip,width=0.32\columnwidth]{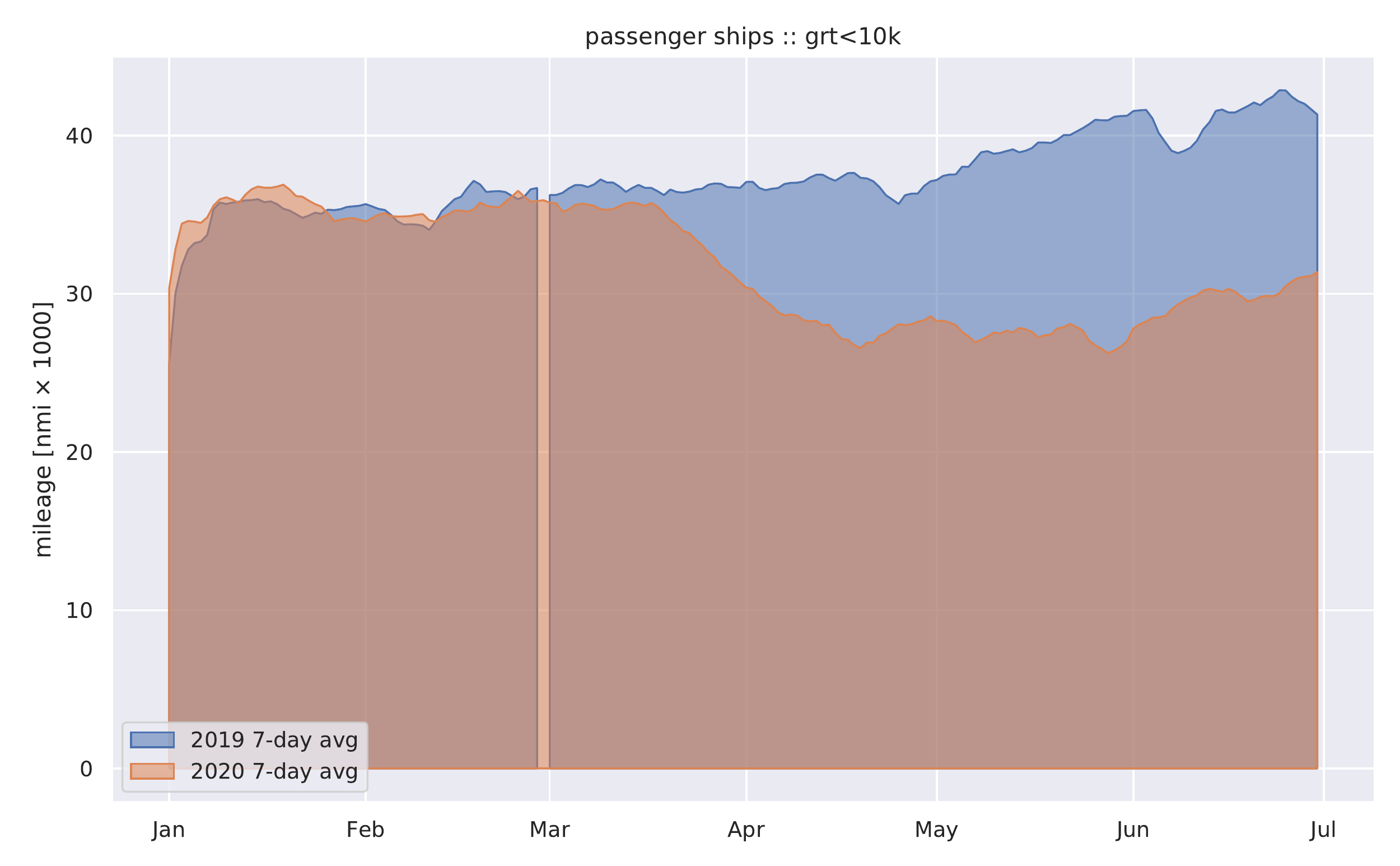}%
        \label{fig:dailymileagearea_passenger_lt_10k}%
        }%
    \hfil%
    \subfloat[][Passenger ships: 10K $<$ \acs{GT} $\leq$ 60K]{%
        \includegraphics[trim=15 10 10 15,clip,width=0.32\columnwidth]{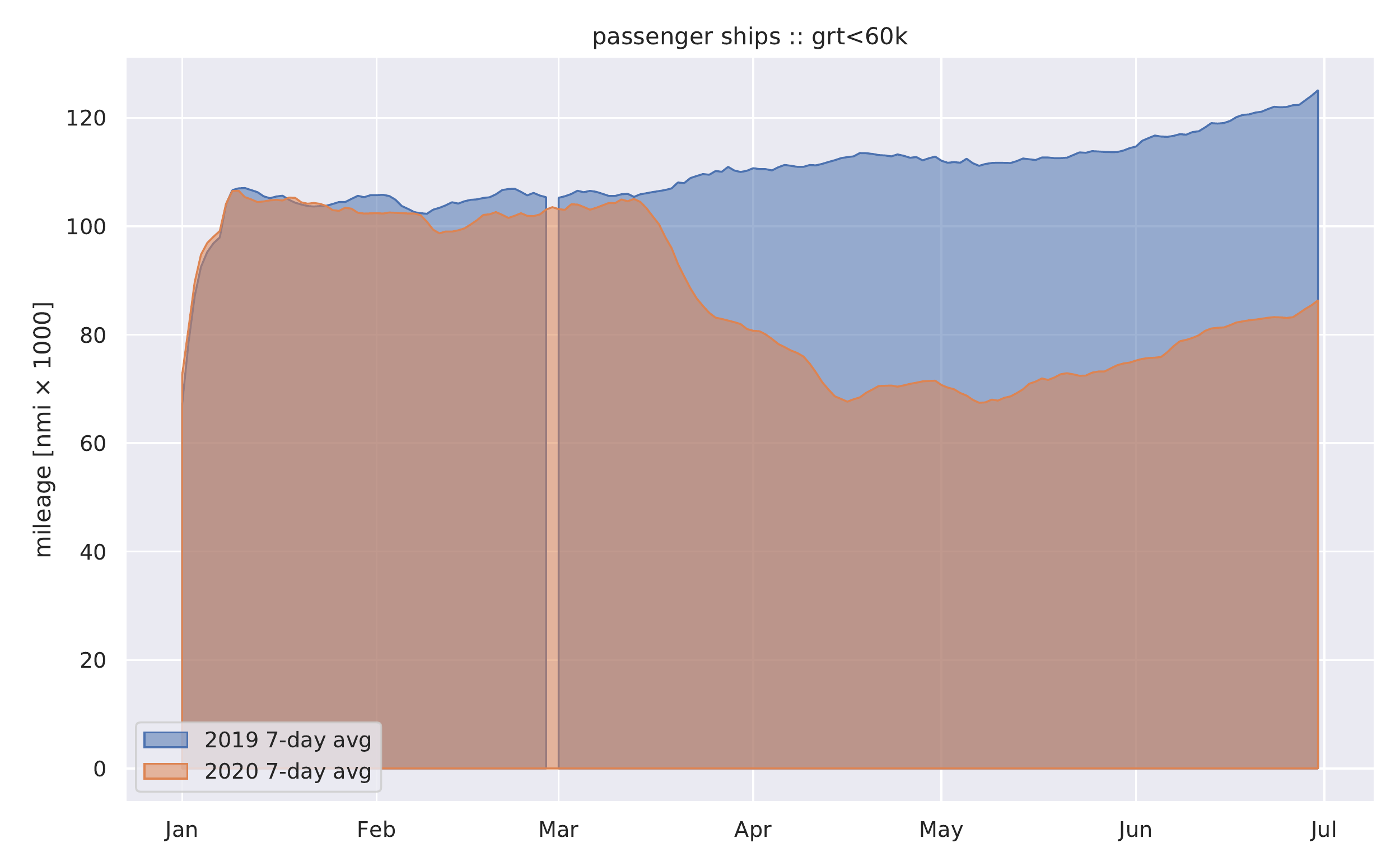}%
        \label{fig:dailymileagearea_passenger_lt_60k}%
        }%
    \hfil%
    \subfloat[][Passenger ships: 60K $<$ \acs{GT} $\leq$ 100K]{%
        \includegraphics[trim=15 10 10 15,clip,width=0.32\columnwidth]{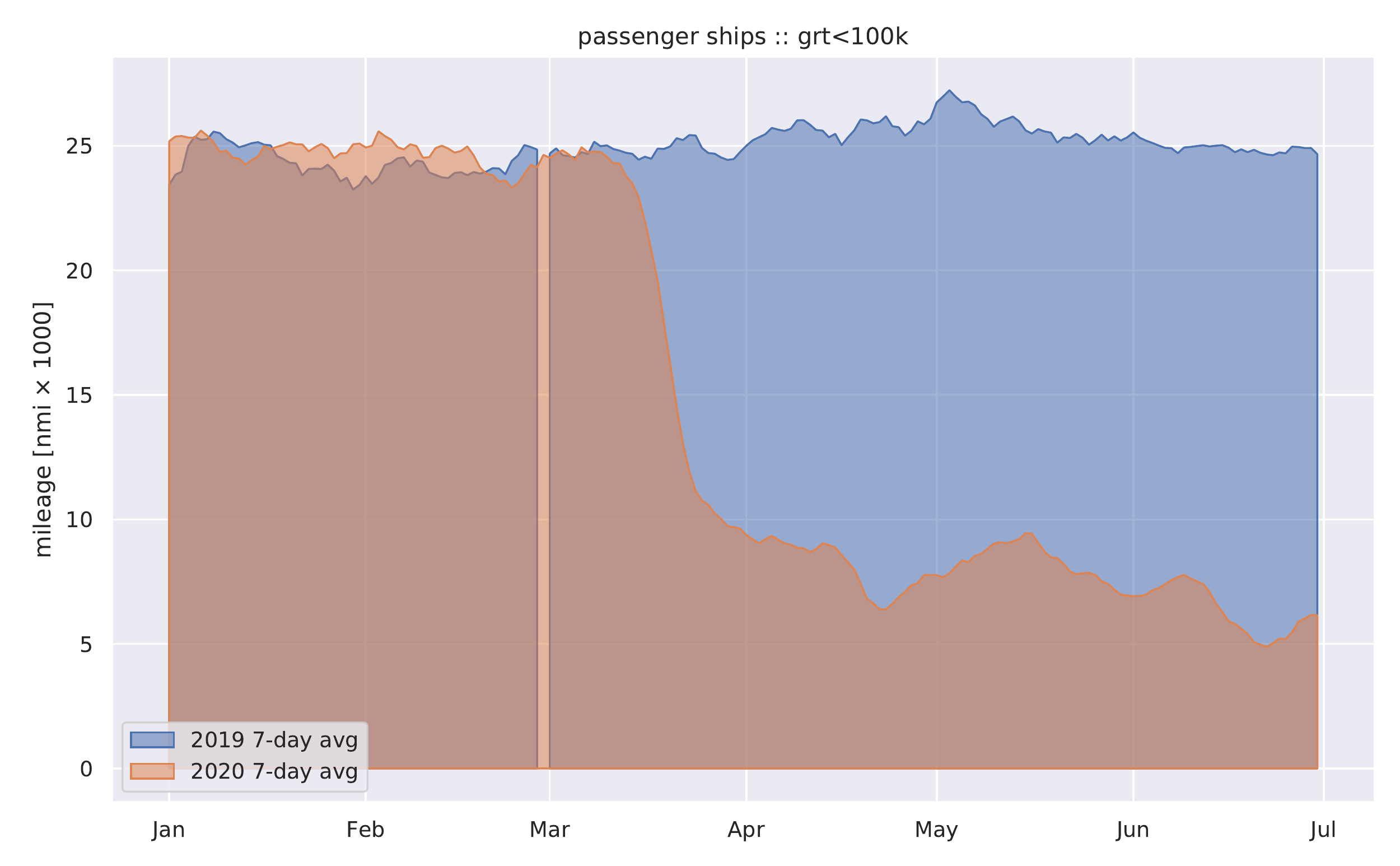}%
        \label{fig:dailymileagearea_passenger_lt_100k}%
        }%
    \\
    \subfloat[][Passenger ships: \acs{GT} $>$ 100K]{%
        \includegraphics[trim=15 10 10 15,clip,width=0.32\columnwidth]{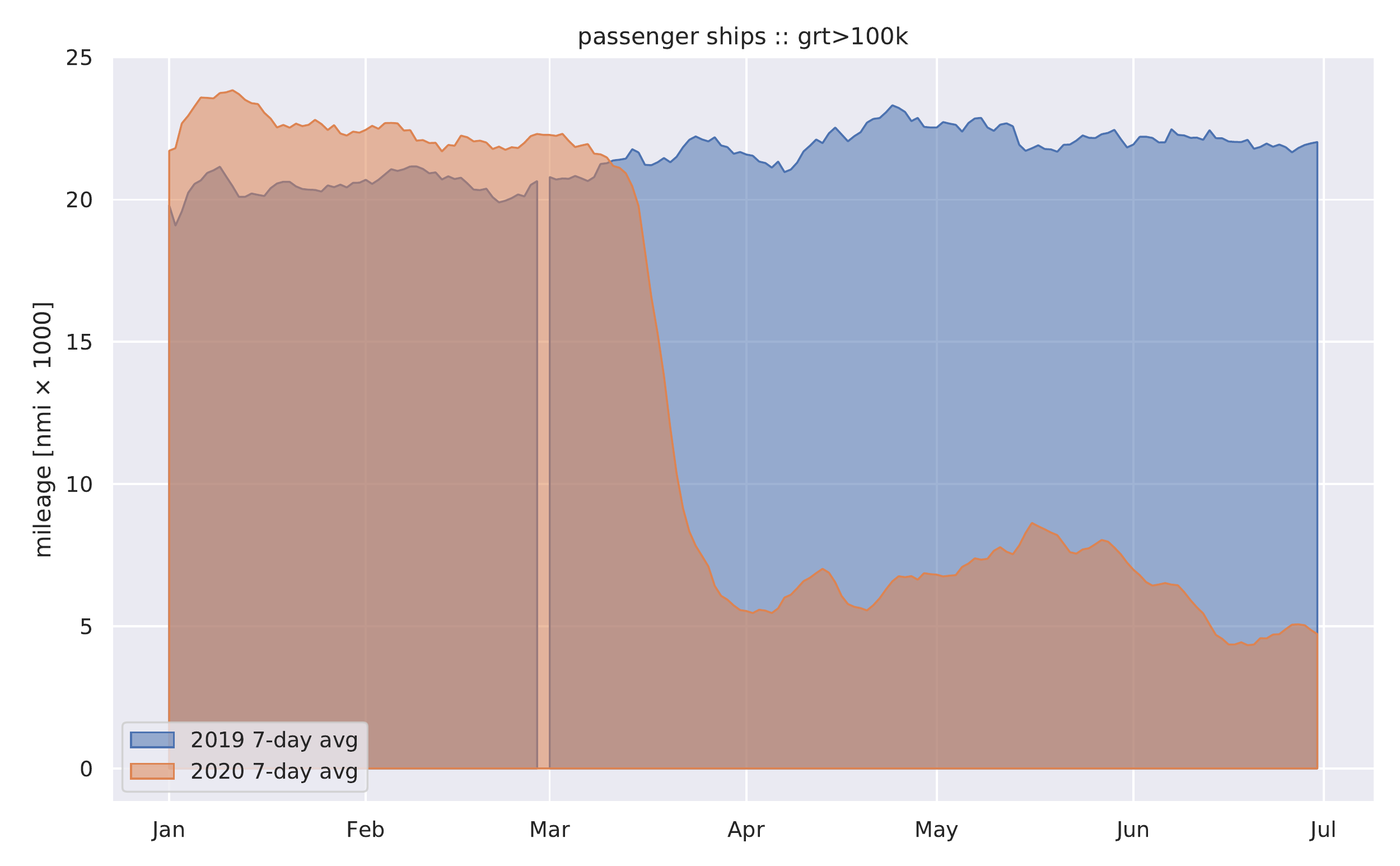}%
        \label{fig:dailymileagearea_passenger_gt_100k}%
        }%
    \hfil%
    \begin{minipage}[b]{0.64\columnwidth}
    \footnotesize{
        \caption{Comparison of daily miles navigated by passenger ships in the first six months of 2020 (orange) versus 2019 (blue) for different ship size categories, ordered by increasing \acf{GT}; the two area charts are overlaid in transparency to highlight trend differences; the discontinuity in the blue data series corresponds to the leap day absence in 2019.}%
        \label{fig:dailymilaeagearea_passenger}%
    }
    \end{minipage}
\end{figure}

For a closer look into each market, we report the daily navigated miles broken down by ship size. In Fig.~\ref{fig:dailymilaeagearea_container}, we compare the daily navigated miles for container ships according to their capacity. Specific segments show a stronger decline compared to others. Across all vessel sizes, with the exception of \acp{ULCV}, from late February or early March, there is a strong decrease of the navigated miles. As before, we must note that, since specific segments and types of ships operate on specific trade routes and regions, there could be regional effects that, in the global indicators, are ``averaged out''; in other words, local trends could be different than the global one. \rev{This is evident from the density map analysis, which show exactly how changes are spatially distributed.} However, if considered globally, the \ac{ULCV} market declined; ship operators cancelled their services, with a consequent \rev{decrease of both active ships and} navigated miles.
%
In Figures~\ref{fig:dailymilaeagearea_dry-bulk} and~\ref{fig:dailymilaeagearea_wet-bulk}, a similar analysis is available for \rev{dry and wet bulk carriers}. Dry bulk shipping refers to the movement of commodities carried in bulk: iron ore, coal, grain, steel products, lumber and other commodities classified as the minor bulks. For this category, only a small decrease of mileage across all vessel sizes is observed. In Fig.~\ref{fig:dailymilaeagearea_wet-bulk}, we report the daily navigated miles for wet bulk shipping broken down by ship capacity. Wet bulk cargoes include petroleum products, crude oil, vegetable oils, chemicals and similar products. The slowing demand in goods’ production and oil consumption had an effect on the mobility of all vessel sizes. The most significantly affected have been the larger tankers, specifically Panamax, Aframax and Suezmax sizes being affected the most; comprehensive information on ship classification by their size is available to the interested reader in the open literature.~\cite{MANDiesel_Container, MANDiesel_Bulk} Due to circumstances unrelated to COVID-19 (i.e., the breakdown of the OPEC alliance---which triggered a \SI{30}{\percent} fall in oil prices in March 2020), the mobility reduction is evident only after April 2020. %
Finally, Fig.~\ref{fig:dailymilaeagearea_passenger} reports the daily navigated miles of passenger vessels, the most affected segment by lockdown measures. The loss in terms of navigated miles in 2020 compared to 2019 is apparent, with larger vessel sizes affected by stronger losses than smaller ones. With respect to 2019, passenger ships larger than 60K GT, which includes large cruise ships, registered a sharp decrease of navigated miles of more than \SI{80}{\percent} since March 2020, when 
several cruise lines, including Carnival cruises, which alone owns more than 100 cruise ships, suspended the operations. The effects are evident from Fig.~\ref{fig:dailymilaeagearea_passenger}, with a sharp drop of navigated miles apparent since March. 

\subsection*{Analysis of daily average speed}
\begin{figure}
    \centering%
    \subfloat[][Container ships: Small feeder]{%
        \includegraphics[trim=15 10 10 15,clip,width=0.32\columnwidth]{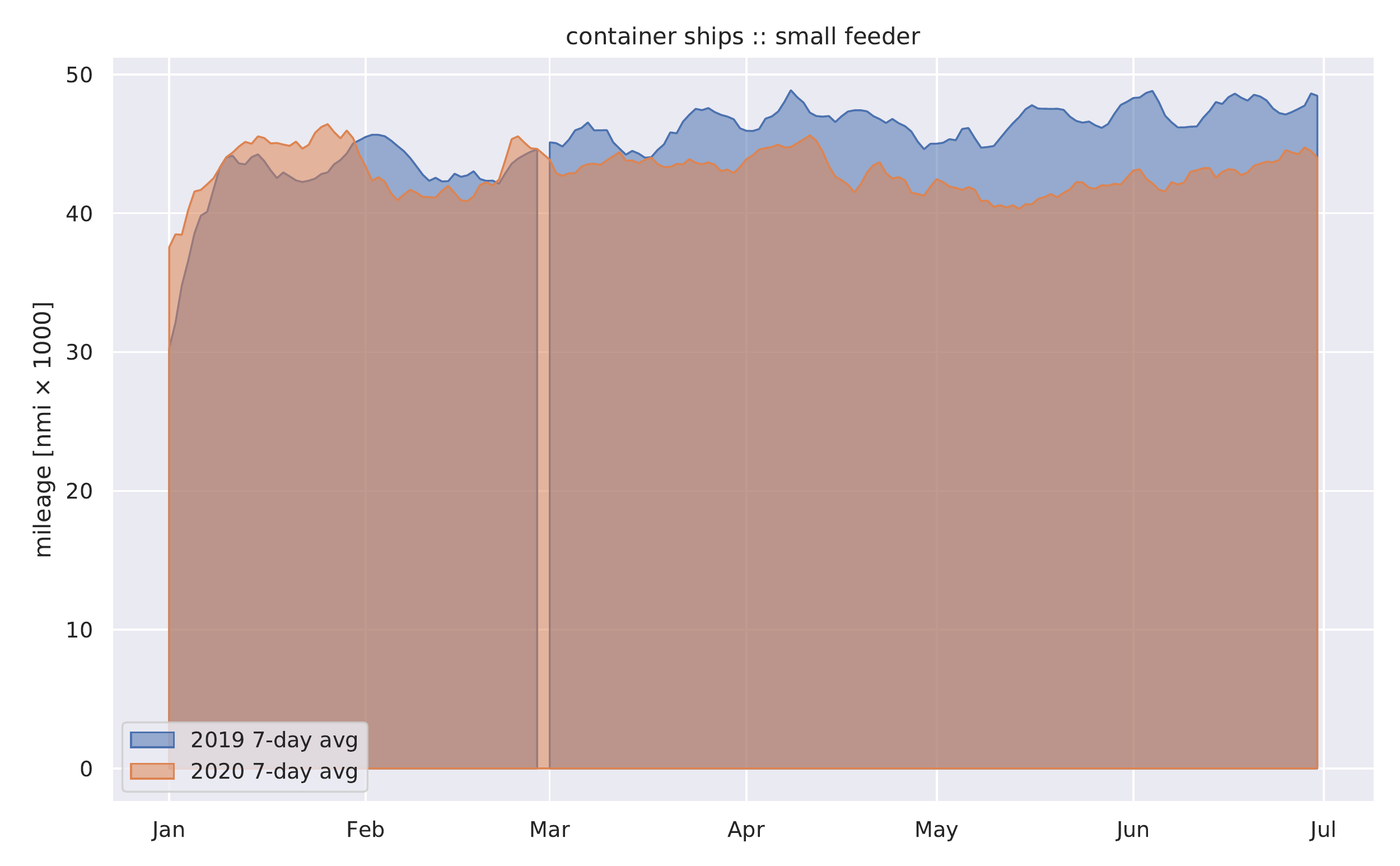}%
        \label{fig:dailymileagearea_container_small-feeder}%
        }%
    \hfil%
    \subfloat[][Container ships: Feeder]{%
        \includegraphics[trim=15 10 10 15,clip,width=0.32\columnwidth]{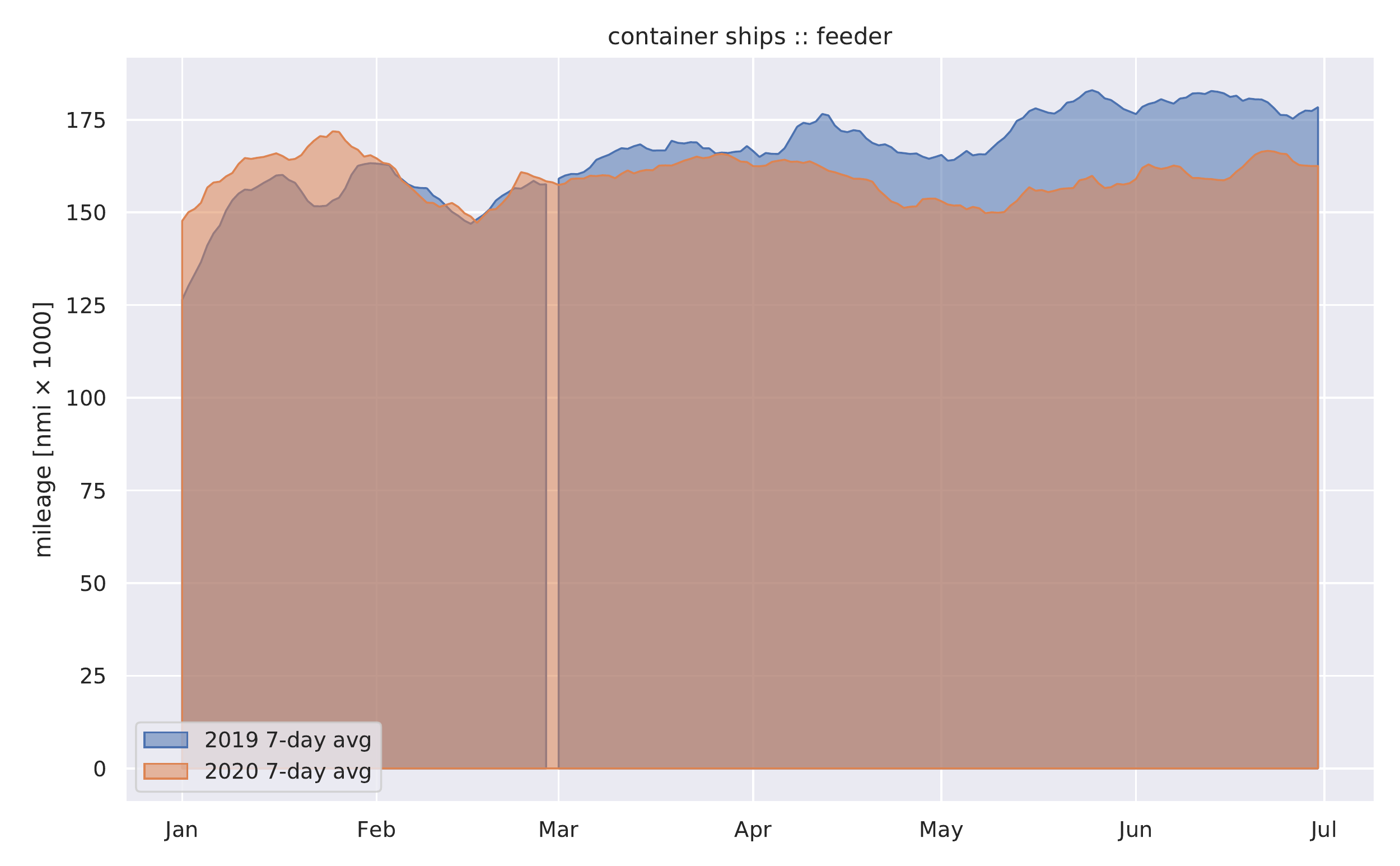}%
        \label{fig:dailymileagearea_container_feeder}%
        }%
    \hfil%
    \subfloat[][Container ships: Feedermax]{%
        \includegraphics[trim=15 10 10 15,clip,width=0.32\columnwidth]{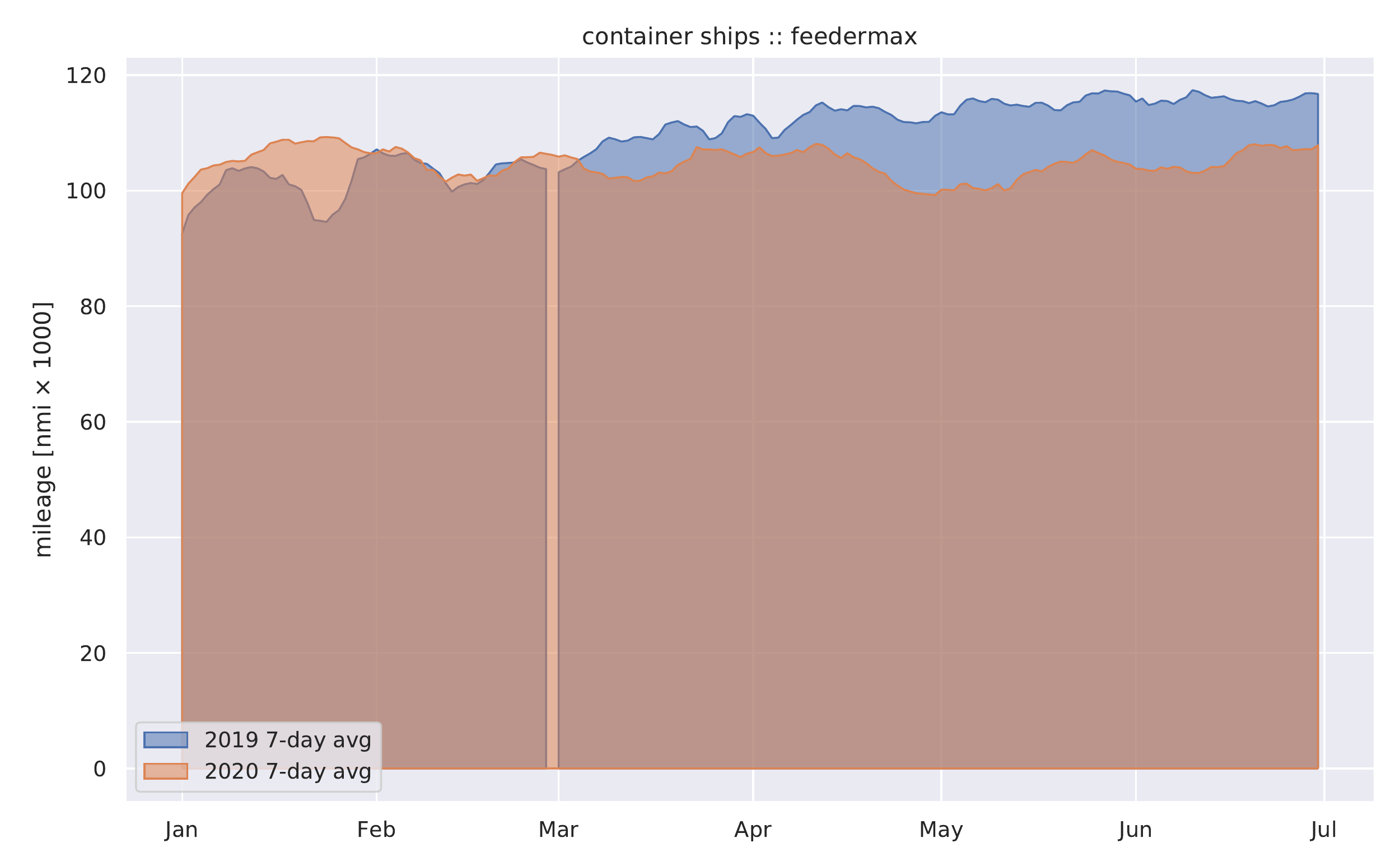}%
        }%
        \label{fig:dailymileagearea_container_feedermax}%
    \\
    \subfloat[][Container ships: Panamax]{%
        \includegraphics[trim=15 10 10 15,clip,width=0.32\columnwidth]{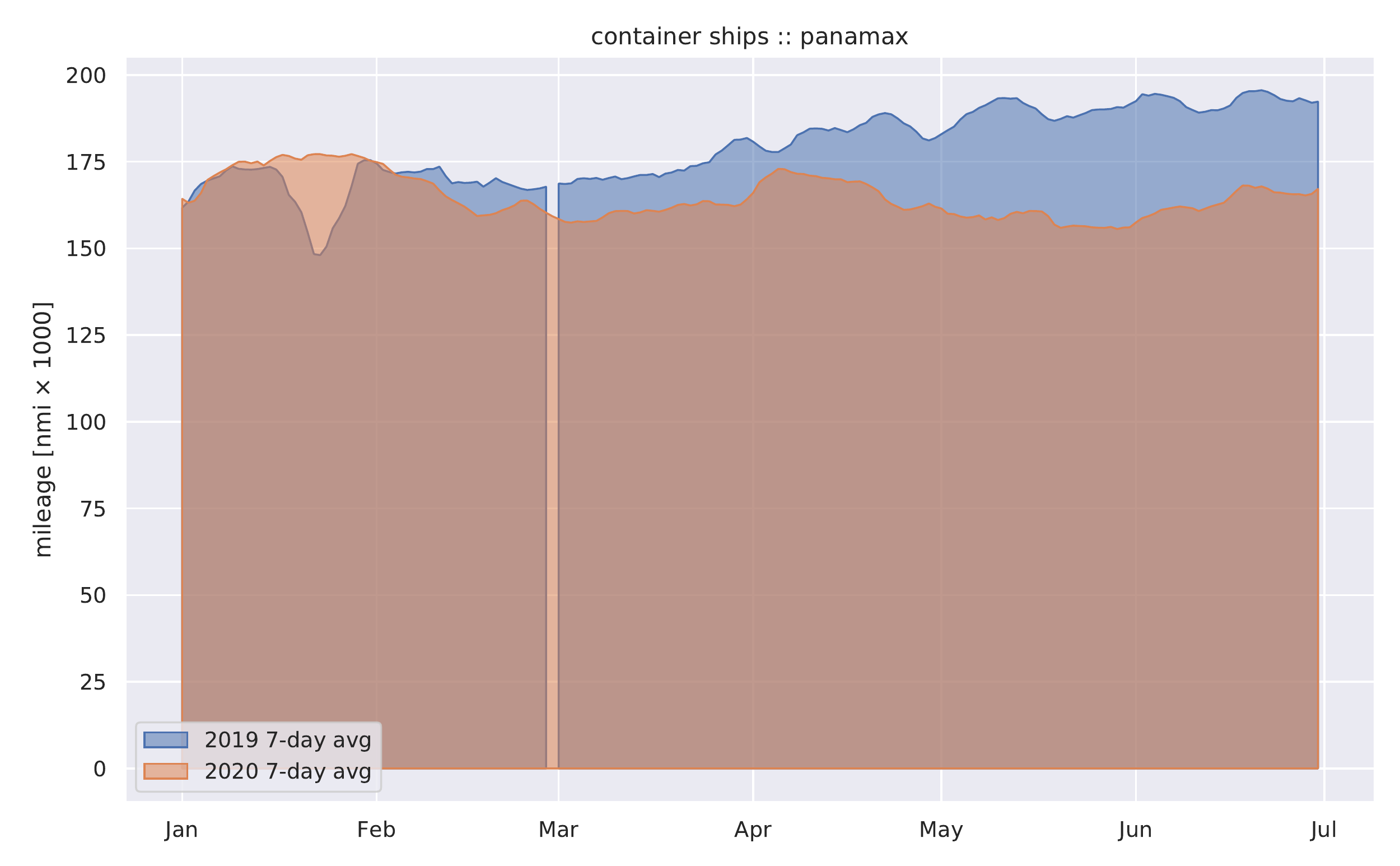}%
        }%
        \label{fig:dailymileagearea_container_panamax}%
    \hfil%
    \subfloat[][Container ships: Post-Panamax]{%
        \includegraphics[trim=15 10 10 15,clip,width=0.32\columnwidth]{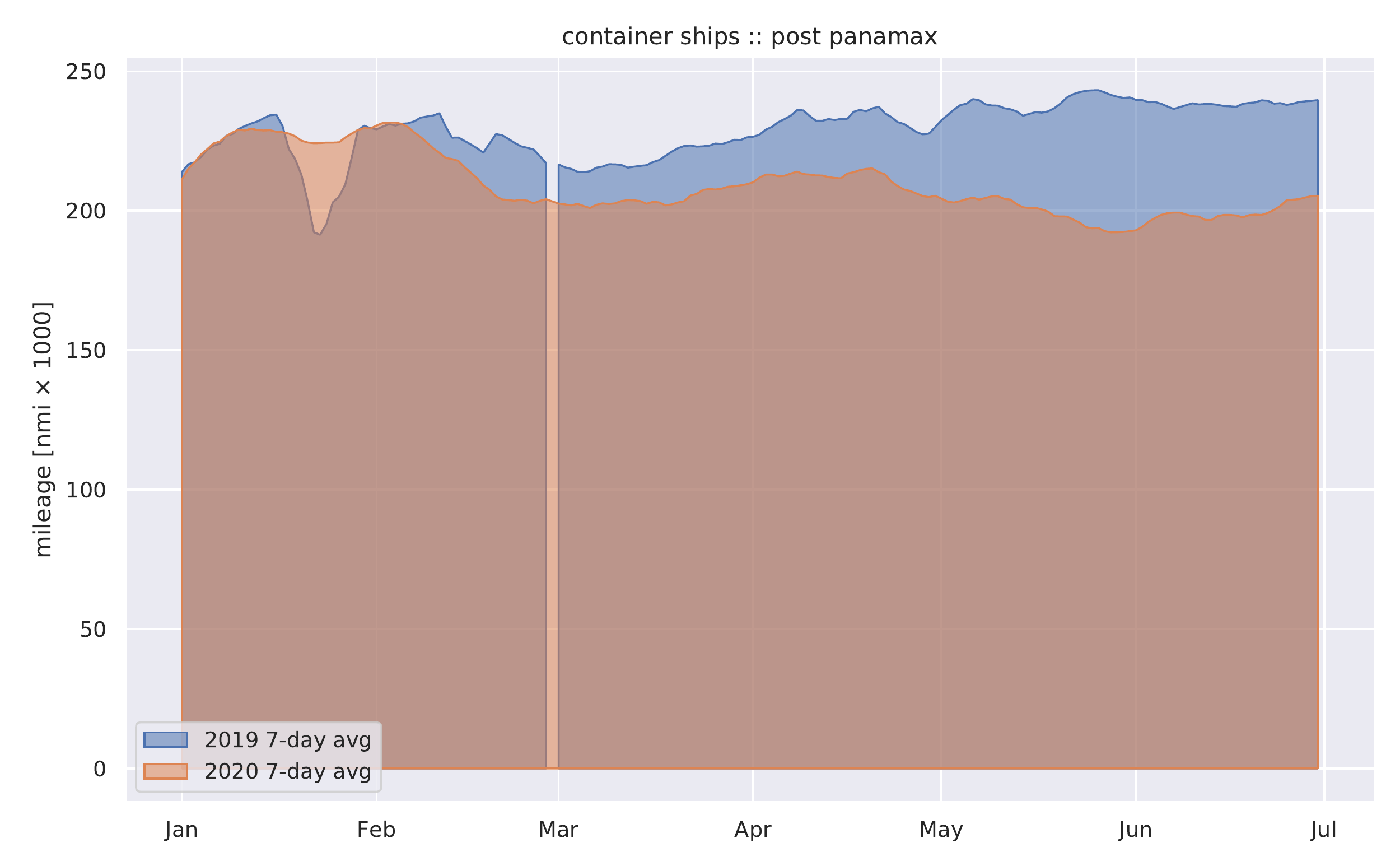}%
        }%
        \label{fig:dailymileagearea_container_post-panamax}%
    \hfil%
    \subfloat[][Container ships: New Panamax]{%
        \includegraphics[trim=15 10 10 15,clip,width=0.32\columnwidth]{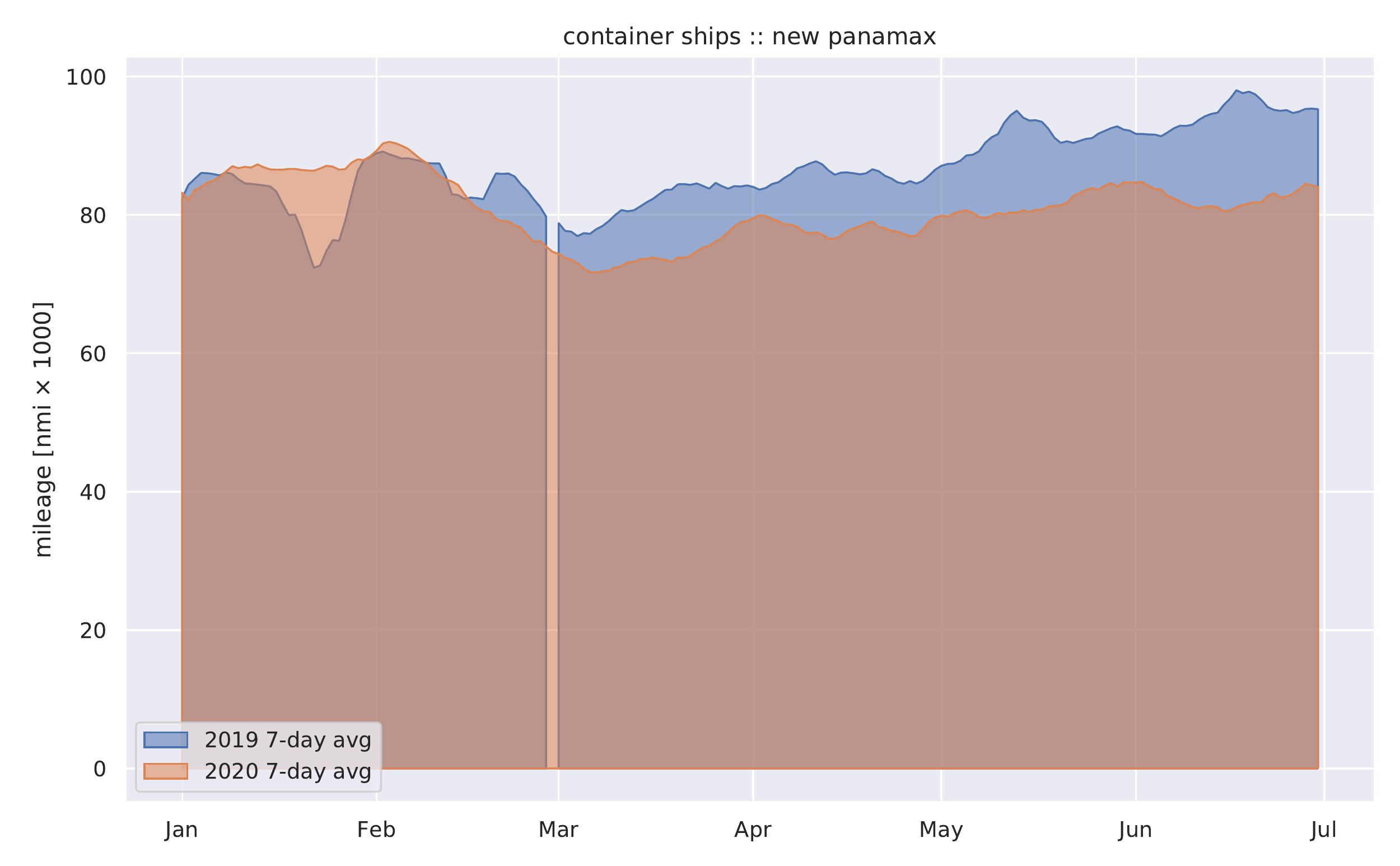}%
        }%
        \label{fig:dailymileagearea_container_new-panamax}%
    \\
    \subfloat[][Container ships: ULCV]{%
        \includegraphics[trim=15 10 10 15,clip,width=0.32\columnwidth]{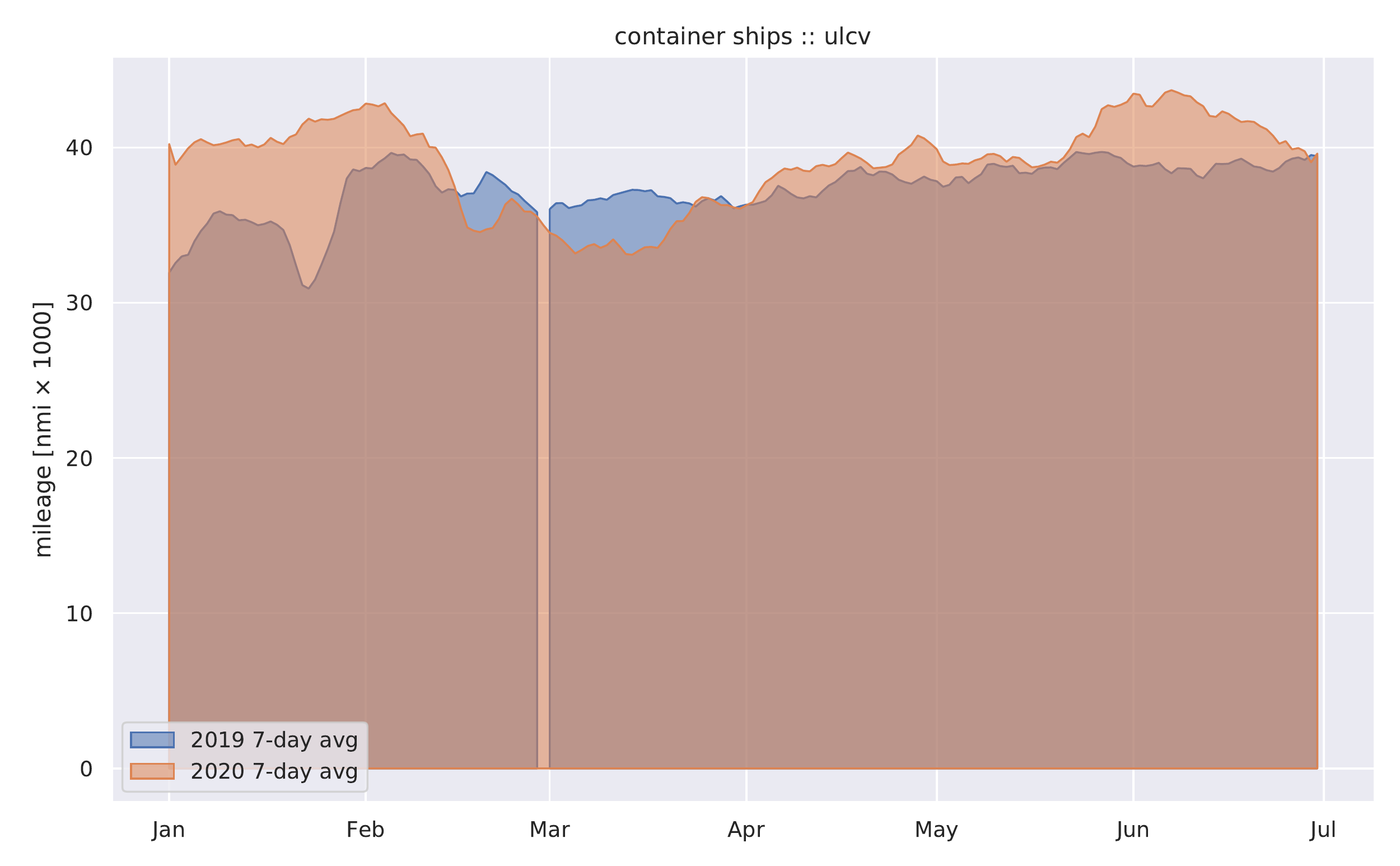}%
        \label{fig:dailymileagearea_container_ulcv}%
        }%
    \hfil%
    \begin{minipage}[b]{0.64\columnwidth}
    \footnotesize{
        \caption{Comparison of daily miles navigated by container ships in the first six months of 2020 (orange) versus 2019 (blue) for different ship size categories, ordered by increasing capacity measured in \acp{TEU}; the two area charts are overlaid in transparency to highlight trend differences; the discontinuity in the blue data series corresponds to the leap day absence in 2019.}%
        \label{fig:dailymilaeagearea_container}%
    }
    \end{minipage}
\end{figure}

\begin{figure}
    \centering%
    \subfloat[][Dry bulk: Handy size]{%
        \includegraphics[trim=15 10 10 15,clip,width=0.32\columnwidth]{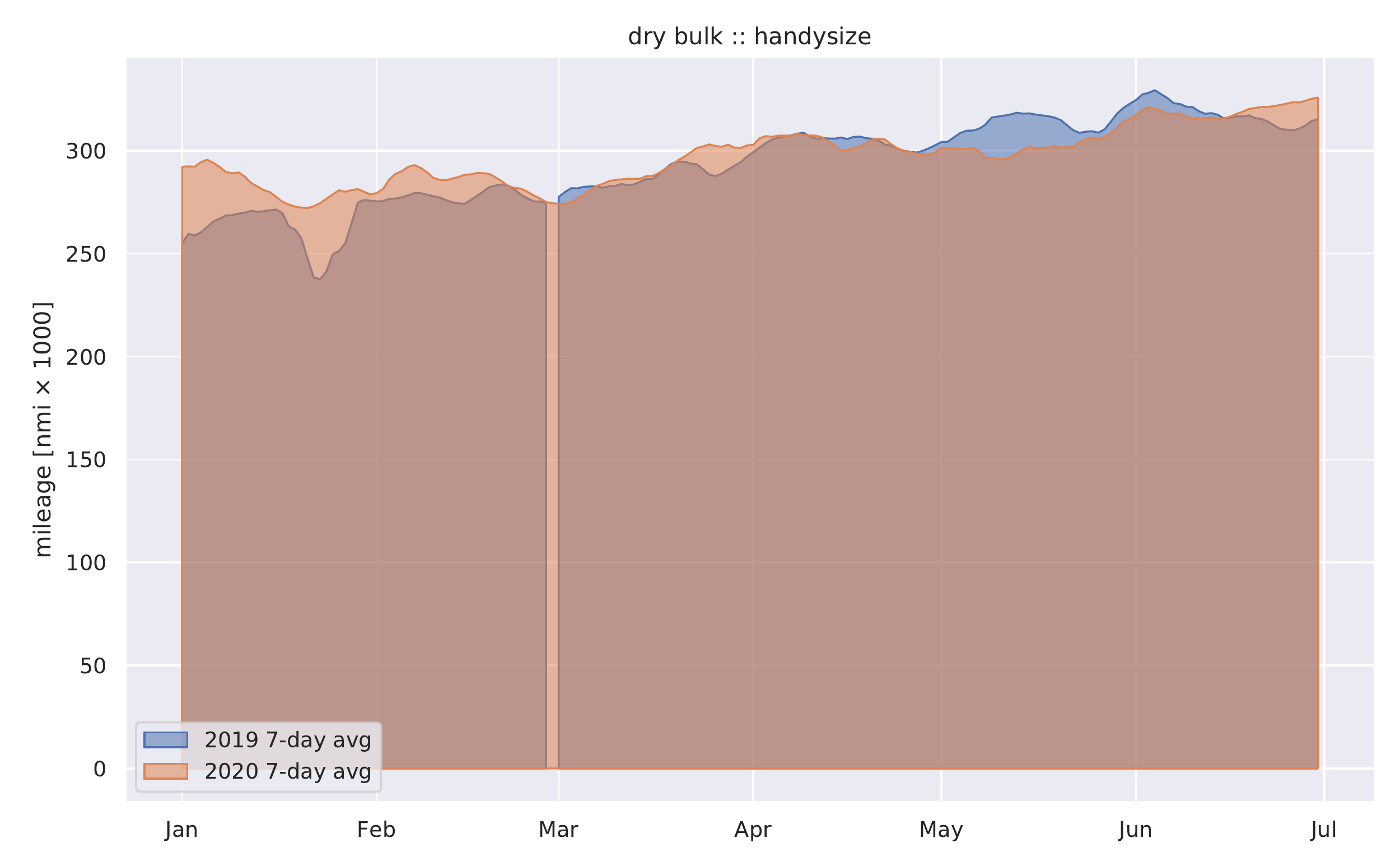}%
        \label{fig:dailymileagearea_dry_bulk_handysize}%
        }%
    \hfil%
    \subfloat[][Dry bulk: Handymax]{%
        \includegraphics[trim=15 10 10 15,clip,width=0.32\columnwidth]{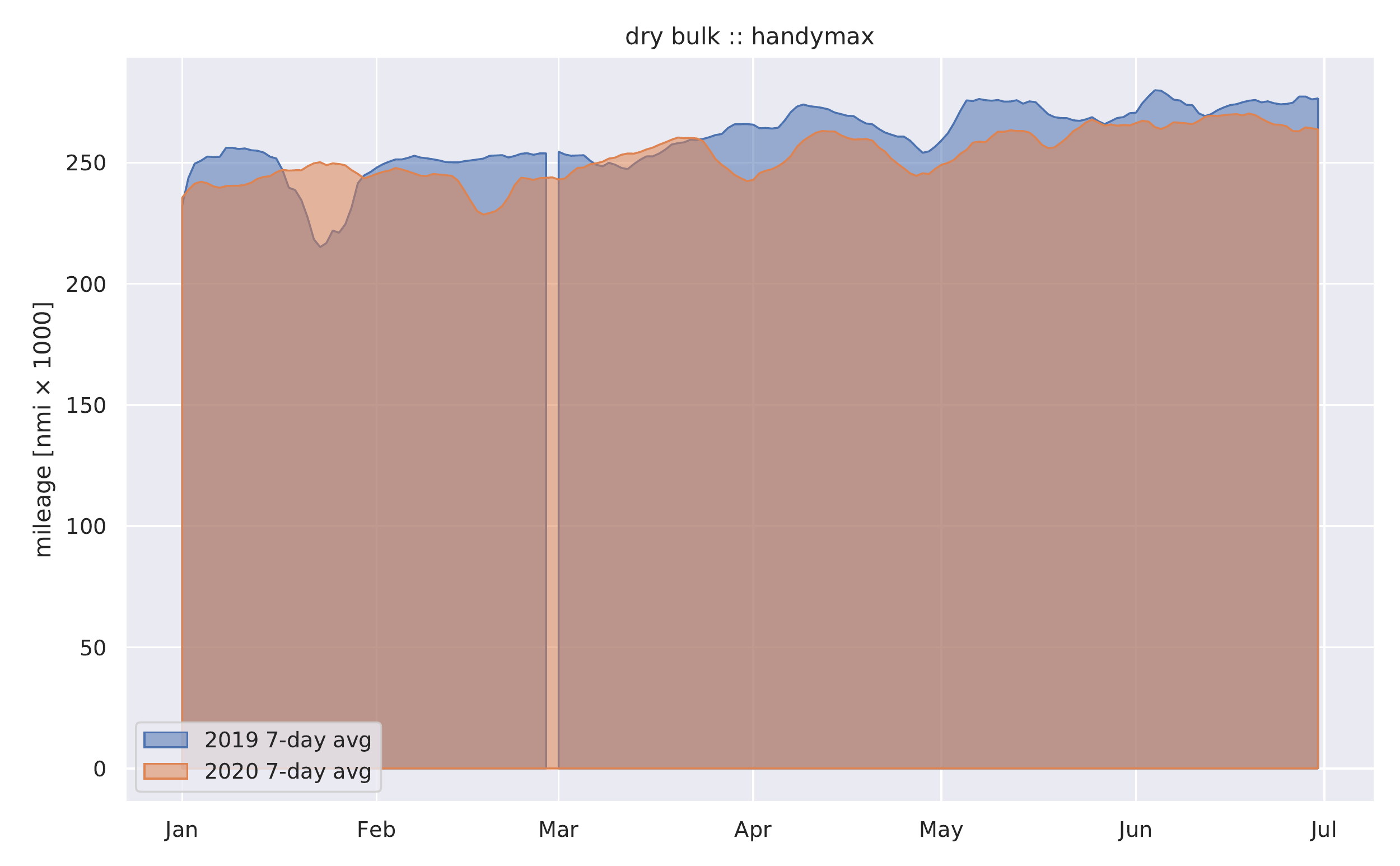}%
        \label{fig:dailymileagearea_dry_bulk_handymax}%
        }%
    \hfil%
    \subfloat[][Dry bulk: Panamax]{%
        \includegraphics[trim=15 10 10 15,clip,width=0.32\columnwidth]{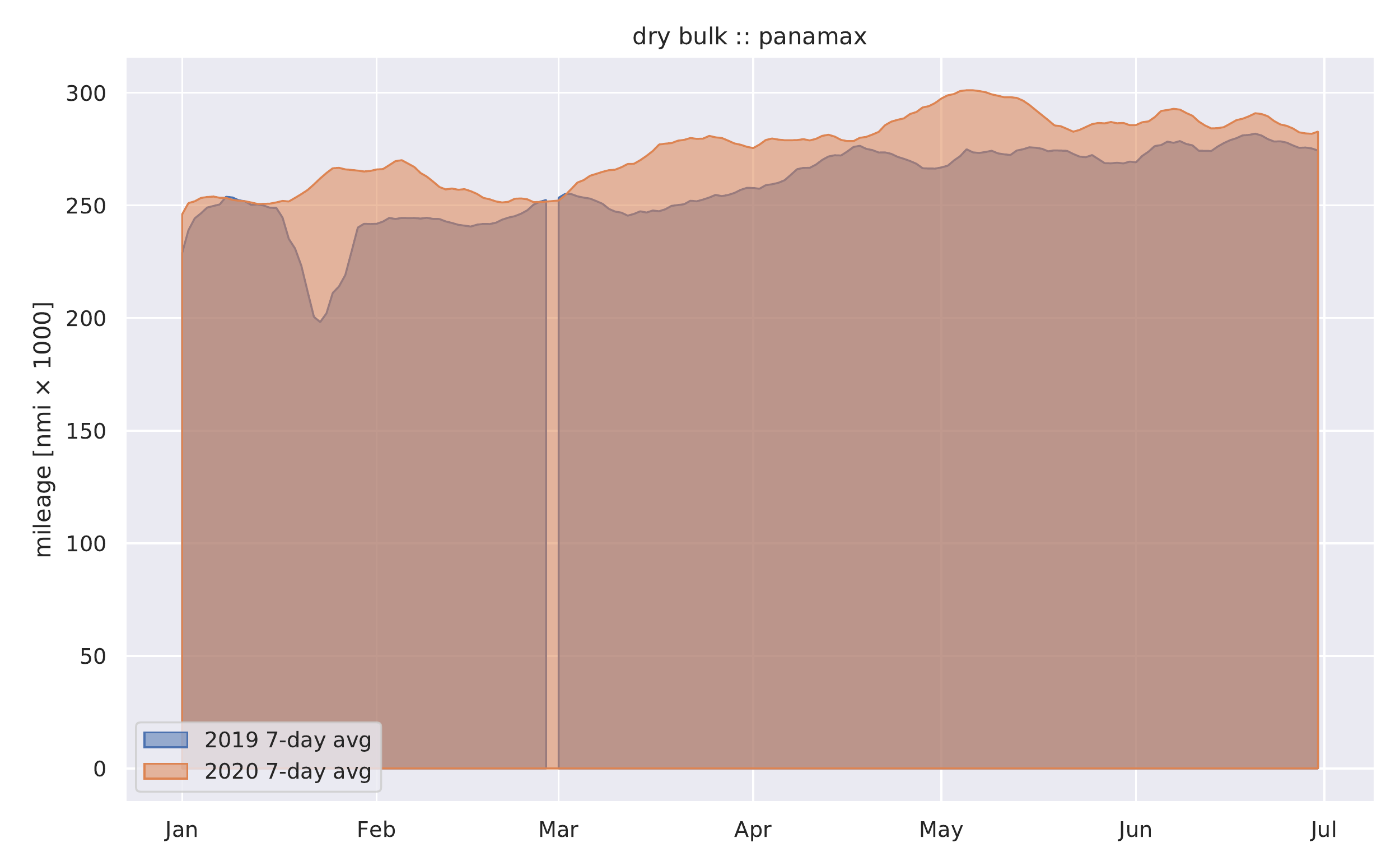}%
        \label{fig:dailymileagearea_dry_bulk_panamax}%
        }%
    \\
    \subfloat[][Dry bulk: Post-Panamax]{%
        \includegraphics[trim=15 10 10 15,clip,width=0.32\columnwidth]{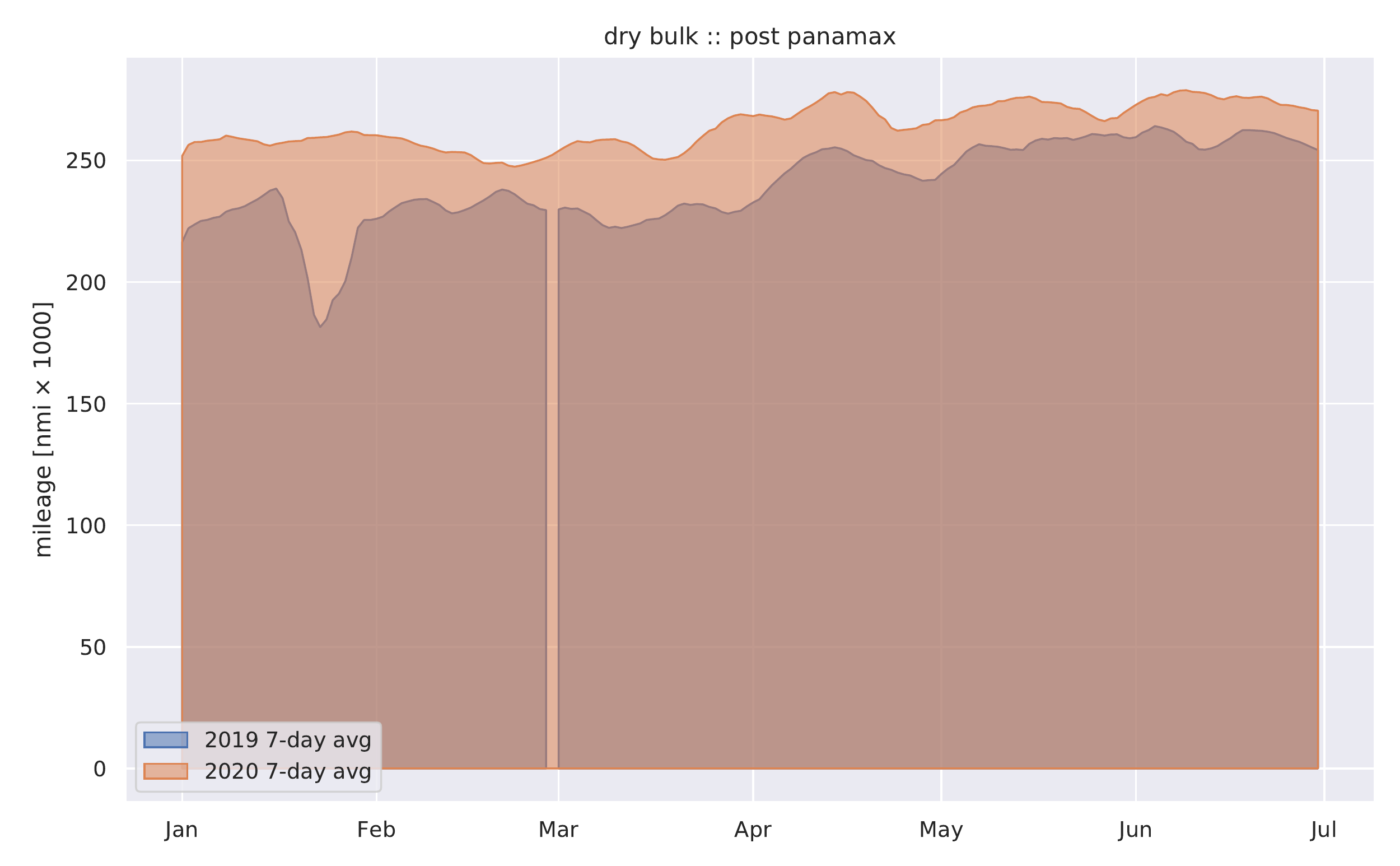}%
        \label{fig:dailymileagearea_dry_bulk_post-panamax}%
        }%
    \hfil%
    \subfloat[][Dry bulk: VLBC]{%
        \includegraphics[trim=15 10 10 15,clip,width=0.32\columnwidth]{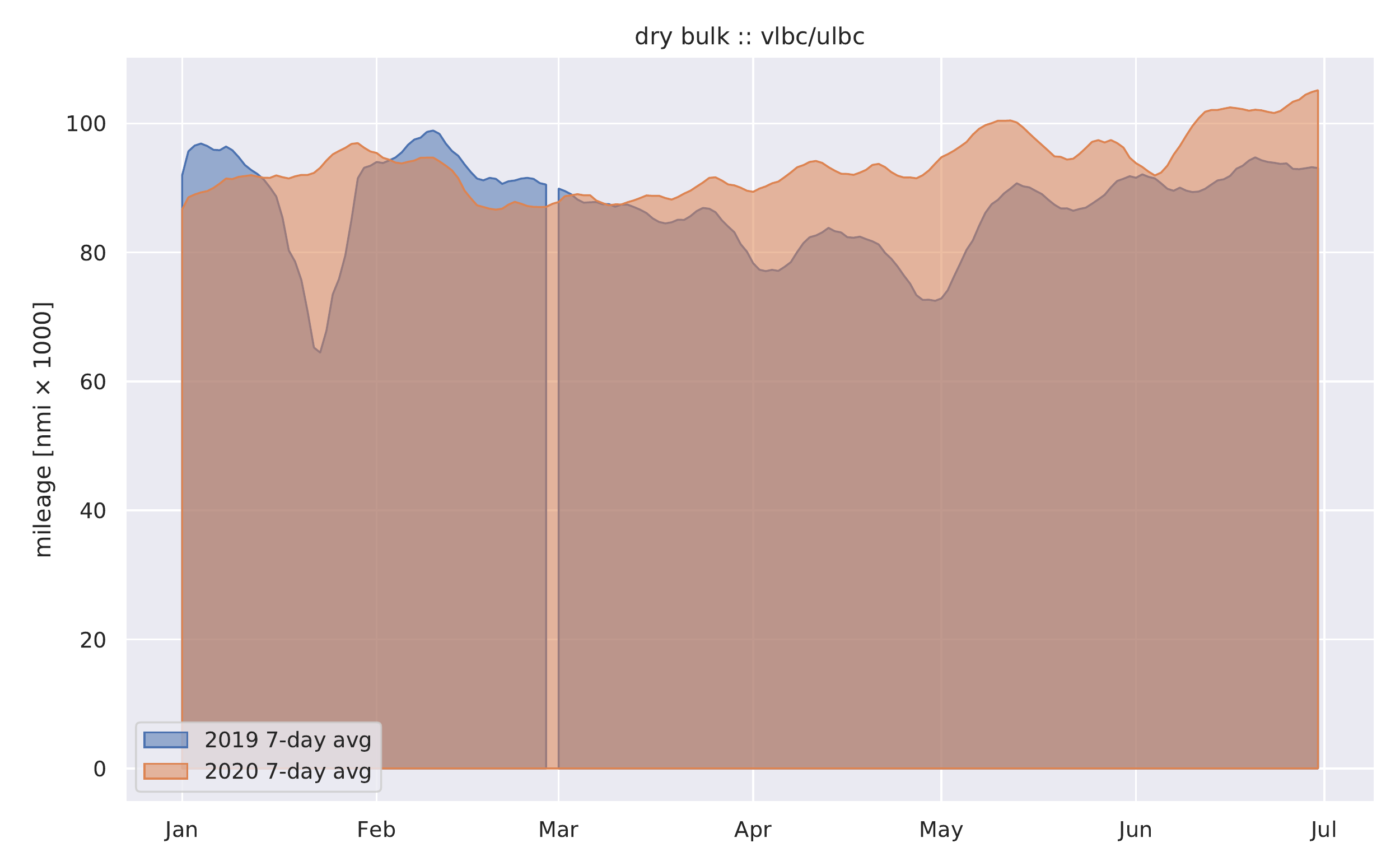}%
        \label{fig:dailymileagearea_dry_bulk_vlbc}%
        }%
    \hfil%
    \subfloat[][Dry bulk: Capesize]{%
        \includegraphics[trim=15 10 10 15,clip,width=0.32\columnwidth]{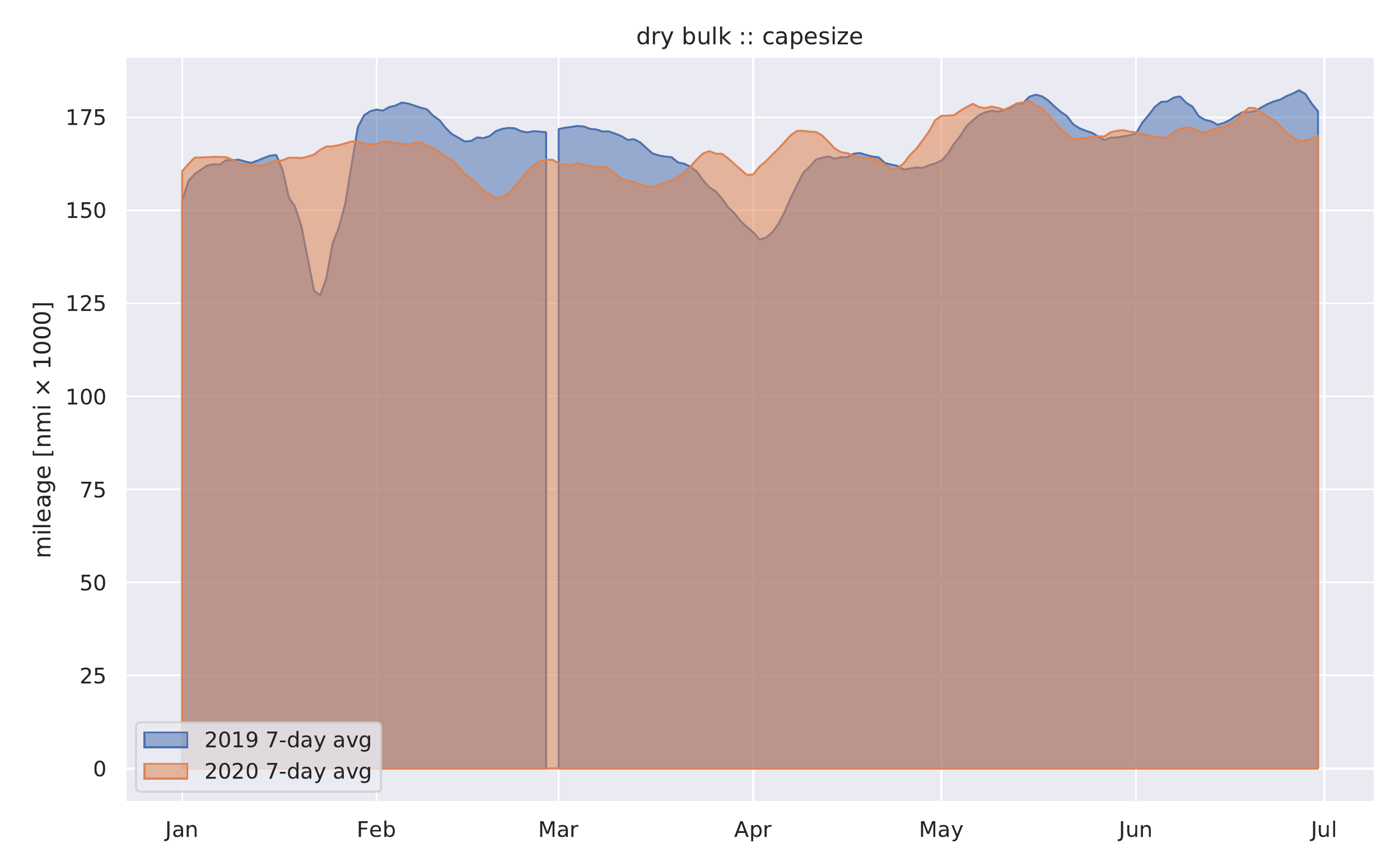}%
        \label{fig:dailymileagearea_dry_bulk_capesize}%
        }
    \caption{Comparison of daily miles navigated by dry bulk ships in the first six months of 2020 (orange) versus 2019 (blue) for different ship size categories, ordered by increasing \ac{DWT}; the two area charts are overlaid in transparency to highlight trend differences; the discontinuity in the blue data series corresponds to the leap day absence in 2019.}%
    \label{fig:dailymilaeagearea_dry-bulk}%
\end{figure}

\begin{figure}
    \centering%
    \subfloat[][Wet bulk: Handy size]{%
        \includegraphics[trim=15 10 10 15,clip,width=0.32\columnwidth]{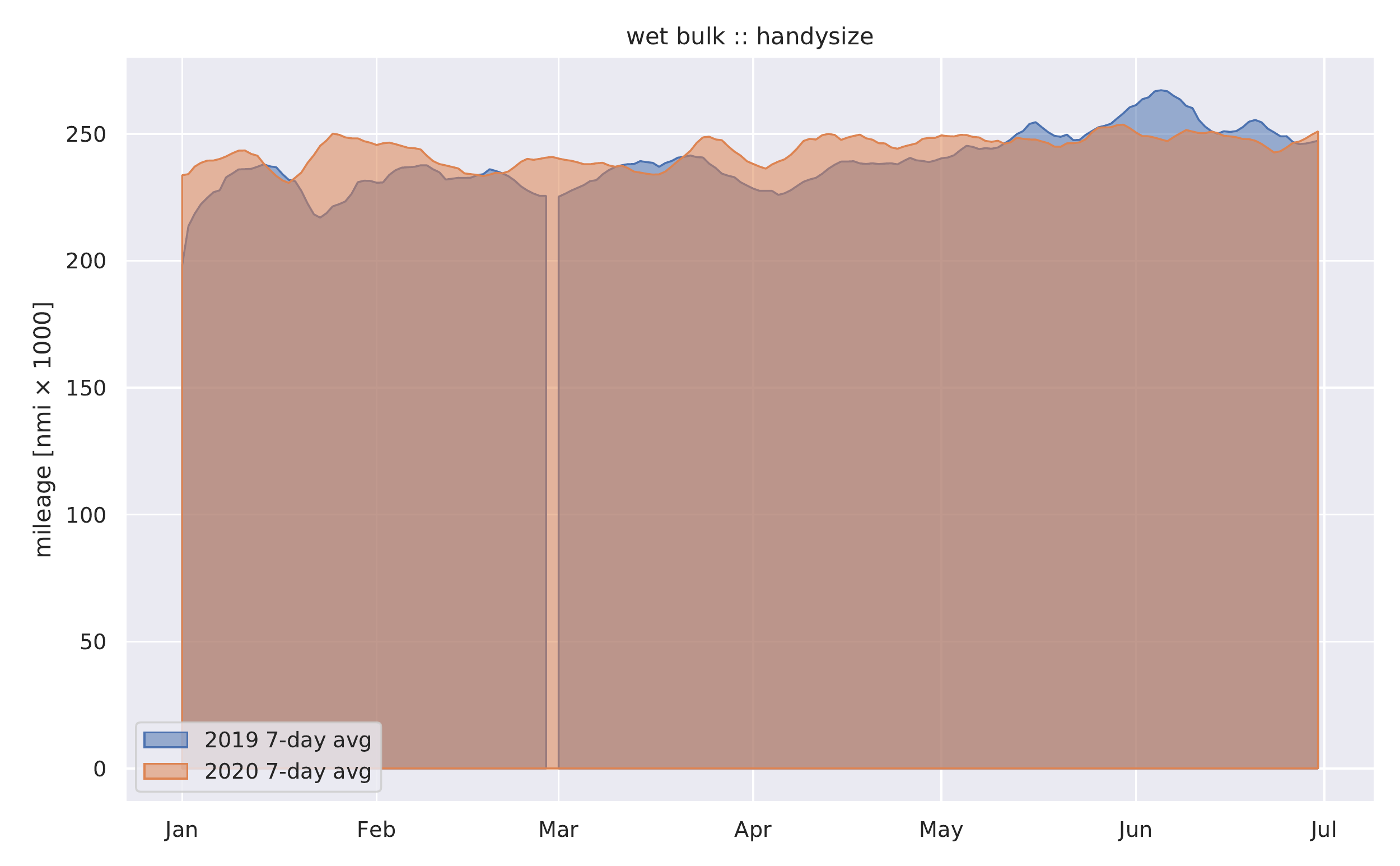}%
        \label{fig:dailymileagearea_wet_bulk_handysize}%
        }%
    \hfil%
    \subfloat[][Wet bulk: Handymax]{%
        \includegraphics[trim=15 10 10 15,clip,width=0.32\columnwidth]{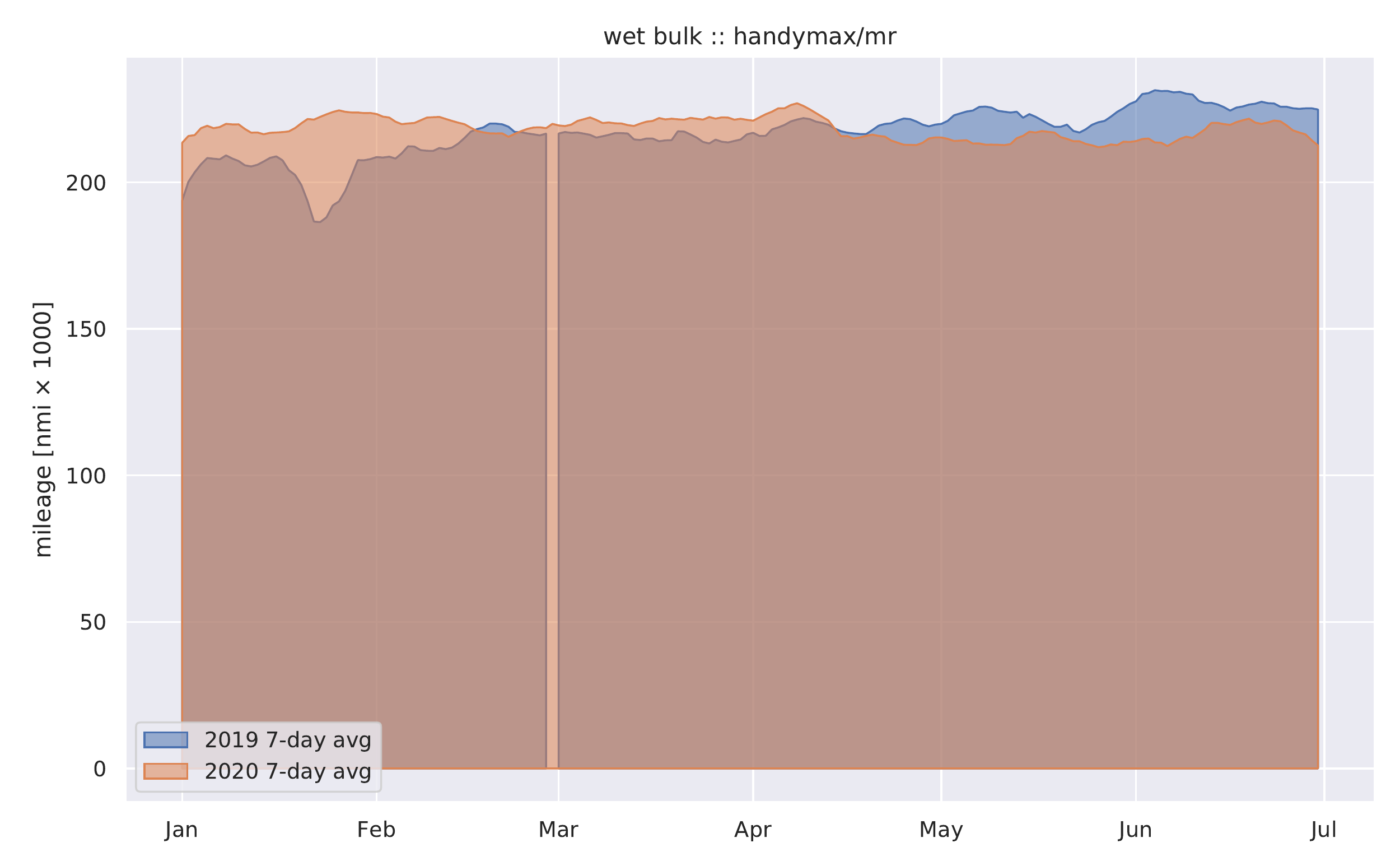}%
        \label{fig:dailymileagearea_wet_bulk_handymax}%
        }%
    \hfil%
    \subfloat[][Wet bulk: Panamax]{%
        \includegraphics[trim=15 10 10 15,clip,width=0.32\columnwidth]{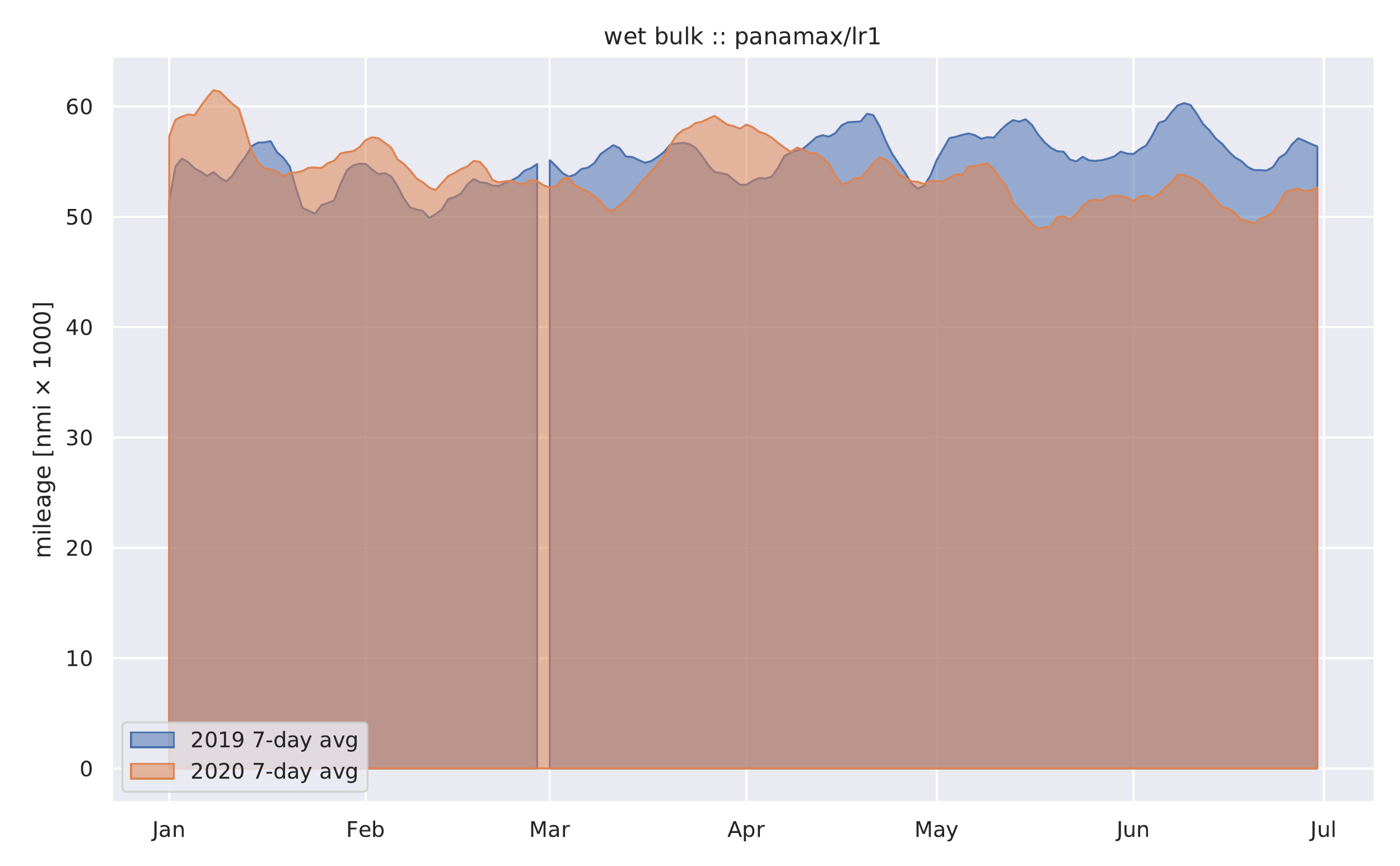}%
        \label{fig:dailymileagearea_wet_bulk_panamax}%
        }%
    \\
    \subfloat[][Wet bulk: Aframax]{%
        \includegraphics[trim=15 10 10 15,clip,width=0.32\columnwidth]{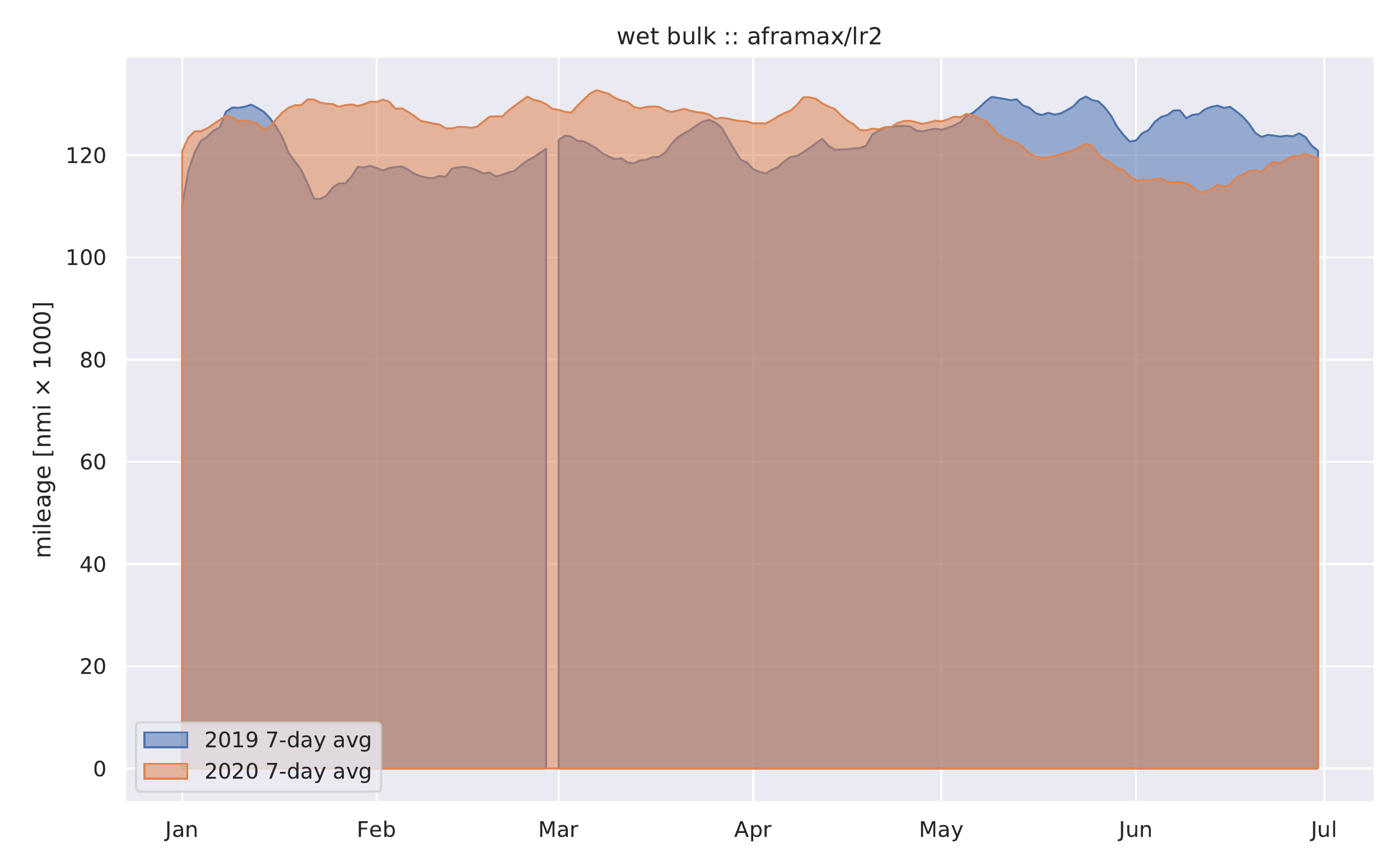}%
        \label{fig:dailymileagearea_wet_bulk_aframax}%
        }%
    \hfil%
    \subfloat[][Wet bulk: Suezmax]{%
        \includegraphics[trim=15 10 10 15,clip,width=0.32\columnwidth]{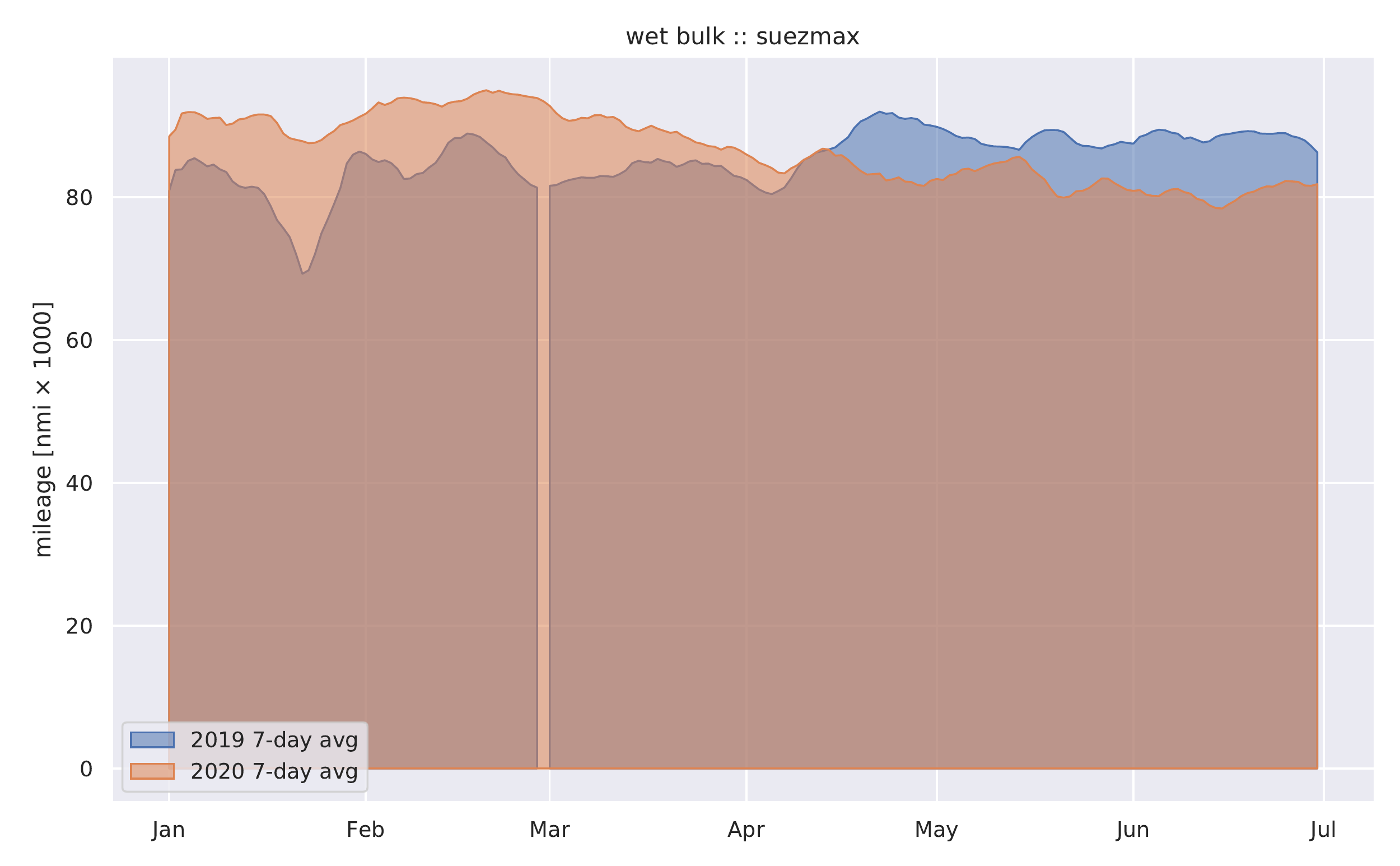}%
        \label{fig:dailymileagearea_wet_bulk_suezmax}%
        }%
    \hfil%
    \subfloat[][Wet bulk: VLCC]{%
        \includegraphics[trim=15 10 10 15,clip,width=0.32\columnwidth]{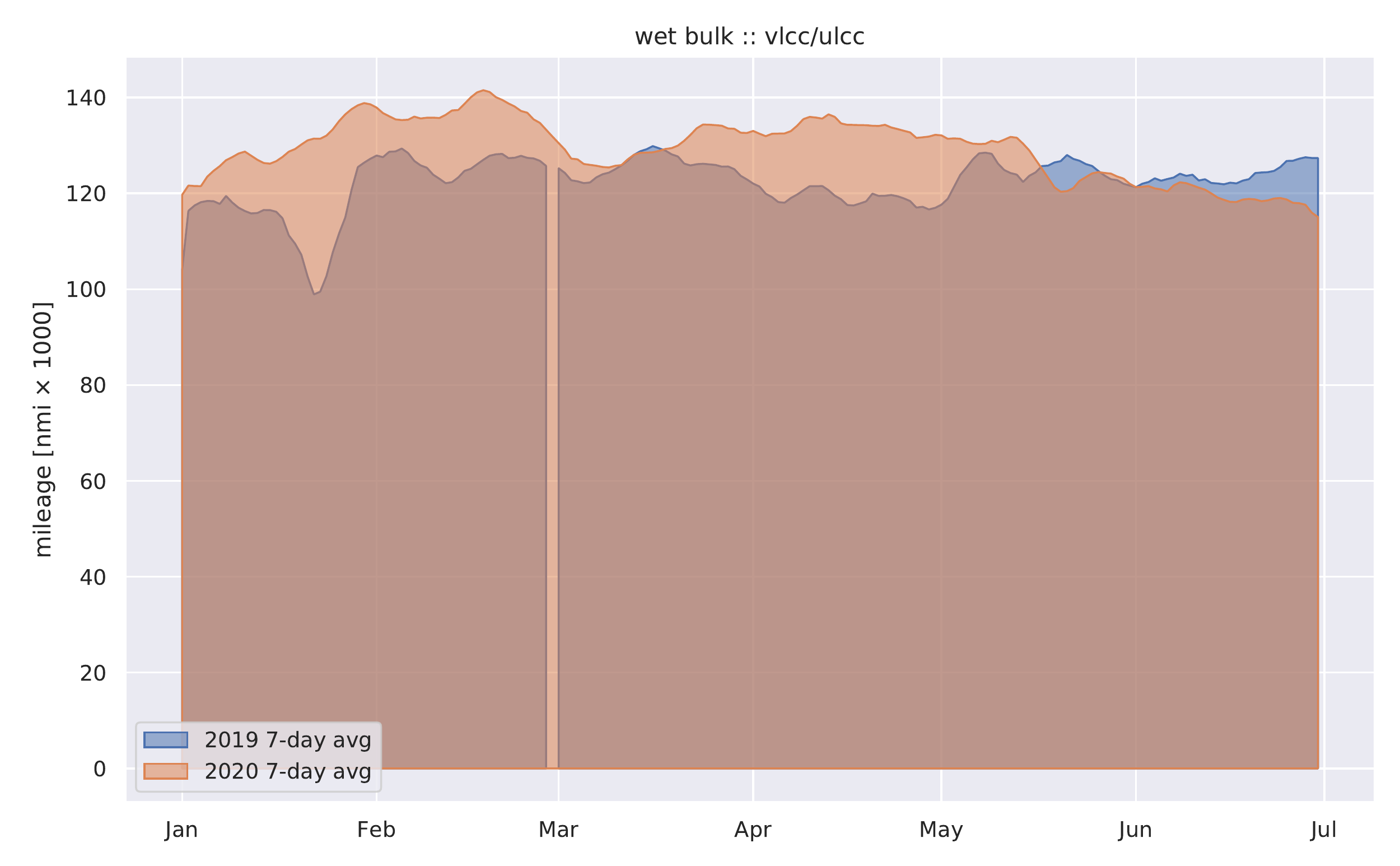}%
        \label{fig:dailymileagearea_wet_bulk_vlcc}%
        }%
    \caption{Comparison of daily miles navigated by wet bulk ships in the first six months of 2020 (orange) versus 2019 (blue) for different ship size categories, ordered by increasing \ac{DWT}; the two area charts are overlaid in transparency to highlight trend differences; the discontinuity in the blue data series corresponds to the leap day absence in 2019.}%
    \label{fig:dailymilaeagearea_wet-bulk}%
\end{figure}

\begin{figure}
    \centering%
    \subfloat[][Container ships: Small feeder]{%
        \includegraphics[trim=15 10 10 15,clip,width=0.32\columnwidth]{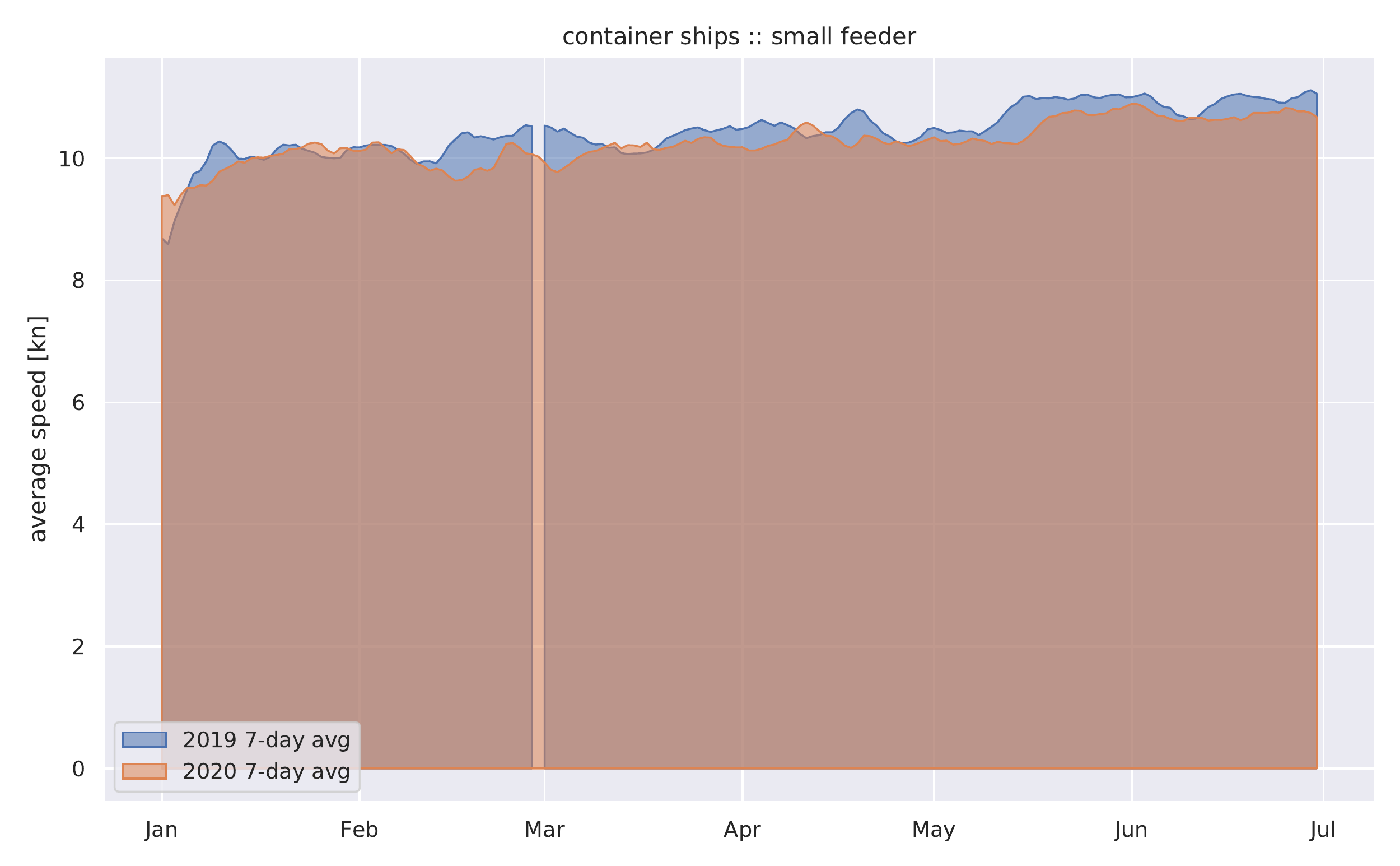}%
        \label{fig:dailyspeedarea_container_small-feeder}%
        }%
    \hfil%
    \subfloat[][Container ships: Feeder]{%
        \includegraphics[trim=15 10 10 15,clip,width=0.32\columnwidth]{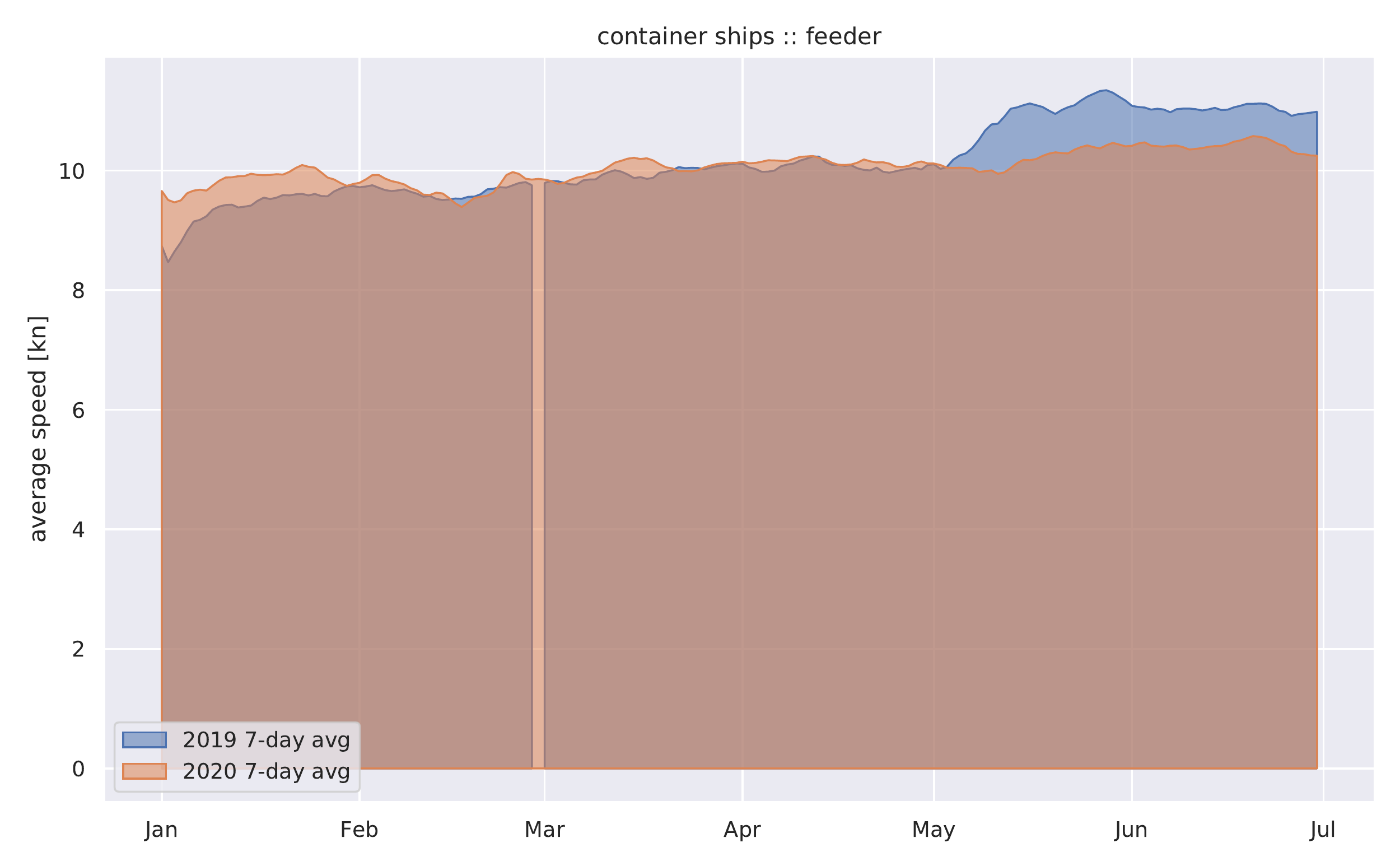}%
        \label{fig:dailyspeedarea_container_feeder}%
        }%
    \hfil%
    \subfloat[][Container ships: Feedermax]{%
        \includegraphics[trim=15 10 10 15,clip,width=0.32\columnwidth]{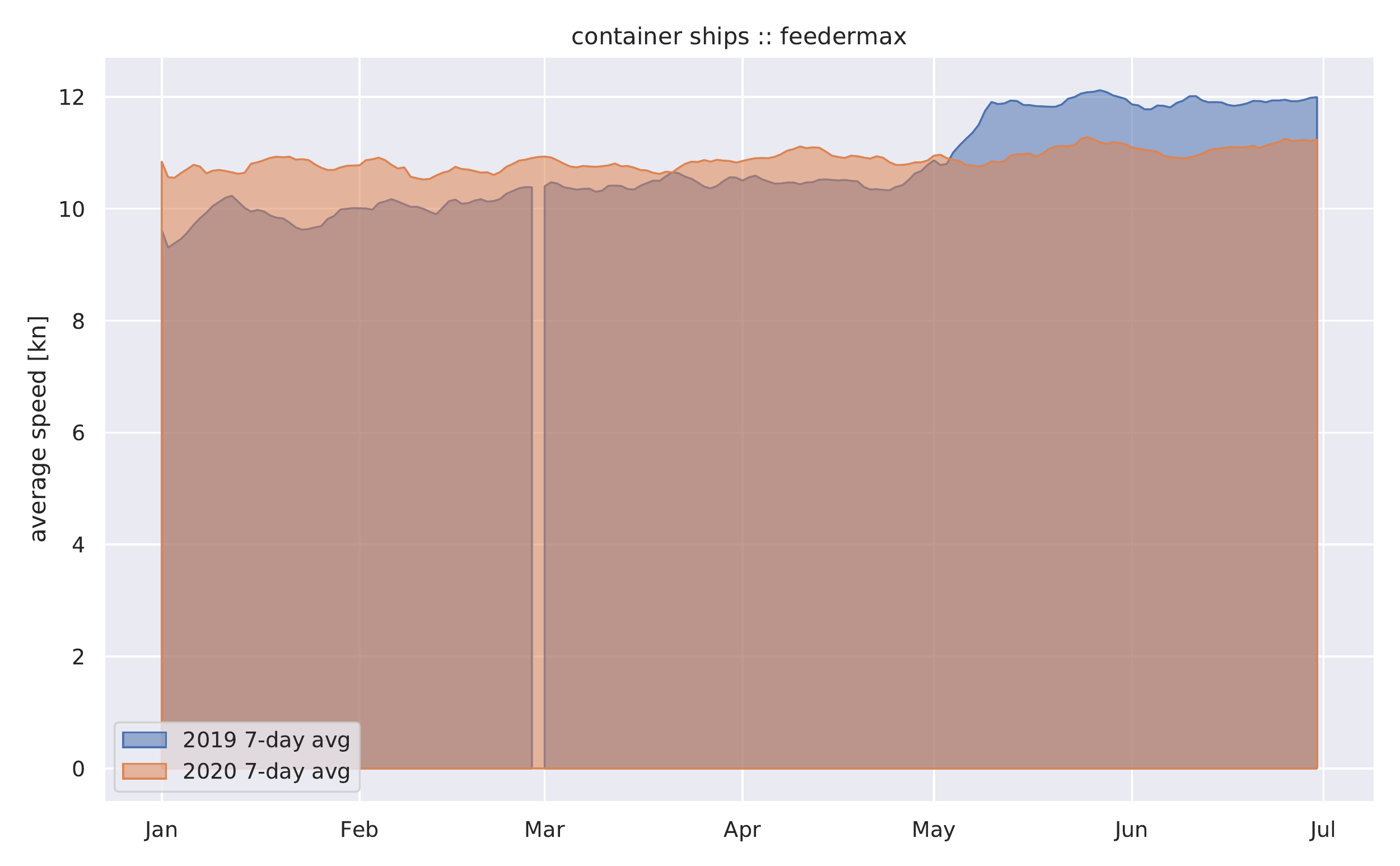}%
        }%
        \label{fig:dailyspeedarea_container_feedermax}%
    \\
    \subfloat[][Container ships: Panamax]{%
        \includegraphics[trim=15 10 10 15,clip,width=0.32\columnwidth]{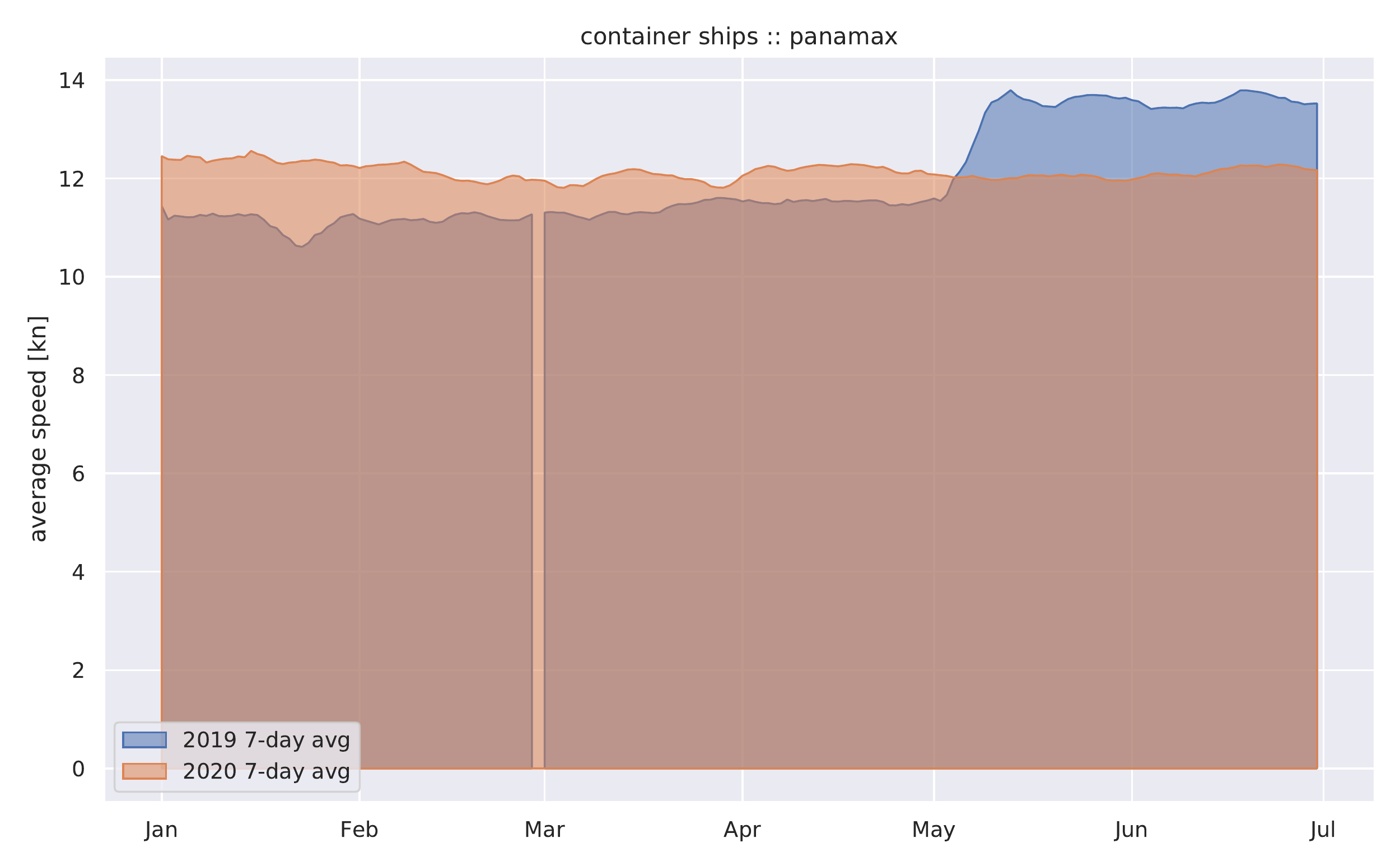}%
        }%
        \label{fig:dailyspeedarea_container_panamax}%
    \hfil%
    \subfloat[][Container ships: Post-Panamax]{%
        \includegraphics[trim=15 10 10 15,clip,width=0.32\columnwidth]{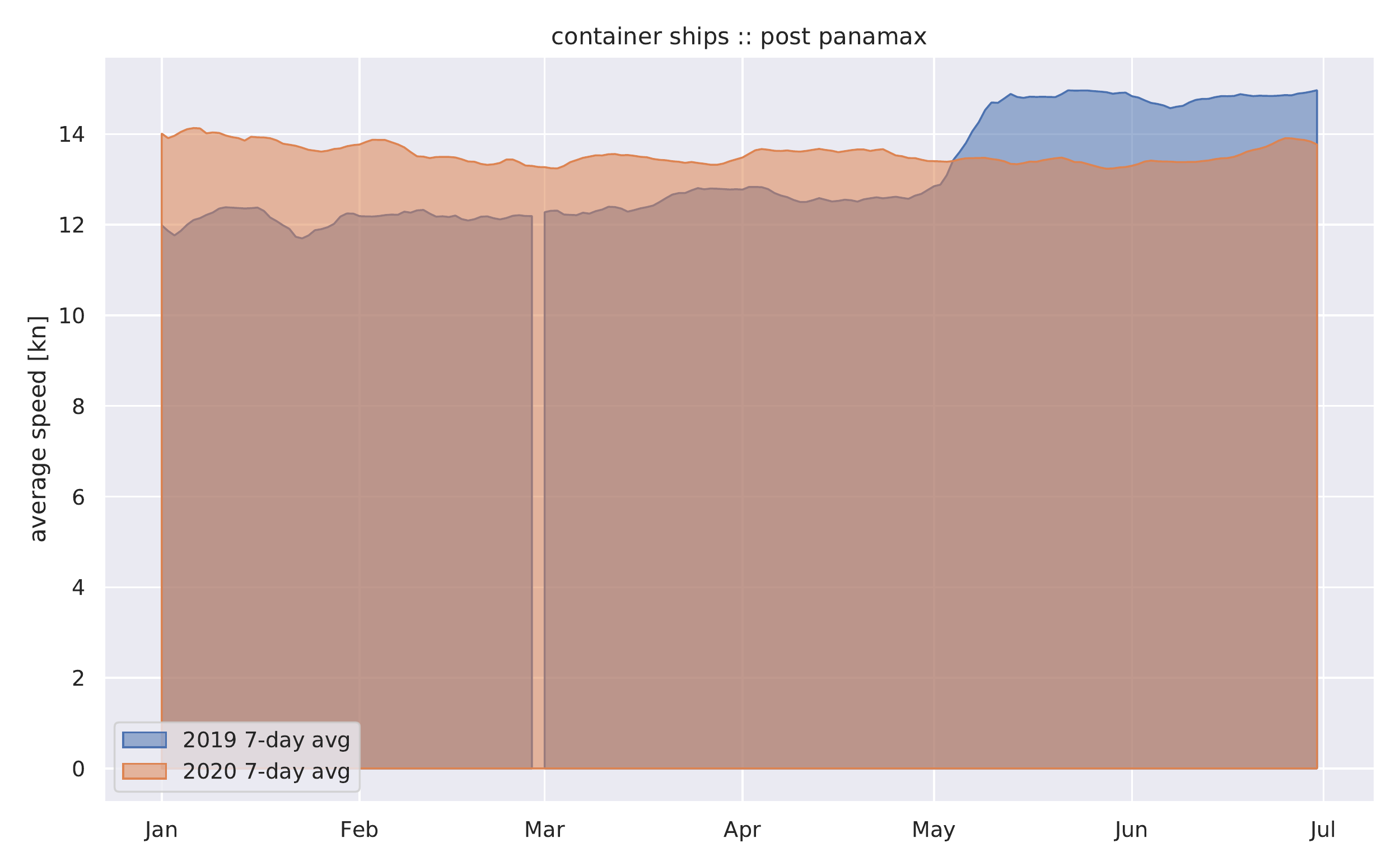}%
        }%
        \label{fig:dailyspeedarea_container_post-panamax}%
    \hfil%
    \subfloat[][Container ships: New Panamax]{%
        \includegraphics[trim=15 10 10 15,clip,width=0.32\columnwidth]{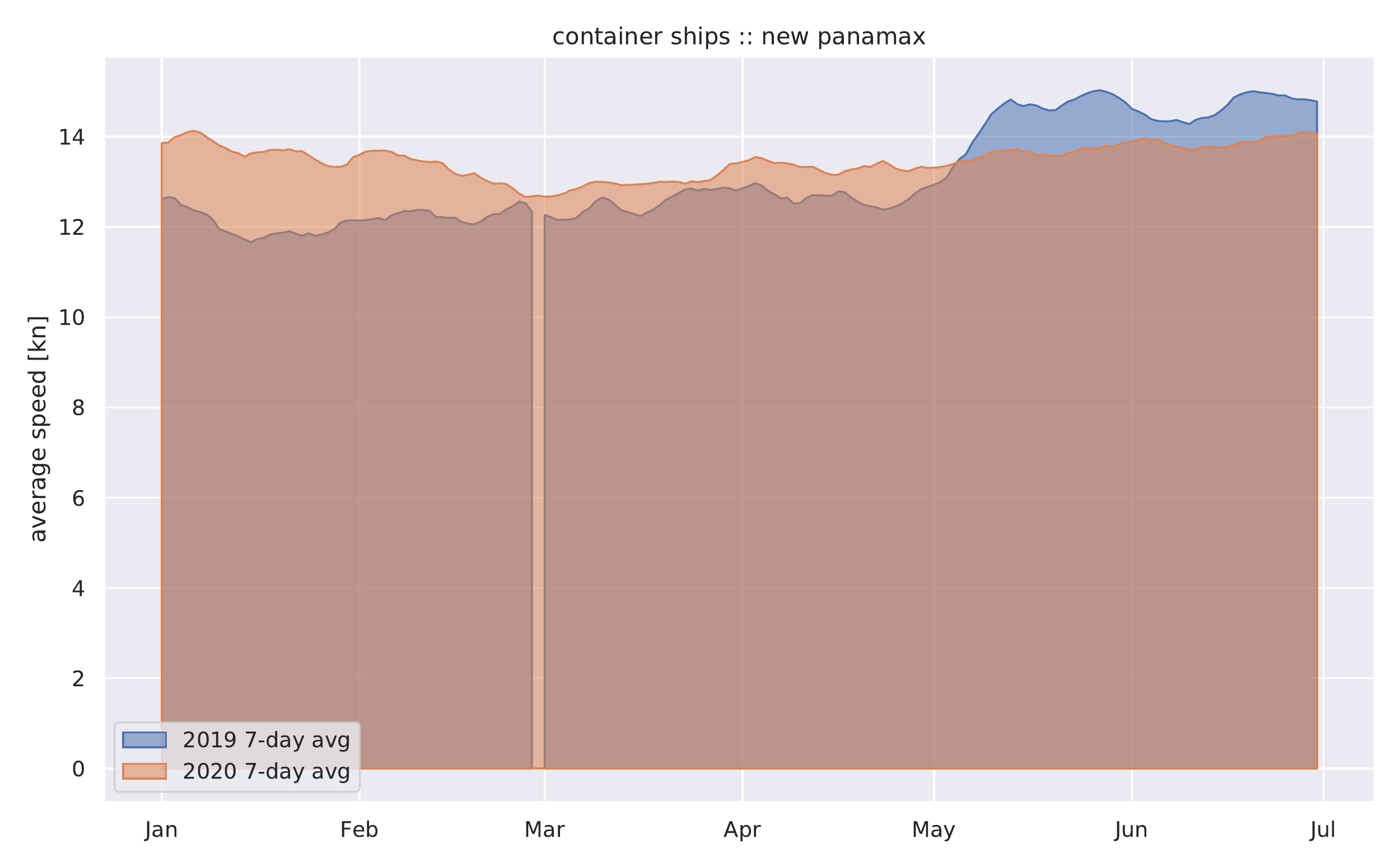}%
        }%
        \label{fig:dailyspeedarea_container_new-panamax}%
    \\
    \subfloat[][Container ships: ULCV]{%
        \includegraphics[trim=15 10 10 15,clip,width=0.32\columnwidth]{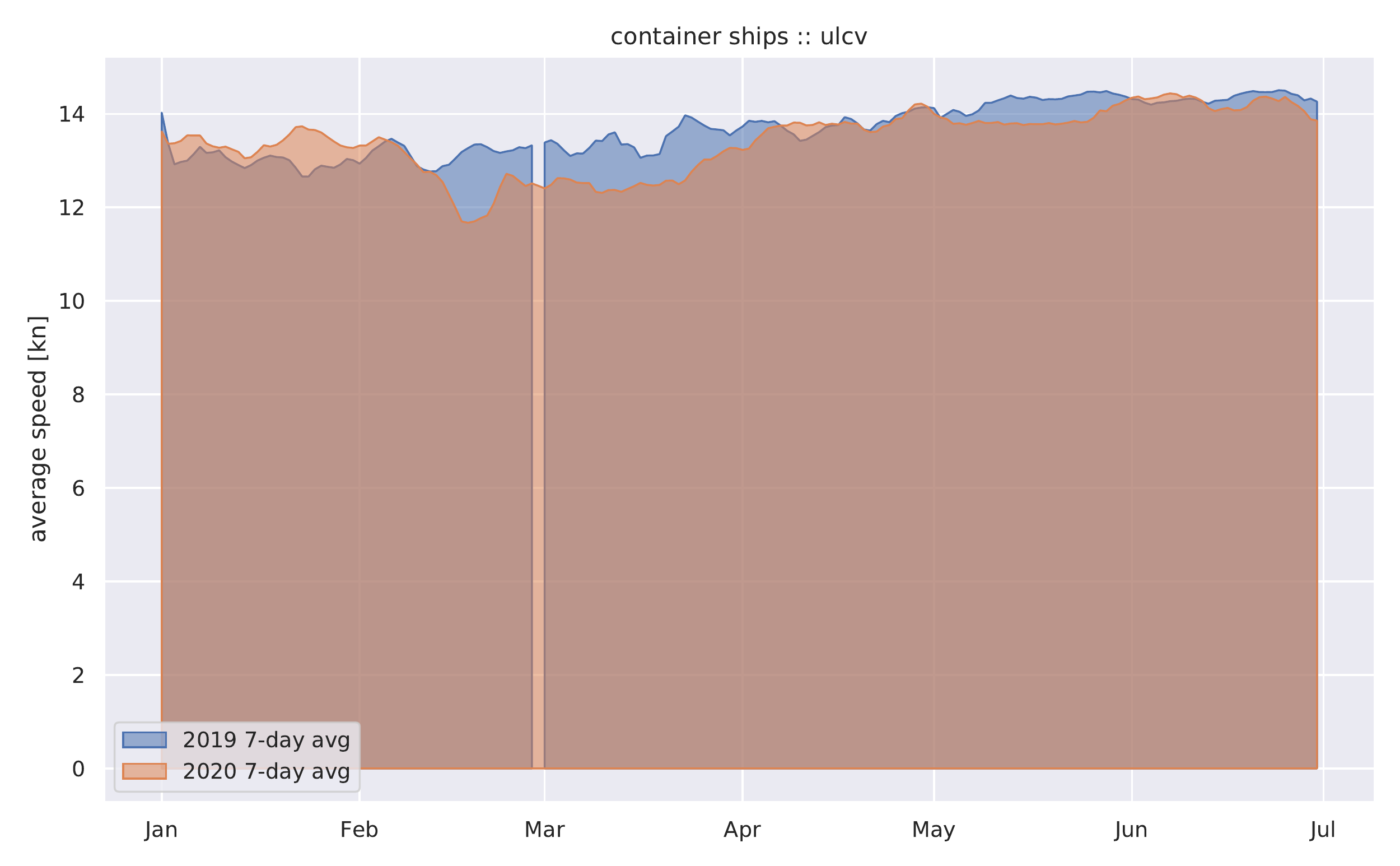}%
        \label{fig:dailyspeedarea_container_ulcv}%
        }%
    \hfil%
    \begin{minipage}[b]{0.64\columnwidth}
    \footnotesize{
        \caption{Comparison of daily average speed of container ships in the first six months of 2020 (orange) versus 2019 (blue) for different ship size categories, ordered by increasing capacity measured in \acp{TEU}; the two area charts are overlaid in transparency to highlight trend differences; the discontinuity in the blue data series corresponds to the leap day absence in 2019.}%
        \label{fig:dailyspeedarea_container}%
    }
    \end{minipage}
\end{figure}

\begin{figure}
    \centering%
    \subfloat[][Dry bulk: Handy size]{%
        \includegraphics[trim=15 10 10 15,clip,width=0.32\columnwidth]{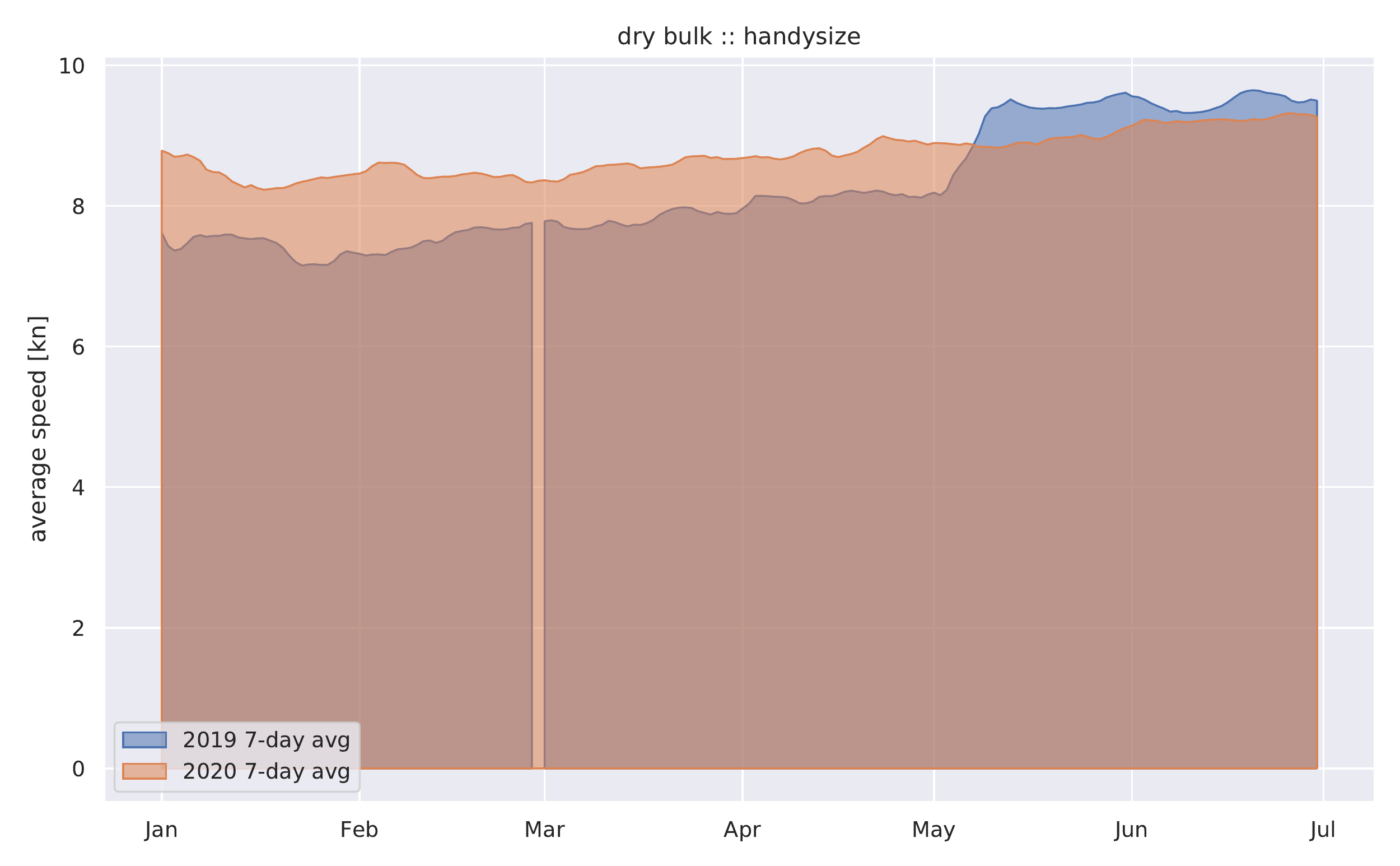}%
        \label{fig:dailyspeedarea_dry_bulk_handysize}%
        }%
    \hfil%
    \subfloat[][Dry bulk: Handymax]{%
        \includegraphics[trim=15 10 10 15,clip,width=0.32\columnwidth]{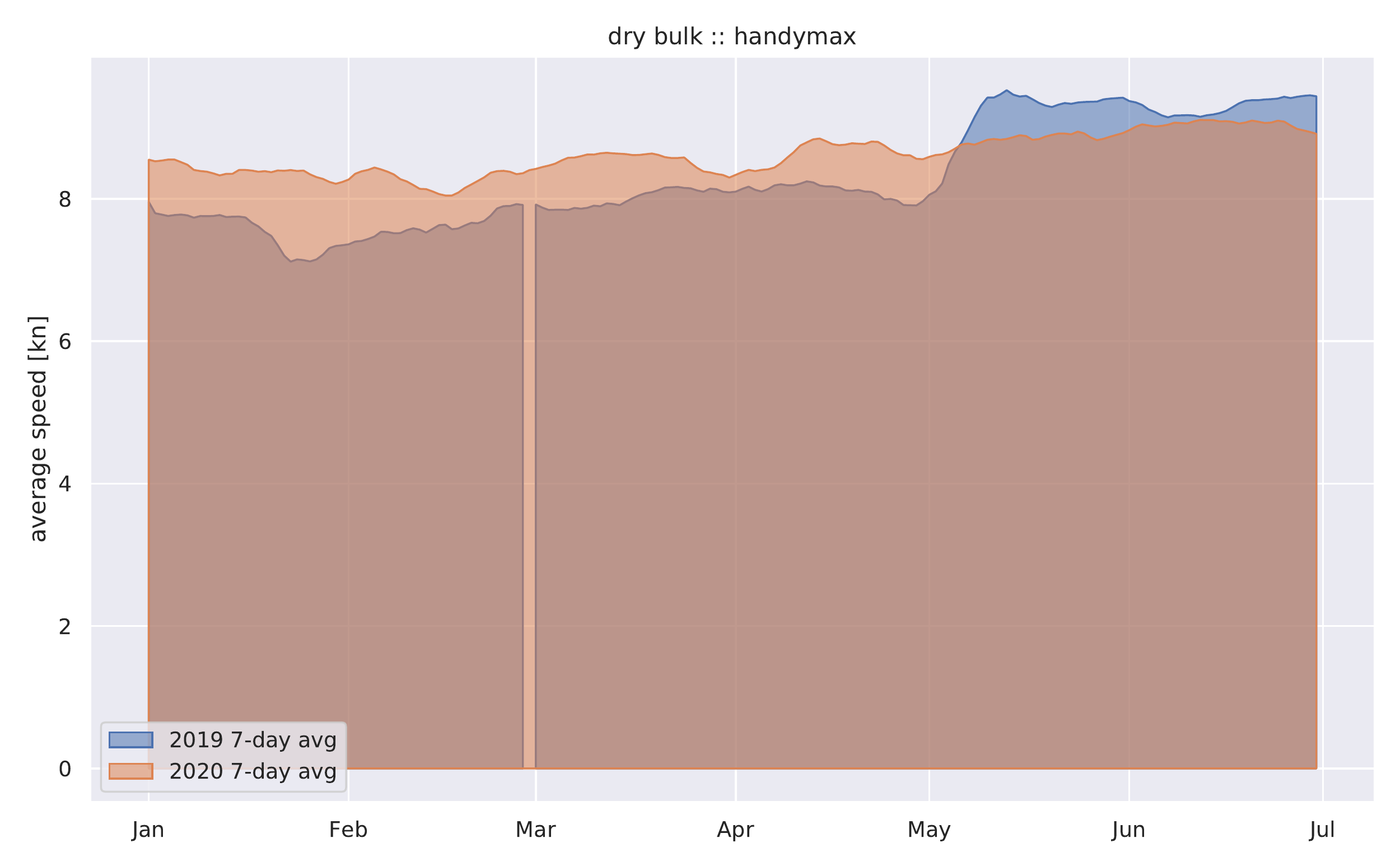}%
        \label{fig:dailyspeedarea_dry_bulk_handymax}%
        }%
    \hfil%
    \subfloat[][Dry bulk: Panamax]{%
        \includegraphics[trim=15 10 10 15,clip,width=0.32\columnwidth]{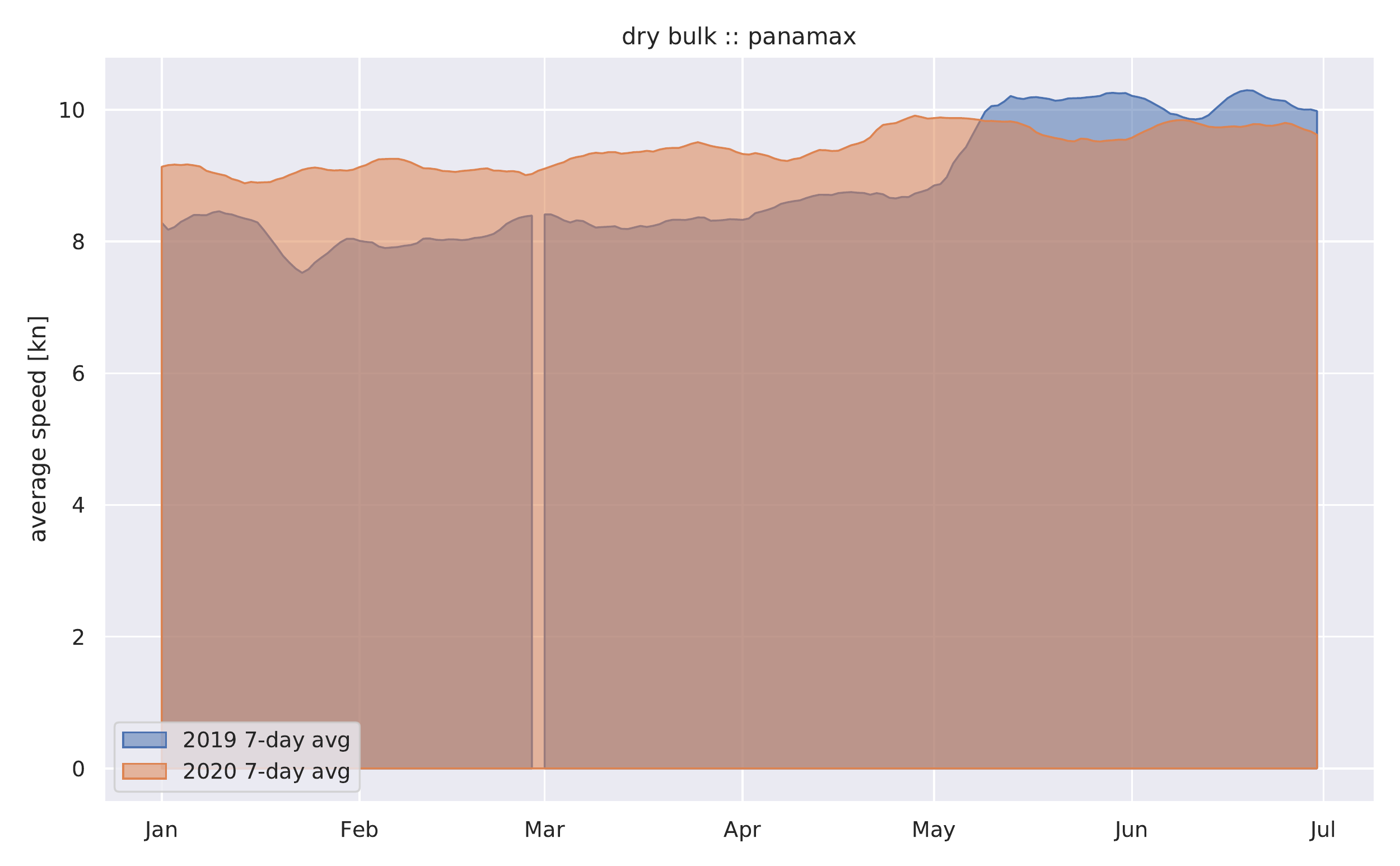}%
        \label{fig:dailyspeedarea_dry_bulk_panamax}%
        }%
    \\
    \subfloat[][Dry bulk: Post-Panamax]{%
        \includegraphics[trim=15 10 10 15,clip,width=0.32\columnwidth]{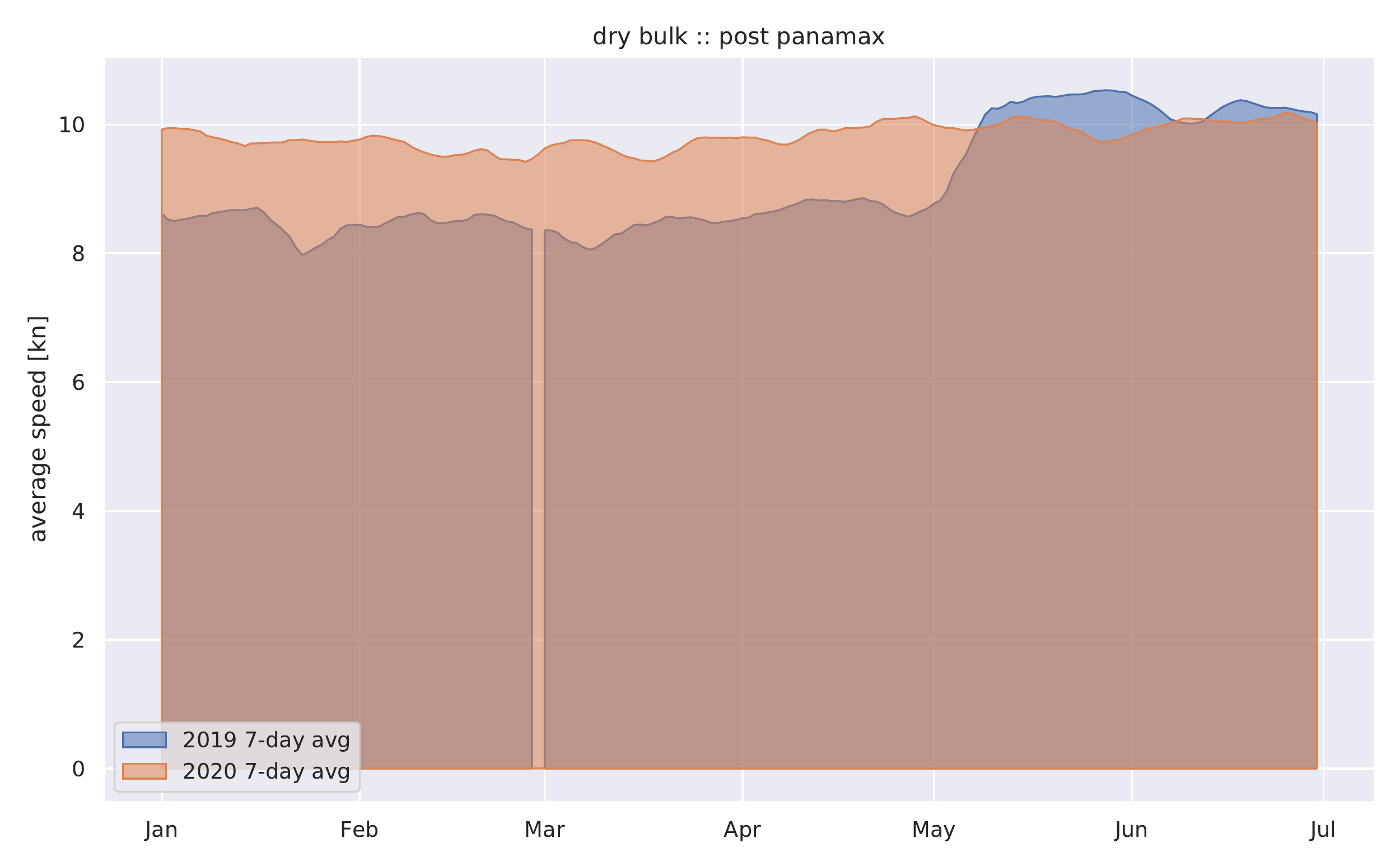}%
        \label{fig:dailyspeedarea_dry_bulk_post-panamax}%
        }%
    \hfil%
    \subfloat[][Dry bulk: VLBC]{%
        \includegraphics[trim=15 10 10 15,clip,width=0.32\columnwidth]{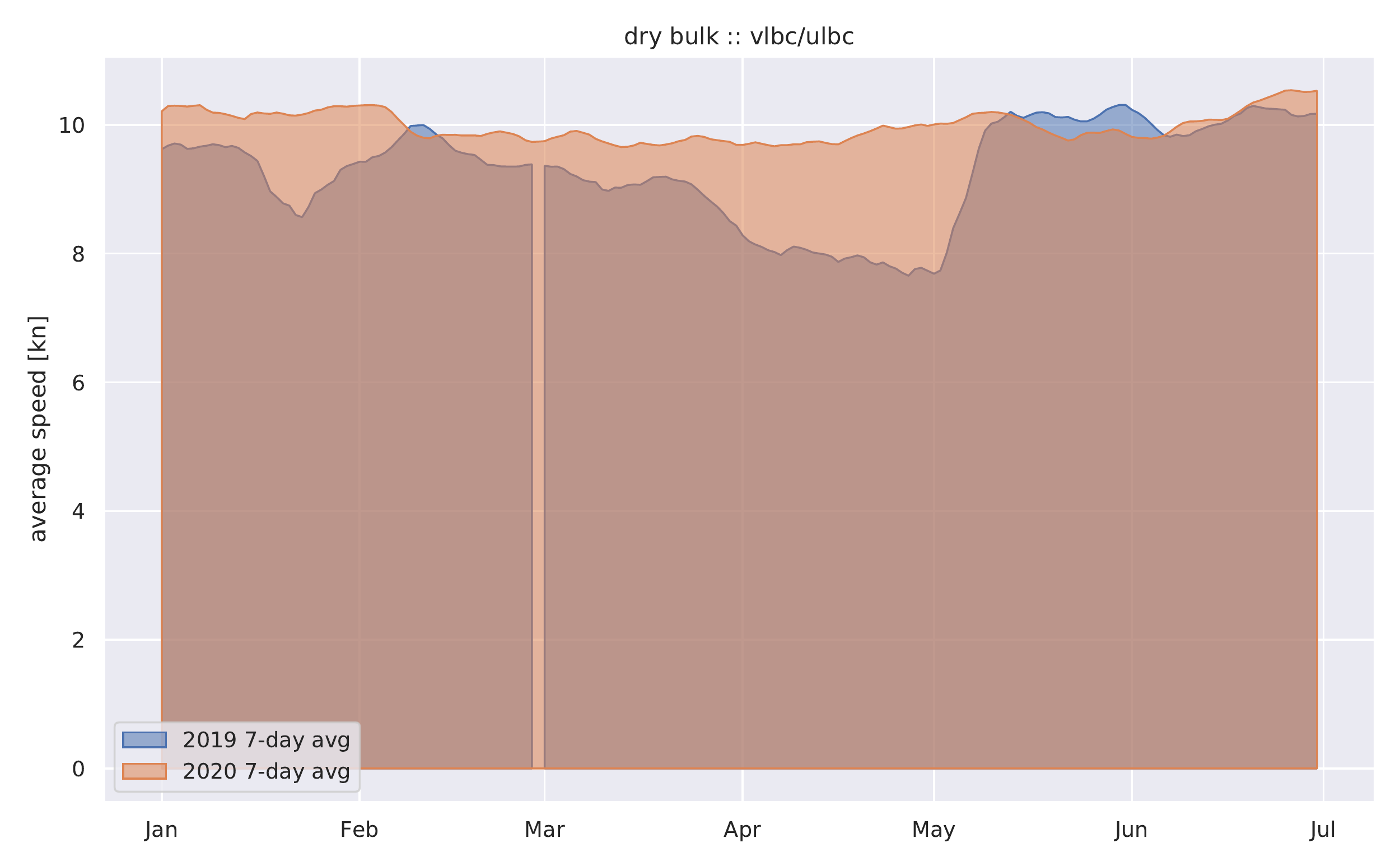}%
        \label{fig:dailyspeedarea_dry_bulk_vlbc}%
        }%
    \hfil%
    \subfloat[][Dry bulk: Capesize]{%
        \includegraphics[trim=15 10 10 15,clip,width=0.32\columnwidth]{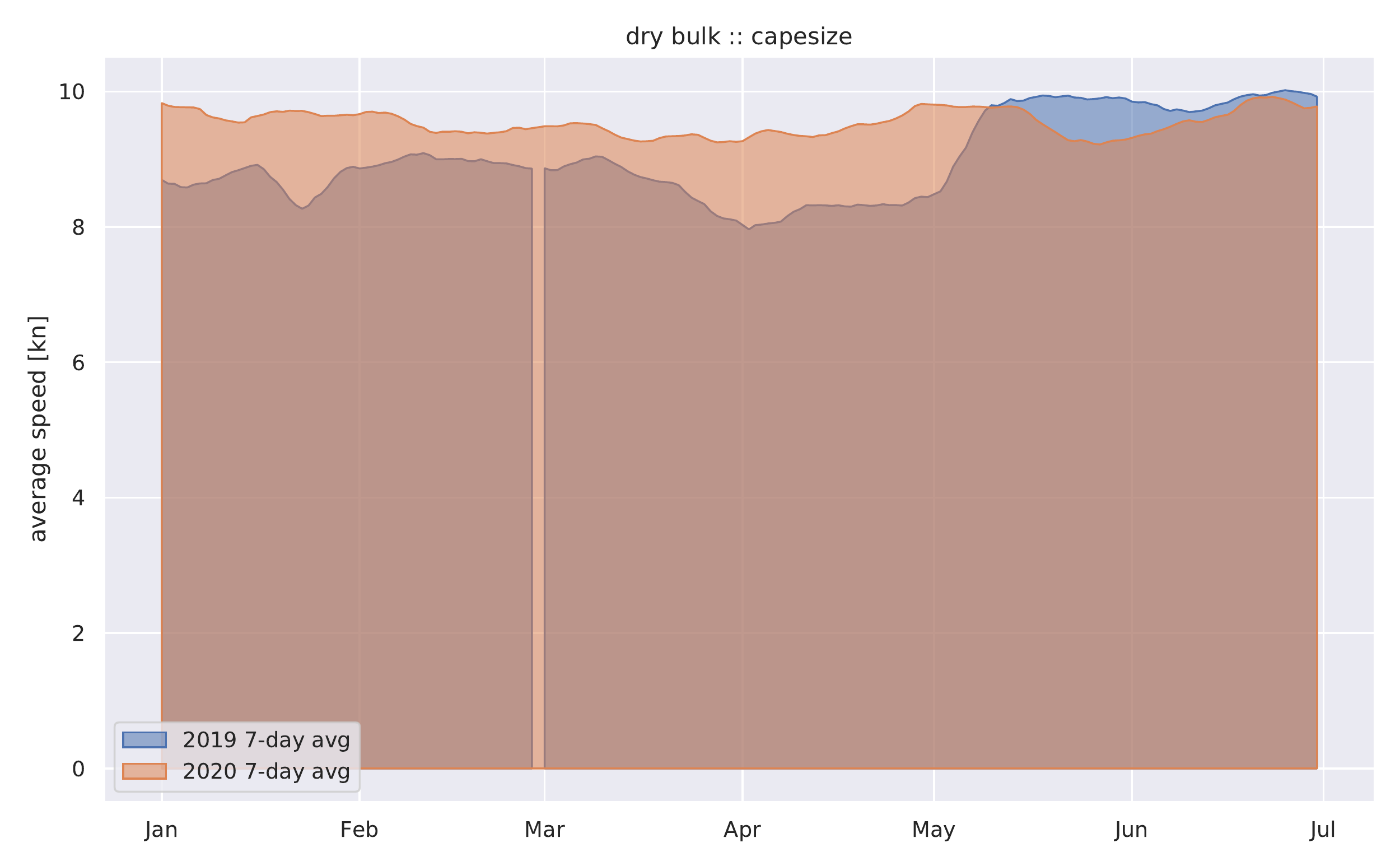}%
        \label{fig:dailyspeedarea_dry_bulk_capesize}%
        }
    \caption{Comparison of daily average speed of dry bulk ships in the first six months of 2020 (orange) versus 2019 (blue) for different ship size categories, ordered by increasing \ac{DWT}; the two area charts are overlaid in transparency to highlight trend differences; the discontinuity in the blue data series corresponds to the leap day absence in 2019.}%
    \label{fig:dailyspeedarea_dry-bulk}%
\end{figure}

\begin{figure}
    \centering%
    \subfloat[][Wet bulk: Handy size]{%
        \includegraphics[trim=15 10 10 15,clip,width=0.32\columnwidth]{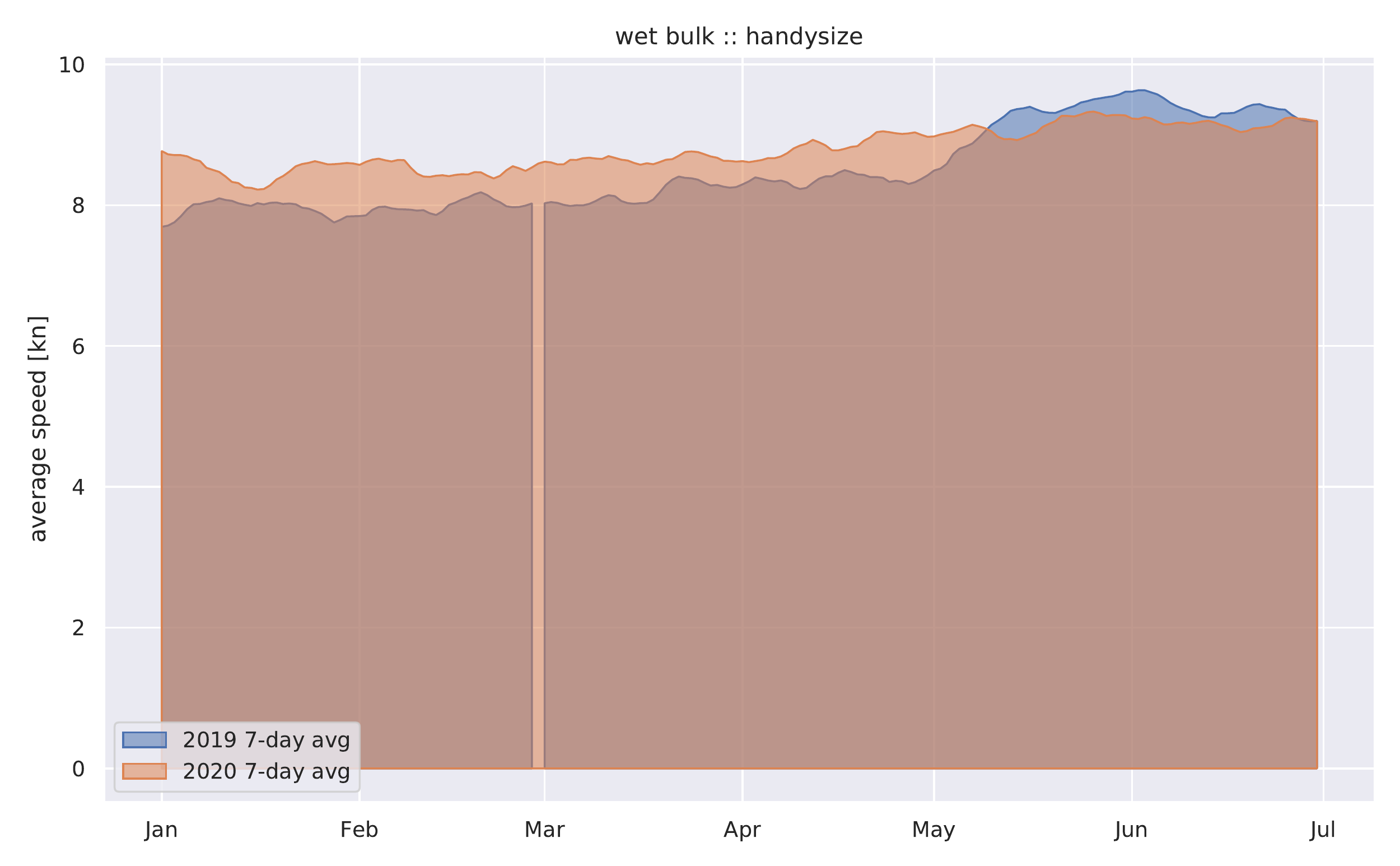}%
        \label{fig:dailyspeedarea_wet_bulk_handysize}%
        }%
    \hfil%
    \subfloat[][Wet bulk: Handymax]{%
        \includegraphics[trim=15 10 10 15,clip,width=0.32\columnwidth]{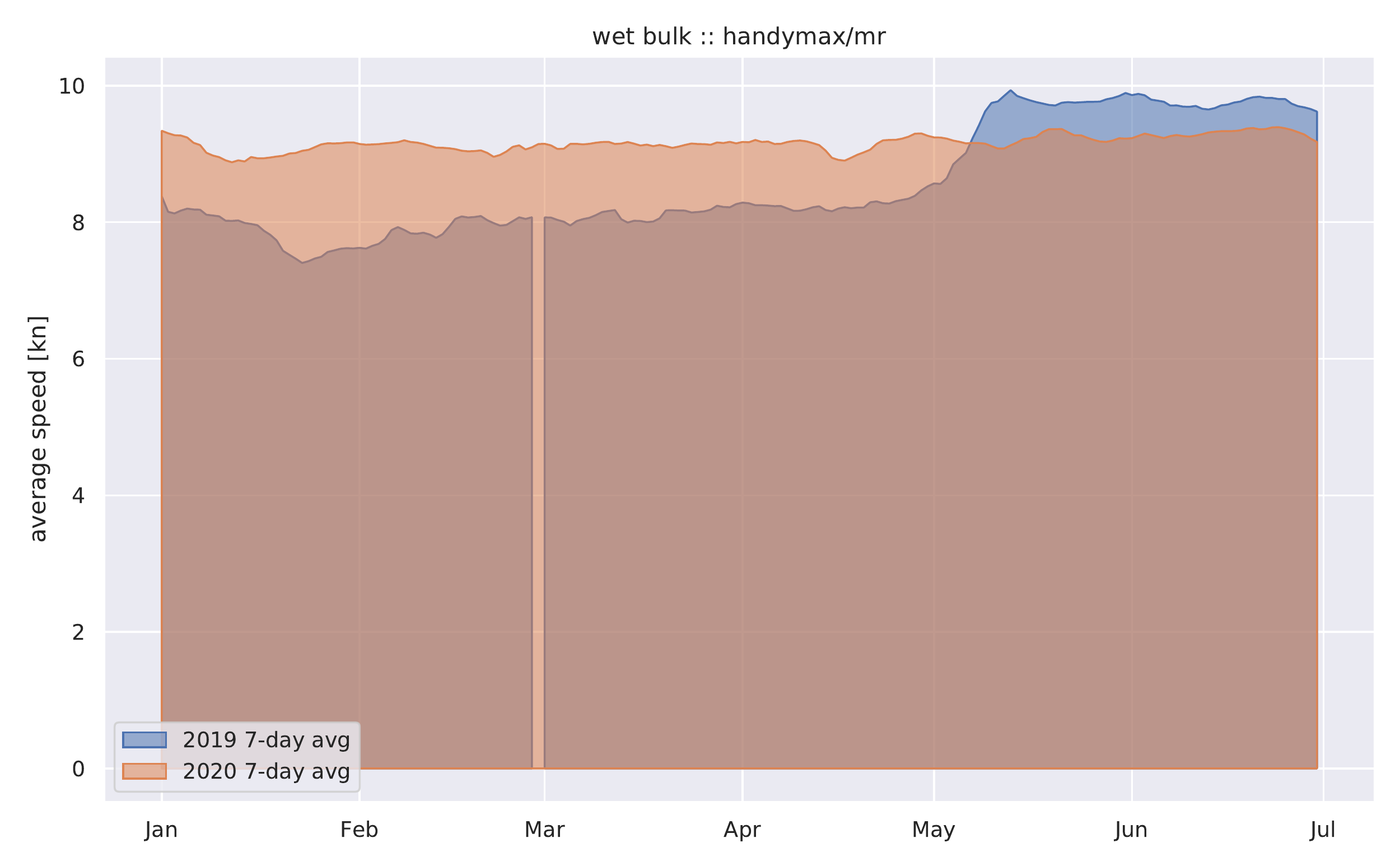}%
        \label{fig:dailyspeedarea_wet_bulk_handymax}%
        }%
    \hfil%
    \subfloat[][Wet bulk: Panamax]{%
        \includegraphics[trim=15 10 10 15,clip,width=0.32\columnwidth]{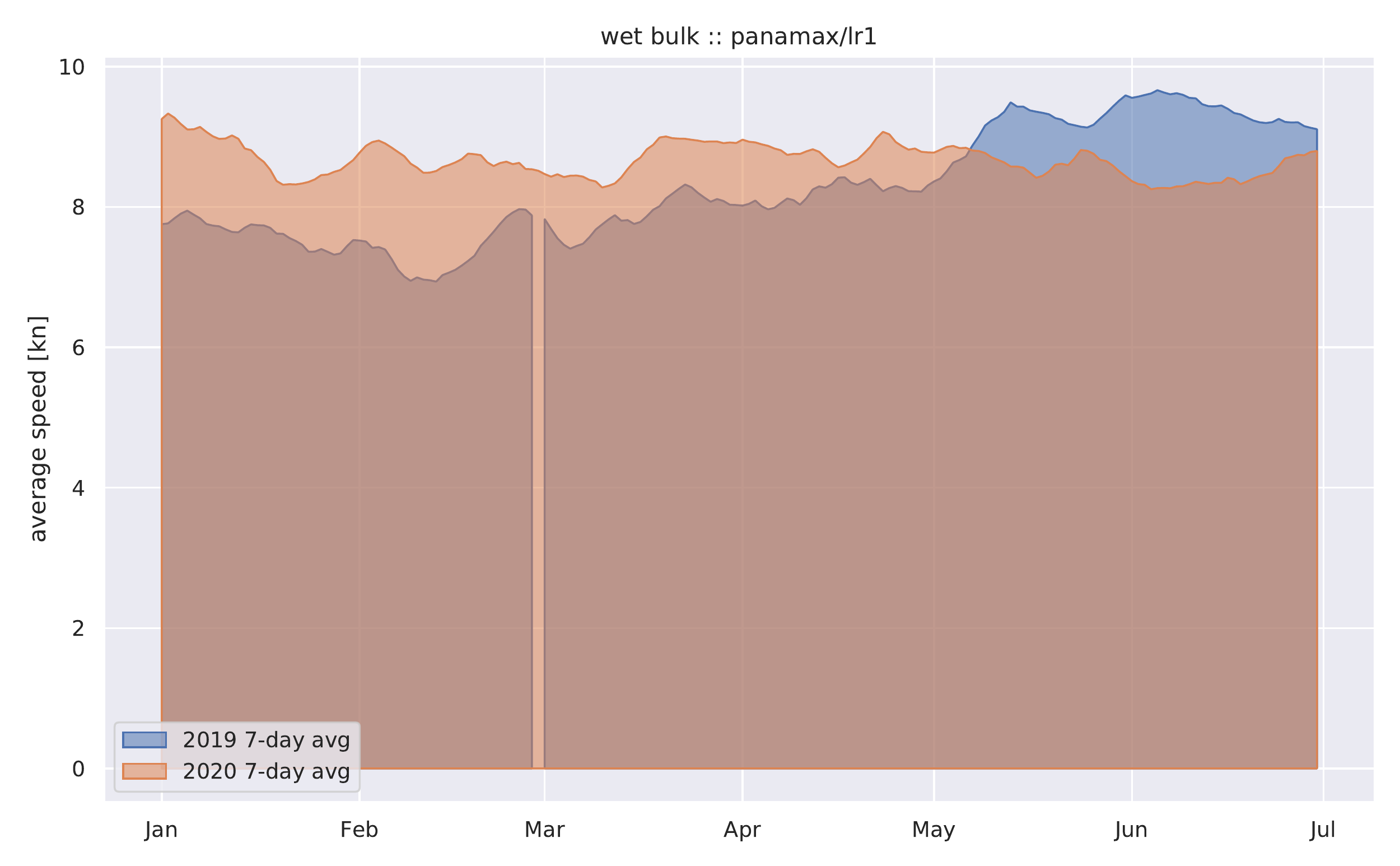}%
        \label{fig:dailyspeedarea_wet_bulk_panamax}%
        }%
    \\
    \subfloat[][Wet bulk: Aframax]{%
        \includegraphics[trim=15 10 10 15,clip,width=0.32\columnwidth]{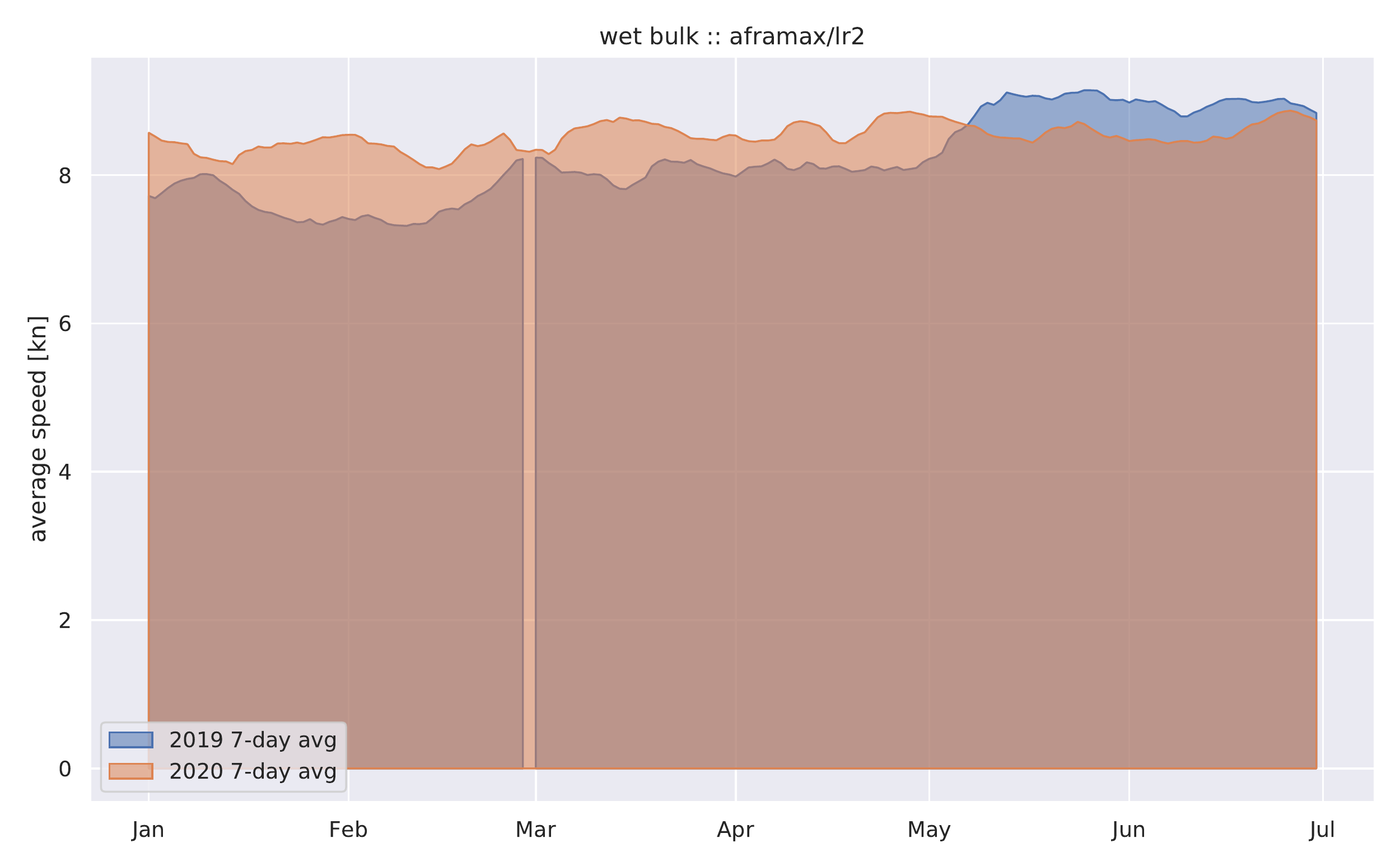}%
        \label{fig:dailyspeedarea_wet_bulk_aframax}%
        }%
    \hfil%
    \subfloat[][Wet bulk: Suezmax]{%
        \includegraphics[trim=15 10 10 15,clip,width=0.32\columnwidth]{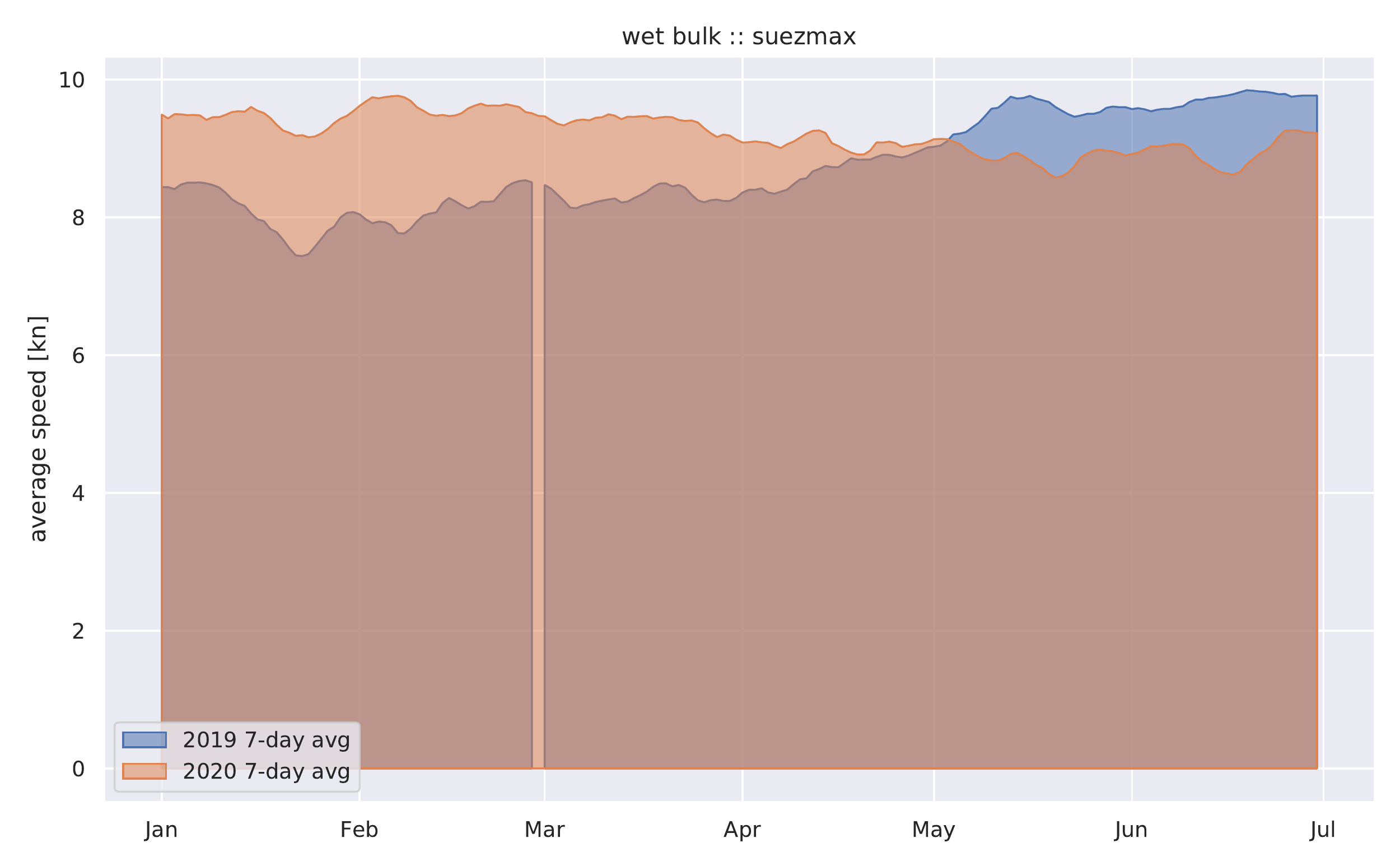}%
        \label{fig:dailyspeedarea_wet_bulk_suezmax}%
        }%
    \hfil%
    \subfloat[][Wet bulk: VLCC]{%
        \includegraphics[trim=15 10 10 15,clip,width=0.32\columnwidth]{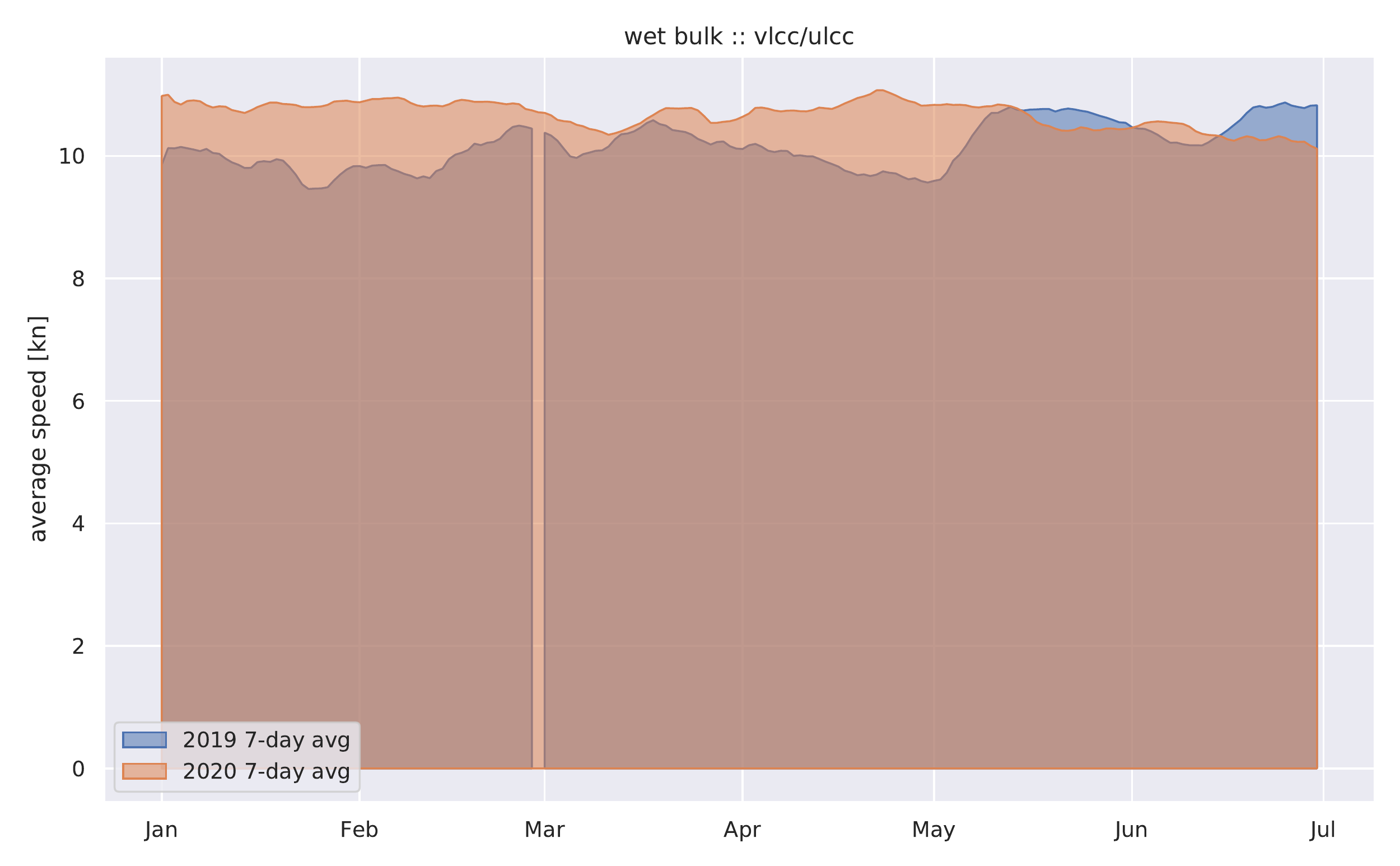}%
        \label{fig:dailyspeedarea_wet_bulk_vlcc}%
        }%
    \caption{Comparison of daily average speed of wet bulk ships in the first six months of 2020 (orange) versus 2019 (blue) for different ship size categories, ordered by increasing \ac{DWT}; the two area charts are overlaid in transparency to highlight trend differences; the discontinuity in the blue data series corresponds to the leap day absence in 2019.}%
    \label{fig:dailyspeedarea_wet-bulk}%
\end{figure}

\begin{figure}
    \centering%
    \subfloat[][Passenger ships: \acs{GT} $\leq$ 10K]{%
        \includegraphics[trim=15 10 10 15,clip,width=0.32\columnwidth]{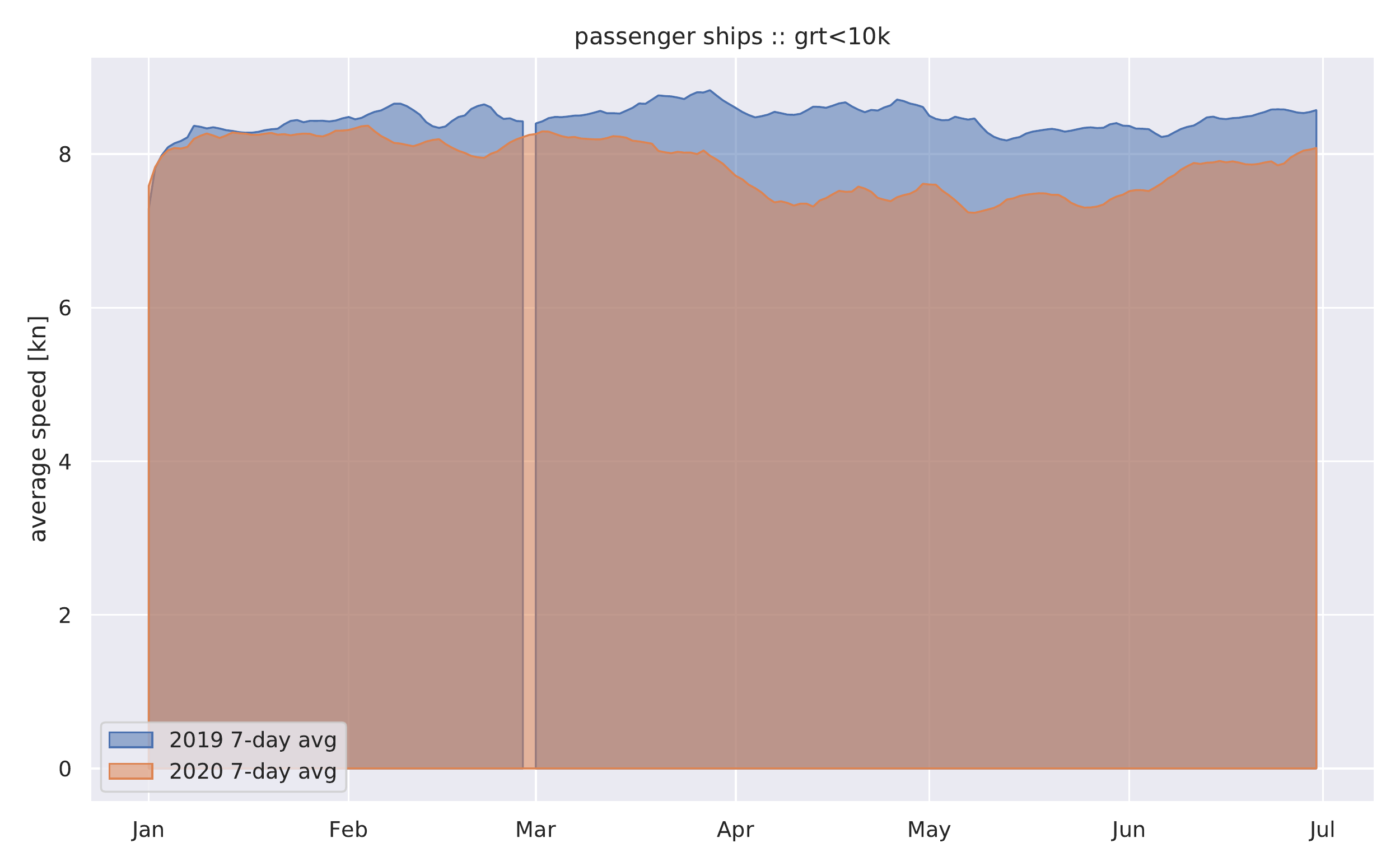}%
        \label{fig:dailyspeedarea_passenger_lt_10k}%
        }%
    \hfil%
    \subfloat[][Passenger ships: 10K $<$ \acs{GT} $\leq$ 60K]{%
        \includegraphics[trim=15 10 10 15,clip,width=0.32\columnwidth]{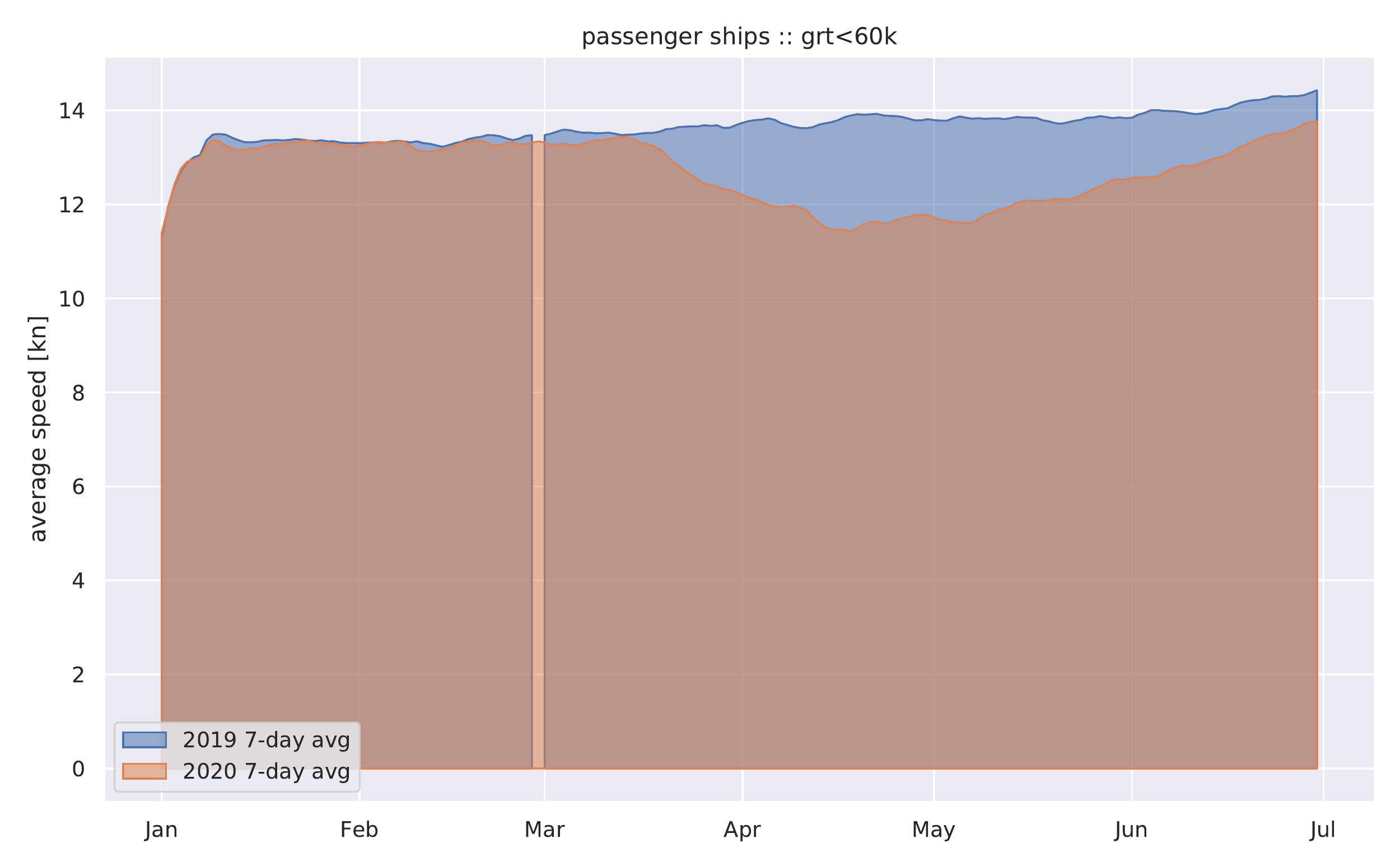}%
        \label{fig:dailyspeedarea_passenger_lt_60k}%
        }%
    \hfil%
    \subfloat[][Passenger ships: 60K $<$ \acs{GT} $\leq$ 100K]{%
        \includegraphics[trim=15 10 10 15,clip,width=0.32\columnwidth]{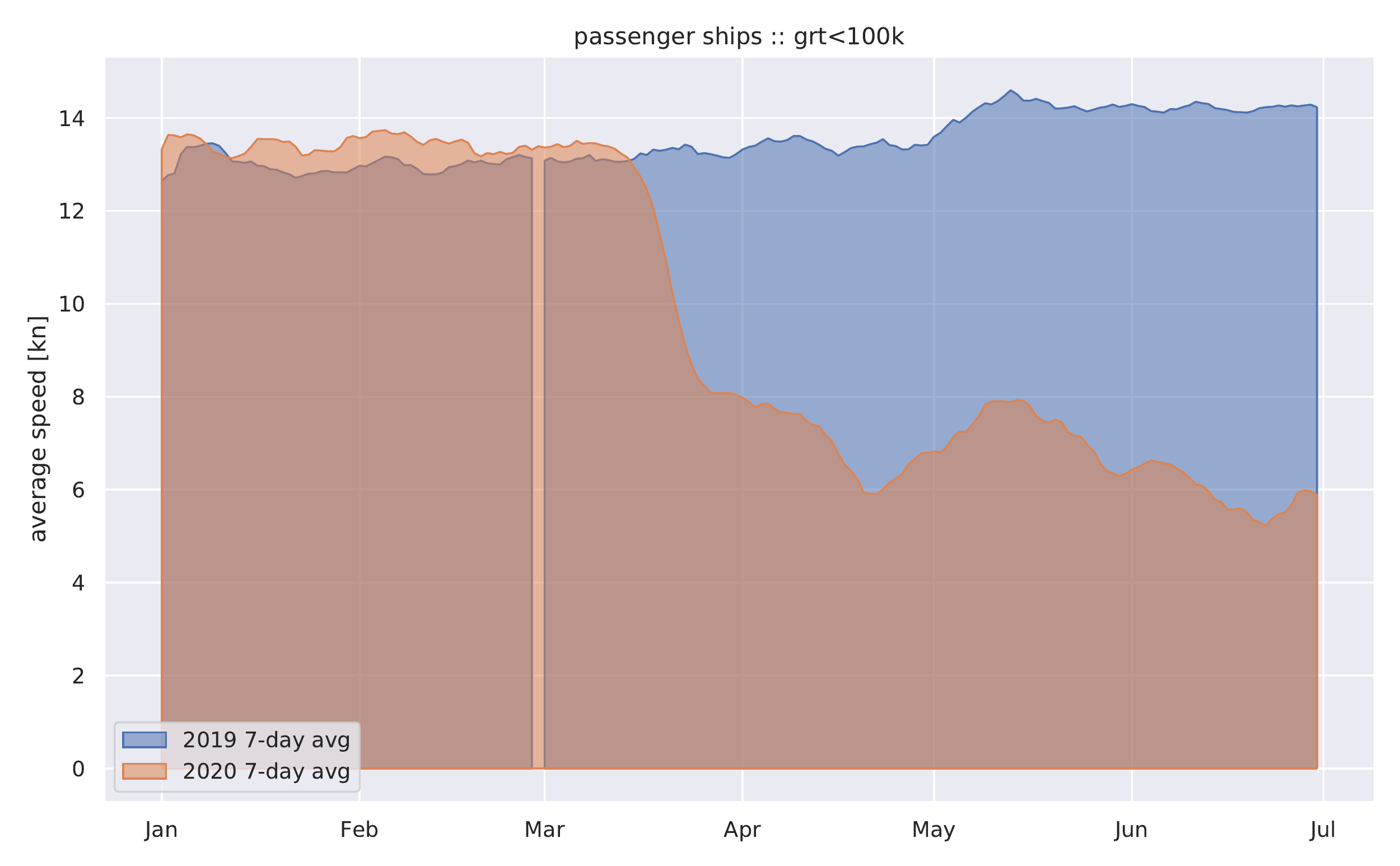}%
        \label{fig:dailyspeedarea_passenger_lt_100k}%
        }%
    \\
    \subfloat[][Passenger ships: \acs{GT} $>$ 100K]{%
        \includegraphics[trim=15 10 10 15,clip,width=0.32\columnwidth]{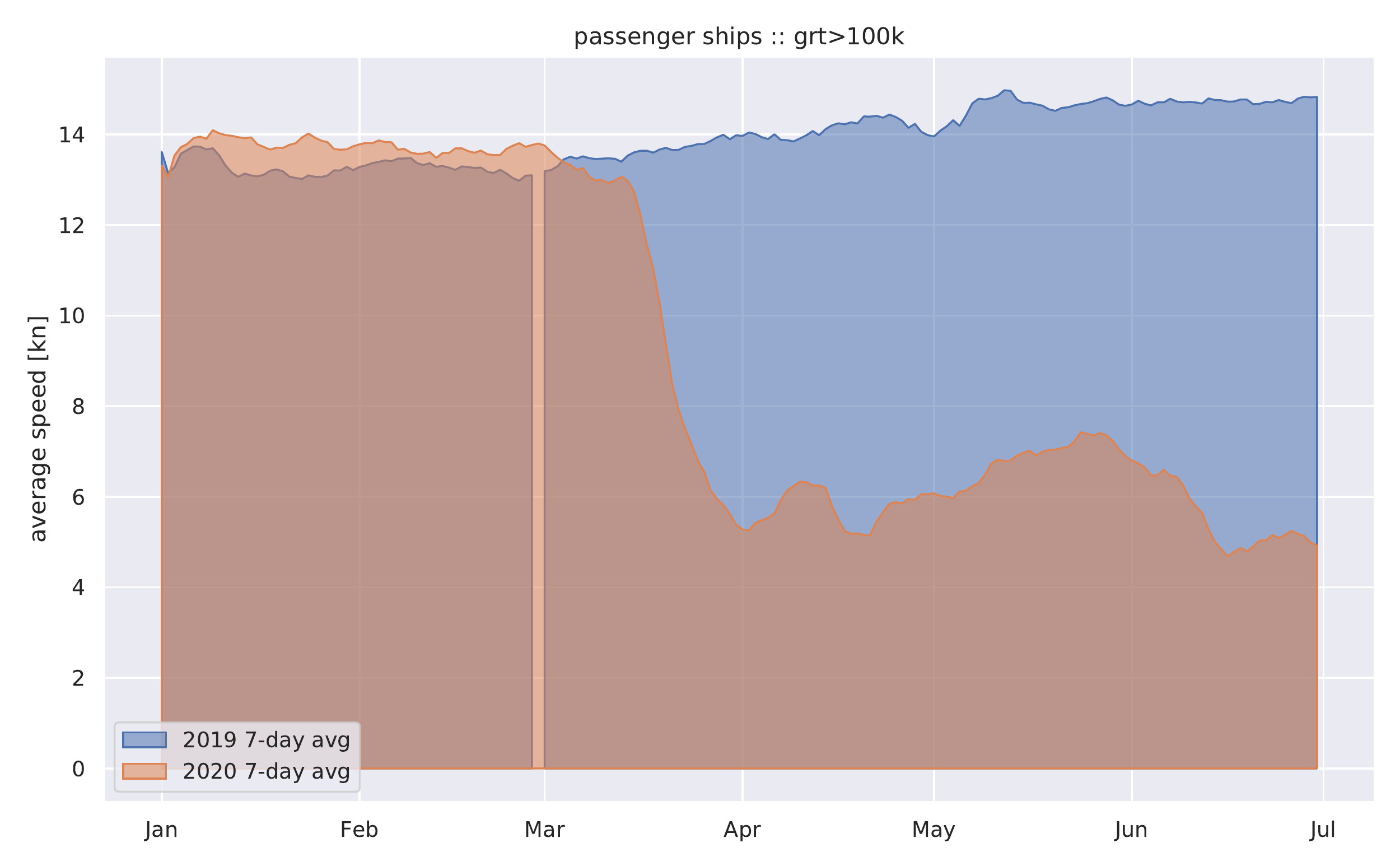}%
        \label{fig:dailyspeedarea_passenger_gt_100k}%
        }%
    \hfil%
    \begin{minipage}[b]{0.64\columnwidth}
    \footnotesize{
        \caption{Comparison of daily average speed of passenger ships in the first six months of 2020 (orange) versus 2019 (blue) for different ship size categories, ordered by increasing \acf{GT}; the two area charts are overlaid in transparency to highlight trend differences; the discontinuity in the blue data series corresponds to the leap day absence in 2019.}%
        \label{fig:dailyspeedarea_passenger}%
    }
    \end{minipage}
\end{figure}

\rev{In Figures~\ref{fig:dailyspeedarea_container}--\ref{fig:dailyspeedarea_passenger}, we report the daily average speed broken down by ship size and category. For the container, dry bulk and wet bulk categories, the average speed did not change significantly from January to June 2020. Instead, passenger ships exhibit a significant decrease in average speed, as reported in Fig.~\ref{fig:dailyspeedarea_passenger}, a behavior that is coherent with the trend of \ac{CNM} reported in Fig.~\ref{fig:dailymilaeagearea_passenger}. Figures~\ref{fig:dailymilaeagearea_container}--\ref{fig:dailymilaeagearea_wet-bulk} show an increase of the average speed for several classes of container, dry bulk and wet bulk in May and June 2019, which is not observed in 2020. Again, this is coherent with the slowdown of the \ac{CNM} in 2020 compared to 2019 for most of ship traffic, except dry bulk traffic, as reported in Figures~\ref{fig:dailymilaeagearea_container}--\ref{fig:dailymilaeagearea_wet-bulk}.}

\rev{The mobility of tanker vessels deserves a dedicated analysis. Several sources reported an increase, up to at least 160 million barrels, of oil held in floating storage on tanker ships, including 60 supertankers, \ac{VLCC} and \ac{ULCC}, which can hold 2 million barrels each.~\cite{REUTERS_storage} The effects of this trend are noticeable in the decrease of \ac{CNM} at the beginning of May, reported in Fig.~\ref{fig:dailymileagearea_wet_bulk_vlcc} 
and in the decrease of active supertankers from April to July, 2020 (not present in 2019), reported in Fig.~\ref{fig:vlcc_activeidle}.}

\subsection*{Shipping traffic density analysis}
\rev{Vessel density maps are data products that show the spatial distribution of ships, and hence of maritime traffic, in specific areas of interest; vessel density maps are widely used to understand shipping operational patterns. There are several ways of computing vessel density maps from AIS data.~\cite{EMODNET_densitymaps} In this work, the method employed to compute density maps amounts to summing up the length of all tracklets that intersect each cell of a pre-defined grid over the area of interest.~\cite{EMODNET_densitymaps} The resulting amount is then divided by the cell's area in squared nautical miles. The reason of this choice is that the produced maps show the density of navigated miles; consequently, their unit of measure is \si{\per\nmi}. In this way, excluded the possible distortion introduced by the projection, \acp{CNM} can be seen as the double integral of the density over an area of interest. Therefore, to all means, such density maps provide a complementary perspective on the \ac{CNM} analysis: while the latter provides a synthetic \emph{quantitative} indication of changes induced to maritime traffic, the former shows exactly \emph{how} these changes are spatially distributed.} 

\rev{In the interest of brevity, we focused this analysis on one month only, from 13 March to 13 April in both 2019 and 2020. Moreover, instead of just showing the density in two comparable time periods, to highlight changes between one year and the other, we compute their difference and color each grid cell according to a sequential colormap: if a cell's value in 2020 is significantly higher than in 2019, the cell's color is bright yellow; conversely, if a cell's value in 2020 is significantly below 2019, its color is dark purple; between these two extremes, cells whose values changed slightly in 2020 with respect to 2019 are coloured in shadings from blue to green. Consequently, regions with a predominance of dark purple cells represent areas where vessels generally navigated much less in 2020 than in the previous year; on the other hand, regions where bright yellow are predominant represent areas with an increase of navigation in 2020 with respect to 2019.} %

\rev{We computed density maps in three areas: worldwide, in the Mediterranean Sea, and in the area around the Suez Canal. The world density map is illustrated in Figures~\ref{fig:density-maps-1}--\ref{fig:density-maps-2}. For container shipping, Fig.~\ref{fig:density-map-container} shows evident signs of reduced activities along important routes, such as that from the Strait of Malacca to the Mediterranean Sea via the Suez Canal, between North America and the Strait of Gibraltar, around West Africa and the Iberian Peninsula, just to mention some. Coherently with Table~\ref{tab:mileage}, the analysis of the density of dry and wet bulk vessels, reported in Figures~\ref{fig:density-map-dry-bulk} and~\ref{fig:density-map-wet-bulk}, reveals instead a general increase of navigation in some regions, e.g., across the Suez Canal and along the East-West route around the Cape of Good Hope, and a decrease in other regions, e.g., along several Atlantic routes that connect Europe to North and South America. In general, all categories of commercial shipping show increased activity in South and East China Seas, evidence of China's effort to return to normality sooner than other countries. Coherently with Table~\ref{tab:mileage}, the density of passenger ship traffic (Fig.~\ref{fig:density-map-passenger}) provides yet another confirmation that this was the most affected segment by the pandemic, with sharp decrease of shipping activities all around the globe. The few light increases in activities noticeable in Fig.~\ref{fig:density-map-passenger}, e.g., between the Cape of Good Hope and West Africa or off the coast of South Australia, are explained with the repatriation operations of crew and passenger of cruise ships.~\cite{IMO_2021}}


\subsubsection*{Suez Canal}
\rev{The Suez Canal allows fast connections between Europe and the Near and Far East. As such, it has a crucial role for trading, and deserves a more focused analysis. According to a new market analysis released by BIMCO,~\cite{SuezCanal} the overall traffic at the Suez Canal has remained resilient. Strong increases in the number of transits by oil tankers and dry bulkers helped to offset the declines in container ship transits according to BIMCO.~\cite{SuezCanal} The density maps illustrated in Fig.~\ref{fig:density-maps-suez} are in complete agreement with BIMCO's analysis and provides further evidence in support of such assessment. Indeed, Figures~\ref{fig:density-map-suez-dry-bulk}-\ref{fig:density-map-suez-wet-bulk} show an evident increase of dry and wet bulkers; at the same time, Fig.~\ref{fig:density-map-suez-container} shows a noticeable decreased activity for container shipping in the same area.
This is mostly a consequence of the increased number of vessels that bypassed the Suez Canal, using instead the route around the Cape of Good Hope, motivated by low bunker prices and lack of demand in European markets.~\cite{Terpilowski} This effect on the global scale is visible in Fig.~\ref{fig:density-map-container}.}

\subsubsection*{Mediterranean Sea}
\rev{The analysis of traffic density differences between 2020 and 2019 in the Mediterranean Sea, reported in Fig.~\ref{fig:density-maps-med}, reveals a situation similar to that already discussed in the area around the Suez Canal. The analysis of \ac{AIS} data for container shipping (Fig.~\ref{fig:density-map-med-container}) shows a mixed situation, with a prevalent reduction on the western side (Tyrrhenian, Ionian, Adriatic, Ligurian, Balearic, and Atlantic Seas), considering that lockdowns in Italy, France and Spain happened before than in northern countries, e.g., Germany. The analysis of the density of dry and wet bulk carriers, reported in Figures~\ref{fig:density-map-dry-bulk} and~\ref{fig:density-map-wet-bulk}, reveals instead a general increase of navigation in most of the regions, except for Italy (Tyrrhenian, Ionian, Adriatic and Ligurian Seas), where a noticeable reduction is observed. The passenger density reported in Fig.~\ref{fig:density-map-med-pass}, as expected, revels a total collapse of the category almost everywhere.}

\section*{Discussion}

The COVID-19 pandemic and containment measures have lead to vast economic disruptions across the world, with near evaporation of both customer demand and industrial activity. Focusing on the first half of 2020 compared with previous years (2016--2019), and based on a global dataset of approximately 1~trillion ship positions received via AIS from more than \num{50000} vessels, the analysis presented in this paper shows that shipping mobility has also been affected negatively. 

We report an unprecedented slowdown in global shipping mobility, which was steadily increasing since 2016, and a noticeable activity decrease for all ship categories in 2020, when compared with projections (assuming the average growth rate of past years). The most affected traffic segment is that of passenger ships, followed by container ships. Effects of the \rev{pandemic} on global ship mobility are observable since March until the end of June 2020, with variations ranging between \SI{-5.62}{\percent} and \SI{-13.77}{\percent} for container ships, between \SI[retain-explicit-plus]{+2.28}{\percent} to \SI{-3.32}{\percent} for dry bulks, between \SI{-0.22}{\percent} and \SI{-9.27}{\percent} for wet bulks, and between \SI{-19.57}{\percent} and \SI{-42.77}{\percent} for passenger ships. 

On the basis of the indicators presented in this paper, we can also conclude that, despite the crisis, shipping was resilient, and in specific markets it was possible to continue operations; in this sense, the analysis highlights the strategic importance of shipping at a global scale. \rev{This statement is supported by thorough quantitative analysis of shipping mobility and operational patterns that shows what happens when the shipping system is subject to an extreme shock, such as that caused by COVID-19 pandemic. The effects of the shock on the system are deep and extend to trade and environmental aspects, even if the \textit{exact} nexus between shipping mobility, trade and environmental emissions is still an open research topic}. Future works could be aimed at validating such connections, and explore quantitatively the relation among mobility, trade, and emissions. Additionally, as more data becomes available our analysis will focus on longer term impacts while attempting to identify signs of a recovery.

\section*{Methods}


\subsection*{Automatic Identification System (AIS)}
Originally designed only for collision avoidance and information exchange between ships, nowadays the AIS is extensively used by operators as the primary means for ship traffic monitoring on a much larger scale than that achievable with conventional coastal surveillance systems. \rev{With AIS}, ships voluntarily broadcast their position, velocity, along with other identification and voyage-related information. The AIS communication protocol is asynchronous and prescribes that different types of messages be transmitted with different frequencies. There are two types of AIS transponders: Class A, for large ships, and class B, for smaller vessels. The International Maritime Organization (IMO) mandates that every ship of more than 300 gross tonnage, all passenger ships and all fishing vessels with a length above 15 meters be equipped with class A transponders. Conversely, class B transponders are designed to bring the benefits of AIS on smaller vessels; indeed, they are smaller and less expensive than class A type transceivers. As such, they can be installed on small ships such as recreational vessels that want to have the benefits of having the AIS even if, for their size, they are not required to fit a transponder onboard.

Over the last few years, \ac{AIS} data have been \rev{extensively} used in research for validation purposes as ``ground-truth'' information, e.g., in maritime surveillance with coastal radars~\cite{braca2,braca3,braca4} as confirmation tracks for radar detections, or for the validation of a target motion model for long-term ship prediction.~\cite{MillefioriBracaTAES2016} 
\ac{AIS} has also been used to show how the data association can be \rev{significantly} improved using the long-term prediction and combining AIS with HF Surface Wave radar (HFSWR) data, or \ac{SAR} data.~\cite{VivoneTGRS2017} But in any case, there is a significant literature that considers AIS the sole source of information for maritime surveillance,~\cite{MillefioriBracaTAES2016,Gaglione_IET,VivoneTGRS2017,Arcieri,Pasquale,Survey_AIS} port traffic analysis,~\cite{millefiori2016distributed,ZHANG2019287} and anomaly detection;~\cite{Vespe,Katsilieris,enrica_TSP,Kontopoulos2020,ristic2008statistical} a common application for historical \ac{AIS} data is also the training of machine learning, including neural networks, algorithms.~\cite{Pasquale,forti2020prediction,ZissisAccess} The interested reader can find in the scientific literature an excellent survey~\cite{Survey_AIS} of AIS data exploitation for safety, anomaly detection, route estimation, collision prediction, and path planning.

\subsection*{\rev{Data processing pipeline}} 
In this paper, we analyze \ac{AIS} data to compute mobility indicators. Besides \ac{AIS}, we made use of information regarding ship characteristics; such as their class, size and type of cargo (i.e., dry or wet bulk), as well as their weight and referred to as \ac{DWT}.
%
%
The input of the processing chain is represented by positional AIS messages, specifically message types 1, 2, and 3, which are \num{168}~bit long and are used by vessels to broadcast their position via class A transceivers. Then, positional information is augmented with information from type 5 AIS messages (\num{424}~bit), which ships use to broadcast their identification, voyage and other static information, such as their size and draught.

The \ac{AIS} dataset we used has a worldwide extent and contains approximately 1~trillion messages, which were broadcast by more than \num{50000} ships, a figure that closely resembles the total number of ships in the world merchant fleet as of January 1, 2019 reported by Statista.~\cite{statista} The ships considered were equipped with class A transceivers and were employed for the transfer dry and wet cargo, containers and passengers. For the purposes of this study, vessels included in the calculations are above \num{10000} \ac{DWT} for dry bulk, wet bulk and container shipping, and above \num{1000} \ac{DWT} for passenger vessels. The total size of the dataset is approximately \SI{55}{\tera\byte} and it is stored in a big-data architecture. The processing is based on a distributed cluster of \num{40} virtual cores and \SI{128}{\giga\byte} of RAM. The overall processing time was less than 4 hours. 

The mobility indicator calculation requires a number of processing steps, detailed as follows. The first step fuses data different \ac{AIS} receivers and converts them into a common format for processing; in this step, positional \ac{AIS} messages are augmented with information on the ship class and type of cargo. Since data may suffer from errors and coverage gaps \rev{(less than \SI{1}{\percent})}, erroneous and redundant data are removed during the second stage. For example, all messages that are found to exceed a 24-hour interval, messages that correspond to infeasible speeds for large vessels (faster than \num{50}~knots), or are not transmitted from the four types of carriers, or are accompanied by invalid identification numbers are discarded. In the third step, \emph{active} and \emph{idle} indicators are computed from the reported navigational status and the ship speed. The ship is assumed to be \emph{idle} if her navigational status is ``at anchor,'' ``not under command,'' ``moored,'' or ``aground,'' or if the ship speed remains below \num{2}~knots. Otherwise, it is considered \emph{active}. In the fourth step we finally compute the \ac{CNM} by first aggregating positional data by identification number and than by taking the sum of the great-circle distance between all consecutive positional messages broadcast by active ships. This eventually brings to the computation of the navigated distance per category and unit of time.

\rev{As mentioned above, a small fraction of data can report unfeasible or non-coherent vessel positions (and/or velocities) as well as time-stamps. However, since the analysis is performed at the trajectory level, we are able to discard the affected contacts and still correctly reconstruct the vessel track. This is possible, e.g., with adaptive filtering and target tracking procedures by which trajectories can be recovered even in presence of heavily cluttered~\cite{Gaglione_IET,Pasquale} or imprecisely time-stamped~\cite{millefiori2015adaptive} data.}

\subsection*{Ship mobility and trade indicators}

\rev{Even today, most economic statistics and forecasting indices focus only on the starting and finish lines of the supply chain, with little consideration about how goods arrive at their final destination. Several economic and financial indices are published by private entities in an attempt to document the status of the shipping industry and foretell broader production and commercial developments; these entities include the \ac{BDI}, ClarkSea Index,
Harper Petersen Index, Hamburg Shipbrokers’ Association New Contex Index and others. Many prominent investment and banking institutions incorporate these indices into their own global trade indicators so as to incorporate the ``shipping'' outlook into their forecasts (such as the Morgan Stanley Global Trade Leading Indicator). The \ac{BDI} for instance, which is considered as a leading economic indicator (issued by the Baltic Exchange), measures on a time-series basis the commodities carried by dry bulk carriers across major shipping routes. Though this index has justified its predictive abilities over the previous decade (i.e., till 2008), it has been losing power after 2010, since China's shipbuilding spree unveiled the index's limitations such as coverage, data uncertainty, and timeliness of data collected manually through surveys and questionnaires.~\cite{odom2010shipping} Additionally, as its indices are based on a “brokers” assessment of a particular route and costs, it is not certain that differentiating fixtures (e.g., cargo type, place of delivery, fuel consumption, ballast bonus etc.) are consistently factored into the assessment. The lack of transparency and openness in the methods used to produce these indices, requires treating them as a ``black box,'' often generating uncertainty regarding their reliability and validity.} 

\rev{The \ac{Eurostat} has recently begun exploring the use of big data~\cite{florescu2014will} and \ac{AIS} data for its maritime statistics. \ac{AIS} data can be used as a pillar for maritime statistics, benefiting statistical organisations as a replacement for data traditionally collected through census or sample surveys. In certain cases, \ac{AIS} data is more accurate than traditional maritime statistics.~\cite{ntakou2014analysis} \ac{AIS} can be used to gain insight into the relationships between maritime and inland waterway transport, intra-port visits, characterising maritime travel routes for specific types of ships, and offering geographical coverage that is currently unavailable. Especially for the maritime domain, \ac{AIS} ``big data'' can offer a much higher temporal and spatial resolution than any sample survey could offer. Most significantly, data such as \ac{AIS} is not restricted to a single country or continent, offering potentially global coverage required for in depth analysis of patterns at a global scale.~\cite{ntakou2014analysis}  In terms of timeliness and frequency, statistics based on big \ac{AIS} data can supplement Official Statistics of very low frequency.~\cite{ntakou2014analysis} The location, capacity of vessels per route and vessel behaviors play a big role. Too many ships in lay up (anchoring a ship in a specific area until the market picks up), number of ships slow steaming across the oceans, congestion at major ports, and increased vessel capacity on specific trade routes are just a few indicators of the state of the market.}

\clearpage
\subsubsection*{Relationship between \ac{CNM} and \acf{PVs}}
\begin{wrapfigure}[35]{r}{0.5\textwidth}
\begin{center}
       \subfloat[][]{%
         \includegraphics[trim=70 98 55 98,clip,width=0.48\textwidth]{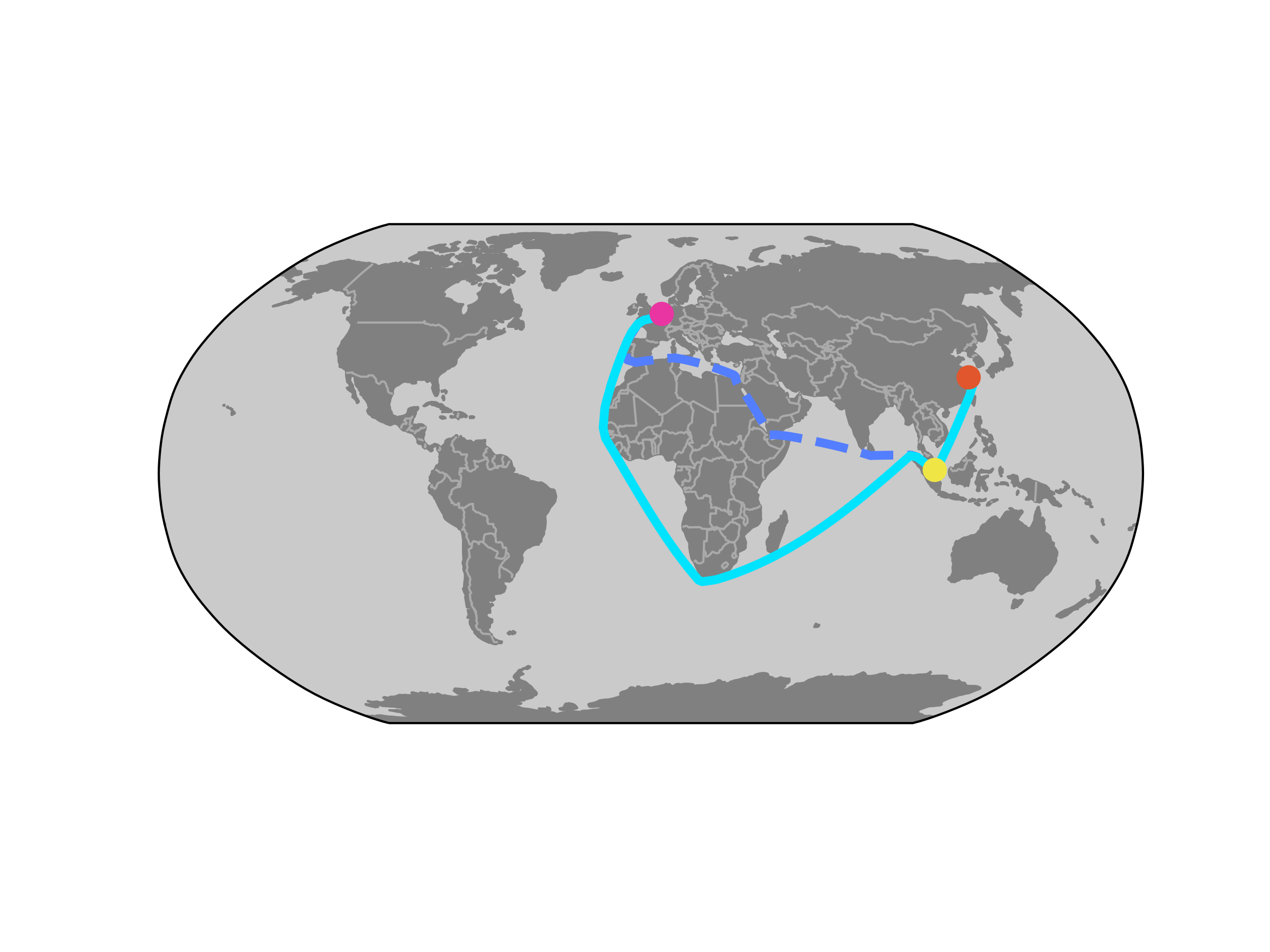}%
         \label{fig:pv-vs-cnm-map}%
       }%
       \hfil%
       \subfloat[][]{%
         \includegraphics[trim=0 0 0 0,clip,width=0.48\textwidth]{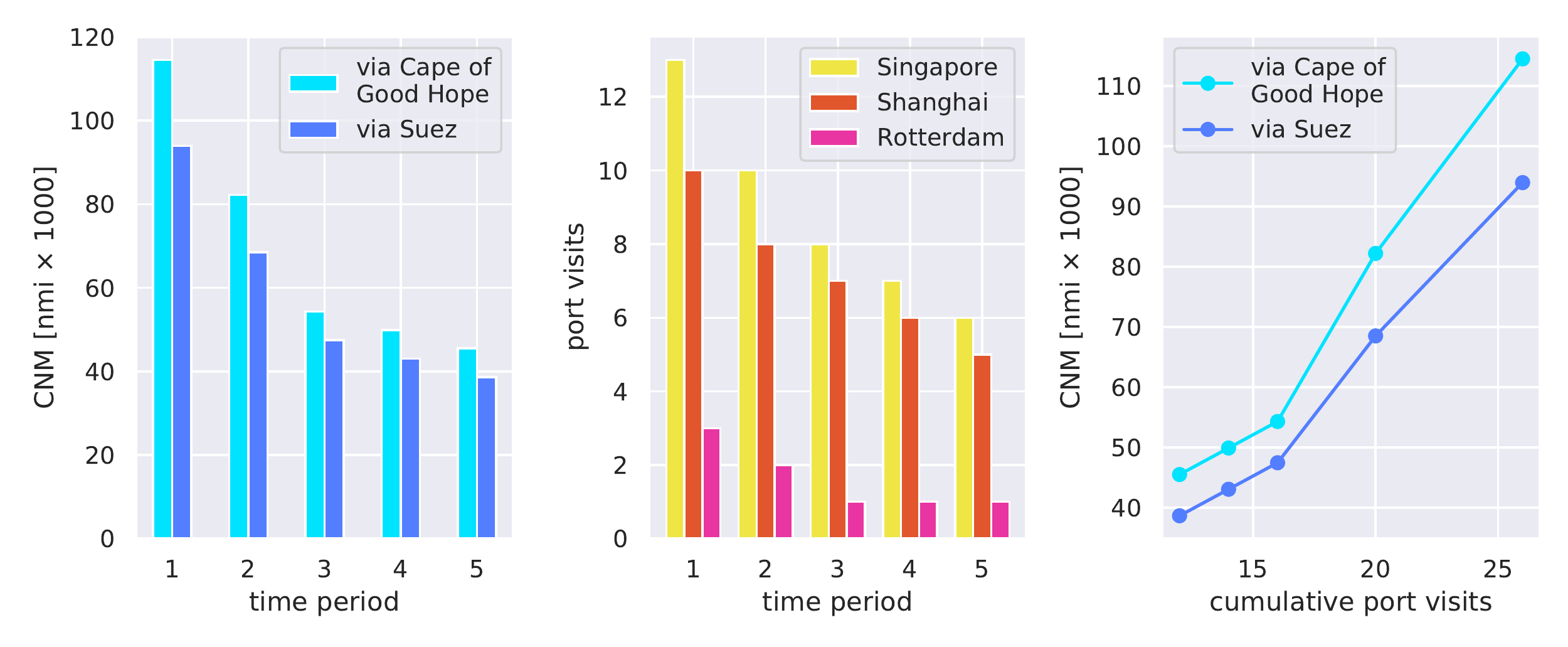}%
         \label{fig:pv-vs-cnm-bars}%
       }%
       \caption{\rev{Relationship between \ac{CNM} and \ac{PVs} in a simulated environment according to equations~(\ref{eq:cnm})-(\ref{eq:pv}). Three ports are considered: Shanghai, Singapore and Rotterdam. Fig.~\protect\subref{fig:pv-vs-cnm-map} shows the location of the ports and their connection on the world map. Two scenarios are implemented: in the first one (solid cyan lines) ships move from Singapore to Rotterdam via the Cape of Good Hope; in the second scenario (dashed blue lines), via the Suez Canal. Fig.~\protect\subref{fig:pv-vs-cnm-bars} reports the outcome of the simulation. The left panel shows the \ac{CNM} indicator over time in the two simulated scenarios; the mid panel depicts the simulated visits in each of the three ports; finally, the right panel reports the empirical relationship between \ac{PVs} and \ac{CNM}.}}%
       \label{fig:pv-vs-cnm-sim}
\end{center}
\end{wrapfigure}

\rev{In light of what above, ship mobility indicators are essential to build reliable and credible trade indicators, as,  indeed, mobility can be seen as a proxy for economic activity. Mobility indicators, such as \ac{PVs}, or port calls, are widely used to assess maritime trade levels, e.g., by UNCTAD in their reports.~\cite{unctad_2017,unctad_2019,unctad_2020,unctad_2020b} 
We based our analysis, instead, on another indicator, the \ac{CNM}. Compared to \ac{PVs}, which only account for the number of journeys connecting one port to another, the \ac{CNM} indicator measures the navigated distance between connected ports. As such, \ac{CNM} is \emph{also} suitable to capture changes in shipping routes and patterns.}

\rev{Clearly, \ac{PVs} and \ac{CNM} are not entirely uncorrelated. On the contrary, it can be shown that there is a high correlation between the two indicators. Indeed, assuming that location of the ports and their connection are known, the relationship between \ac{PVs} and \ac{CNM} can be derived as follows.}
\rev{Let us define the network of port connections, more formally the graph~\cite{godsil2001algebraic} ${\cal G} = \left( {\cal P},{\cal E} \right)$, whose nodes ${\cal P} = \left\{ p_i \right\}_{i=1}^{N}$ correspond to the ports under consideration, and whose edges ${\cal E} \subseteq {\cal P} \times {\cal P}$ represent the possible connections between pairs of ports.
Let $V_{pp^\prime}$ be the number of port visits to $p$ from $p^\prime$, and $D_{pp^\prime}$ the distance between $p$ and $p^\prime$ in nautical miles, with $(p,p^\prime) \in {\cal E}$. Assuming that the navigated distance between two ports is fixed, the relationship between the \ac{CNM} the \ac{PVs} at port $p$, defined as $V_{p}$, is  
\begin{align}
\textrm{CNM} &= \sum_{(p,p^\prime) \in {\cal E}} V_{pp^\prime} D_{pp^\prime}, \label{eq:cnm}\\
V_{p} &= \sum_{p^\prime \in {\cal P}} V_{pp^\prime}.
\label{eq:pv}
\end{align}}
\rev{Alternatively, we can think of $D_{pp^\prime}$ as the mean of a distribution that takes into account the variability of the distance, and the generalization of the relationship is immediate, having defined with $D_{pp^\prime}(k)$ the actual distance navigated between port $p$ and $p^\prime$ on the $k$ -th journey:
\begin{equation}
\textrm{CNM} =  \sum_{(p,p^\prime) \in {\cal E}} \sum_{k=1}^{V_{pp^\prime}} D_{pp^\prime}(k).
\end{equation}}
\rev{These expressions can be simplified further when considering only two ports ($N=2$). 
This simplification is useful to reveal the underlying relationship between \ac{CNM} and \ac{PVs}, which is linear, as it amounts to multiplying the distance $D=D_{12}$ between the two ports by the cumulative number of visits $V=V_{12}+V_{21}$, that is 
\[
\textrm{CNM} = V\,D.
\]
}
\rev{In Fig.~\ref{fig:pv-vs-cnm-sim} we report a simulation example to compare \ac{PVs} and \ac{CNM} according to equations~(\ref{eq:cnm})-(\ref{eq:pv}). We consider three ports: Shanghai, Singapore and Rotterdam. Fig.~\ref{fig:pv-vs-cnm-map} shows the port locations on the map and the connections among them, which are inspired to real shipping routes between North Europe and the Far East. Two scenarios are envisioned in the simulation; in the first one, ships going from Singapore to Rotterdam choose the route around the Cape of Good Hope (cyan); in the second scenario, they use the Suez Canal (dashed blue). Fig.~\ref{fig:pv-vs-cnm-bars} illustrates the outcome of the simulation. The aim is to reproduce the situation that was observed during the lockdown, when a non-negligible number of vessels opted to longer route around the Cape of Good Hope, motivated by low bunker prices and lack of demand in European markets.~\cite{Terpilowski} The chart in the middle shows the \ac{PVs} in each of the three ports, which decrease over time but do not change from one scenario to the other. The left panel in Fig.~\ref{fig:pv-vs-cnm-bars} shows the computed CNM, which obviously change between the two scenarios. Finally, the panel on the right column of Fig.~\ref{fig:pv-vs-cnm-bars} show the empirical relationship between \ac{PVs} and \ac{CNM}, which is approximately linear, as already seen in the formulae above. The simulation in a such a controlled setting conveys a clear message: being highly correlated, \ac{PVs} and \ac{CNM} are both representative of ship mobility; at the same time, the \ac{CNM} is a more informative indicator, which allows to reveal effects that are not observable in the \ac{PVs} indicator.} 

\rev{At the same time, the computation of \ac{PVs} from \ac{AIS} is subject to a series of assumption and approximations. The \ac{AIS} performs reasonably good job to track the position, speed and heading etc. of a vessel. However, the data broadcast by ships does not, by itself, provide any specific information of when a port visit is made. For such an analysis, the data needs to be enriched with other data, such as geospatial data on the location of ports and specific port boundaries. These positions of ports and their operational boundaries are calculated and produced from the data itself and may not be accurate or change as new terminals are built or expanded, or new anchorage areas created. Indeed, \ac{PVs} statistics often focus only on a subset of ports, the most important ones, for each country.}

\rev{Conversely, the computation of \ac{CNM} from \ac{AIS} leverages on the most reliable part of information that is broadcast by ships: their position. Therefore, not only \ac{CNM} have to be considered a more robust indicator for shipping mobility, but also a more informative and capillary one, as it reveals effects that would not be revealed by \ac{PVs}, and is not limited to a subset of ports whose location and boundaries are known, but rather potentially has the same coverage than \ac{AIS}.}


\bibliography{mybib_upd_clean} 
	
\section*{Acknowledgements}
The authors would like to thank Prof. Roar Adland for the initial feedback and comments.

	
\section*{Author contributions statement}
	
L.M.M., P.B. and D.Z. conceptualized the methodology and all the investigations and wrote large sections of the paper. L.M.M., P.B., D.Z. and G.S. were involved in the interpretation of the results. L.M.M. was responsible for the visualization and presentation of the mobility indicators and handled the editing of the manuscript. P.B. was responsible for the literature review and general setting of the manuscript. G.S. was responsible for data curation and the computation of mobility indicators. S.M., P.K.W. and S.C. contributed to the writing of the manuscript. All authors discussed and commented on the methods, results and content of the manuscript; all authors also reviewed the manuscript.

	

\section*{Competing interest}
The authors declare no competing interests.

	
	

	
\end{document}